\documentclass[12pt]{article}
\usepackage{amsmath,amsfonts,amssymb}
\usepackage{pstricks}
\numberwithin{equation}{section}

\newcommand{\nc}{\newcommand}
\nc{\fh}{\hat{f}}
\nc{\muh}{\hat{\mu}}
\nc{\nuh}{\hat{\nu}}
\nc{\bib}{\bibitem}
\nc{\al}{\alpha}
\nc{\mch}{\mathrm{ch}}
\nc{\g}{\gamma}
\nc{\G}{\Gamma}
\nc{\D}{\Delta}
\nc{\eps}{\epsilon}
\nc{\la}{\lambda}
\nc{\La}{\Lambda}
\nc{\var}{\varphi}
\nc{\pa}{\partial}
\nc{\nn}{\nonumber \\ }
\nc{\hf}{\frac{1}{2}}
\nc{\dz}{\frac{dz}{2\pi i}}
\nc{\bin}[2]{\left(\!\!\!\begin{array}{c} {#1}\\ {#2} \end{array}\!\!\!\right)}
\nc{\be}{\begin{equation}}
\nc{\ee}{\end{equation}}
\nc{\bea}{\begin{eqnarray}}
\nc{\eea}{\end{eqnarray}}
\nc{\bra}[1]{\langle {#1}|}
\nc{\ket}[1]{|{#1}\rangle}
\def\Re{\mathop{\rm Re}\nolimits}
\def\Im{\mathop{\rm Im}\nolimits}
\nc{\chit}{\raisebox{0.25ex}{$\chi$}}
\nc{\Bb}{\mbox{\boldmath $B$}}
\nc{\Fb}{\mbox{\boldmath $F$}}
\nc{\Hb}{\mbox{\boldmath $H$}}
\nc{\Ib}{\mbox{\boldmath $I$}}
\nc{\Jb}{\mbox{\boldmath $J$}}
\nc{\Rb}{\mbox{\boldmath $R$}}
\nc{\Tb}{\mbox{\boldmath $T$}}

\nc{\Hc}{{\cal H}}
\nc{\Rc}{{\cal R}}
\nc{\Lc}{{\cal L}}
\nc{\Vc}{{\cal V}}
\nc{\rbar}{\bar{r}}
\nc{\qbar}{\bar{q}}
\nc{\kbar}{\bar{k}}
\def\floor#1{\lfloor #1\rfloor}
\nc{\sbin}[2]{\left\{\!\!\!\begin{array}{c} {#1}\\ {#2} \end{array}\!\!\!\right\}}
\nc{\sbinlr}[2]{\Big\langle\!\!\begin{array}{c} {#1}\\ {#2} \end{array}\!\!\Big\rangle}
\def\vvdots{\mathinner{\mkern1mu\raise1pt\vbox{\kern7pt\hbox{.}}\mkern2mu
  \raise4pt\hbox{.}\mkern2mu\raise7pt\hbox{.}\mkern1mu}}
\nc{\gauss}[2]{\left[\!\!\begin{array}{c} {#1}\\ {#2} \end{array}\!\!\right]_{\!q}}
\nc{\gaussbar}[2]{\left[\!\!\begin{array}{c} {#1}\\ {#2} \end{array}\!\!\right]_{\!\qbar}}
\nc{\perm}[2]{\left[\!\!\begin{array}{c} {#1}\\ \\ {#2} \end{array}\!\!\right]}
\nc{\bino}[2]{\left(\!\!\begin{array}{c} {#1}\\ {#2} \end{array}\!\!\right)}
\def\half {\mbox{$\textstyle \frac{1}{2}$}}
\def\vec#1{\mbox {\boldmath $#1$}}
\def\svec#1{\mbox {\scriptsize\boldmath $#1$}}
\definecolor{lightblue}{rgb}{.7,.7,1}
\definecolor{purple}{rgb}{1,0,1}
\definecolor{lightlightblue}{rgb}{.85,.85,1}

\def\leftarc#1{\psarc[linecolor=blue,linewidth=1.5pt]#1{.5}{90}{270}}
\def\rightarc#1{\psarc[linecolor=blue,linewidth=1.5pt]#1{.5}{-90}{90}}
\def\loopa{
\psframe[linewidth=.25pt](0,0)(1,1)
\psarc[linewidth=1.5pt,linecolor=blue](1,0){.5}{90}{180}
\psarc[linewidth=1.5pt,linecolor=blue](0,1){.5}{-90}{0}
}
\def\loopb{
\psframe[linewidth=.25pt](0,0)(1,1)
\psarc[linewidth=1.5pt,linecolor=blue](0,0){.5}{0}{90}
\psarc[linewidth=1.5pt,linecolor=blue](1,1){.5}{180}{270}
}

\def\facegrid#1#2{
\psframe[fillstyle=solid,fillcolor=lightlightblue,linewidth=0pt]#1#2
\psgrid[gridlabels=0pt,subgriddiv=1]#1#2}

\def\conn#1#2{
\psframe[linewidth=1pt,fillstyle=solid,fillcolor=lightlightblue]#1#2}
\definecolor{sky}{rgb}{.4,.8,1}
\definecolor{orange}{rgb}{1,.4,0}
\definecolor{green}{rgb}{0,1,0}
\definecolor{brightblue}{rgb}{0,1,1}
\definecolor{mediumblue}{rgb}{.34,.446,1}
\definecolor{greyblue}{rgb}{.345,.345,.7424}
\definecolor{bglightblue}{rgb}{.625,.845,.917}
\definecolor{graphblue}{rgb}{.719,.918,.996}
\definecolor{lightlightblue}{rgb}{.85,.85,1}
\definecolor{lightpurple}{rgb}{1,.65,1}
\definecolor{WeekColor}{rgb}{1,.7,.9}
\definecolor{lightblue}{rgb}{.55,.55,1}
\definecolor{midblue}{rgb}{.7,.7,1}
\definecolor{lightlightblue}{rgb}{.85,.85,1}
\definecolor{lightestblue}{rgb}{.96,.96,1}

\def\qbin#1#2#3{{\genfrac{[}{]}{0pt}{}{#1}{#2}}_{#3}}
\def\sqbin#1#2#3{\mbox{$\genfrac{[}{]}{0pt}{}{#1}{#2}$}_{#3}}
\def\sc#1{\mbox{\scriptsize $#1$}}
\def\sm#1{\mbox{\small $#1$}}
\def\disp{\displaystyle}

\begin{document}

\topmargin -5mm
\oddsidemargin 5mm

\begin{titlepage}
\setcounter{page}{0}

\vspace{8mm}
\begin{center}
{\huge {\bf Solvable Critical Dense Polymers\\[6pt] on the Cylinder}}

\vspace{10mm}
{\Large Paul A. Pearce,\ \ J{\o}rgen Rasmussen,\ \ Simon P. Villani}\\[.3cm]
{\em Department of Mathematics and Statistics, University of Melbourne}\\
{\em Parkville, Victoria 3010, Australia}\\[.4cm]
P.Pearce@ms.unimelb.edu.au, \ J.Rasmussen@ms.unimelb.edu.au
\\[.05cm] 
S.Villani@ms.unimelb.edu.au

\end{center}

\vspace{10mm}
\centerline{{\bf{Abstract}}}
\vskip.4cm
\noindent
A lattice model of critical dense polymers is solved exactly on a cylinder with finite circumference. 
The model is the first member ${\cal LM}(1,2)$ of the Yang-Baxter integrable series of 
logarithmic minimal models. The cylinder topology allows for non-contractible loops with 
fugacity $\alpha$ that wind around the cylinder or for an arbitrary number $\ell$ of defects
that propagate along the full length of the cylinder. 
Using an enlarged periodic Temperley-Lieb algebra, we set up commuting transfer 
matrices acting on states whose links are considered distinct 
with respect to connectivity around the front or back of the cylinder.  
These transfer matrices satisfy a functional equation in the form of an inversion identity. 
For even $N$, this involves a non-diagonalizable braid operator $\vec J$ and 
an involution ${\vec R}=-(\vec J^3-12\vec J)/16=(-1)^{\svec F}$ with eigenvalues $R=(-1)^{\ell/2}$. 
This is reminiscent of supersymmetry with a pair of defects interpreted as a fermion. 
The number of defects $\ell$ thus separates the theory into 
Ramond ($\ell/2$ even), Neveu-Schwarz ($\ell/2$ odd) and $\mathbb{Z}_4$ ($\ell$ odd) sectors.
For the case of loop fugacity $\alpha=2$, 
the inversion identity is solved exactly sector by sector for the eigenvalues in finite geometry. 
The eigenvalues are classified by the physical combinatorics 
of the patterns of zeros in the complex spectral-parameter plane.
This yields selection rules for the physically relevant solutions to 
the inversion identity. 
The finite-size corrections are obtained from Euler-Maclaurin formulas.
In the scaling limit, we obtain the conformal partition functions as sesquilinear forms and 
confirm the central charge $c=-2$ and conformal
weights $\Delta,\bar\Delta=\Delta_t=(t^2-1)/8$. Here $t=\ell/2$ and $t=2r-s\in\mathbb{N}$ in the 
$\ell$ even sectors with Kac labels $r=1,2,3,\ldots; s=1,2$ while $t\in\mathbb{Z}-\half$ in the 
$\ell$ odd sectors. Strikingly, the $\ell/2$ odd sectors exhibit a ${\cal W}$-extended 
symmetry but the $\ell/2$ even sectors do not. Moreover, the naive trace summing over all 
$\ell$ even sectors does not yield a modular invariant.

\end{titlepage}
\newpage
\renewcommand{\thefootnote}{\arabic{footnote}}
\setcounter{footnote}{0}

\tableofcontents

\newpage
\section{Introduction}

Familiar materials such as plastics, nylon, polyester and plexiglass 
are made from polymers. Polymers~\cite{polymers} consist of very long chain molecules 
with a large number of repeating structural units called monomers. 
Polymers exist in low- or high-temperature phases which are 
characterised as either dense or dilute. Polymers are {\em dense\/} 
if they fill a finite (non-zero) fraction of the available volume in 
the thermodynamic limit. 

The modern era of two-dimensional polymer theory began in the late eighties~\cite{Saleur87,Duplantier,SaleurSUSY} 
when Saleur and Duplantier initiated the study of polymers as a conformal field theory (CFT). 
From the viewpoint of lattice 
statistical mechanics, polymers are of interest as a prototypical 
example of a system involving (extended) non-local degrees of 
freedom. It might be expected that the non-local nature of these 
degrees of freedom has a profound effect on the associated CFT obtained in the continuum scaling limit. Indeed, 
the associated CFT is in fact {\em logarithmic\/}~\cite{Gurarie} in the sense that, for certain 
representations, the Virasoro dilatation 
generator $L_0$ is non-diagonalizable and exhibits 
non-trivial $2\times 2$ Jordan blocks.

In this paper, we consider a Yang-Baxter integrable model of critical dense polymers on 
a cylinder, both on the lattice and in the continuum scaling limit. 
In fact, this model is the first member ${\cal LM}(1,2)$ of the infinite series of 
logarithmic minimal models ${\cal LM}(p,p')$~\cite{PRZ}. 
An alternative approach to logarithmic CFT based on quantum spin chains appears in \cite{RS0701}.
The fugacity of contractible loops for ${\cal LM}(1,2)$ is $\beta=0$. 
Previously~\cite{PR0610}, we considered this model on finite-width strips with various 
boundary conditions. The associated integrals of motion and Baxter's $Q$ matrix have 
been considered in~\cite{Nigro09}.
The solvable critical dense polymer model on a {\em cylinder} is built from a (locally)
planar version~\cite{Jones} of the periodic Temperley-Lieb algebra~\cite{perTL}.
Because of the non-local degrees of freedom, logarithmic theories are sensitive
to the topology. So changing the topology from a strip to a cylinder has profound effects. 
Specifically, a cylinder topology allows for non-contractible loops with fugacity $\alpha$
that wind around the cylinder 
or for defects that propagate along the full length of the cylinder. 
These can have dramatic effects on the properties of the model. 
The number of defects $\ell$ is a quantum number that separates the theory into 
Ramond ($\ell/2$ even), Neveu-Schwarz ($\ell/2$ odd) and $\mathbb{Z}_4$ ($\ell$ odd) sectors. 
Following \cite{SaleurSUSY,ReSa01}, we use the terminology of supersymmetry even though 
we do not claim any superconformal symmetry in our model.

Remarkably, as for the square lattice Ising model~\cite{BaxBook}, 
the commuting single-row transfer 
matrices of this model satisfy a simple functional equation in the 
form of an inversion identity. This enables us to solve for the exact eigenvalues of the 
transfer matrices on a finite lattice for $\alpha=2$. The conformal spectra are readily 
accessible from finite-size corrections. In particular, in the 
continuum scaling limit, we obtain the partition functions as sesquilinear forms and confirm 
the central charge $c=-2$ and conformal
weights $\Delta,\bar\Delta=\Delta_t=(t^2-1)/8$. Here $t=\ell/2$ and $t=2r-s\in\mathbb{N}$ in the 
$\ell$ even sectors with Kac labels $r=1,2,3,\ldots; s=1,2$ while $t\in\mathbb{Z}-\half$ in the 
$\ell$ odd sectors. On the strip, this model admits a ${\cal W}$-extended conformal 
algebra and is identified~\cite{PRR0803} with {\em symplectic fermions}~\cite{Kau00}. 
Strikingly, on the cylinder, we find that this extended symmetry only holds in the sum over 
$\ell/2$ odd sectors and not in the sum over $\ell/2$ even sectors.
Moreover, the naive trace summing over all 
even $N$ sectors does not yield the known modular invariant~\cite{ModInv} of the $c_{1,2}$ triplet model.

The layout of this paper is as follows. In Section~\ref{SecLattice}, we define the solvable  
critical dense polymer lattice model. We also discuss the periodic Temperley-Lieb (TL) algebra, 
its enlargement by adding the shift operators $\Omega$ and $\Omega^{-1}$, and its relation to the 
{\em cylinder} TL algebra which is a direct generalization of the planar TL
algebra of Jones~\cite{Jones}. In Section~\ref{SecCylTransfer}, we define the 
single-row transfer matrices 
directly in the cylinder TL algebra. We also define the vector spaces of link 
states on which these transfer matrices act and relate these to the 
cases of {\em distinct} (DC) and {\em identified} (IC) connectivities~\cite{PRGN02}. 
The inversion identities are derived in Section~\ref{SecInversion}, first in the setting of the 
cylinder TL algebra and then as matrix inversion identities. Certain details are deferred to
Appendix~\ref{AppJ}.
The inversion identities are solved sector by sector for the transfer matrix eigenvalues 
in Section~\ref{SecSolution}, while the finite-size corrections are extracted in Section~\ref{EulerM}. 
In Section~\ref{SecSelection}, the physically relevant solutions are obtained empirically and 
encoded by applying physical combinatorics supplemented by selection rules. 
Finitized conformal partition functions are obtained as sesquilinear forms in finitized characters~\cite{FinChar}. 
The conformal partition functions arising in the continuum scaling limit are
discussed in Section~\ref{SecConformal}. We conclude with 
some remarks and directions for future research in Section~\ref{SecConclusion}.

\section{Lattice Model}
\label{SecLattice}

\subsection{Critical dense polymers}

To model critical dense polymers on a cylinder with a finite circumference, we consider a square 
lattice on a strip, with $N$ columns and $M$ rows of faces, and identify the left and right edges 
as shown in Figure~1. 
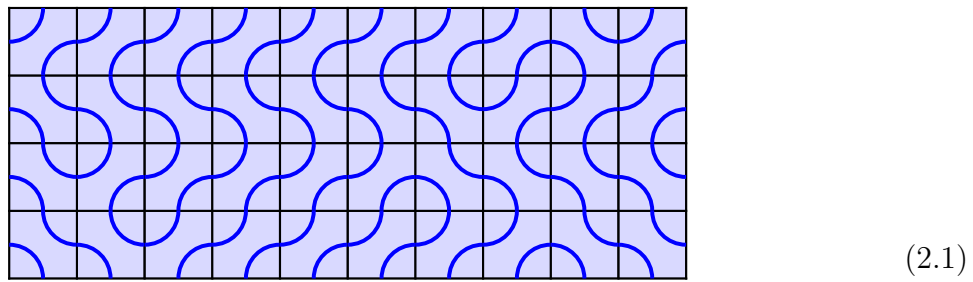
\begin{figure}[h]
\psset{unit=.9cm}
\setlength{\unitlength}{.9cm}
\be
\begin{pspicture}[shift=-.45](0,-.3)(10,4.4)
\facegrid{(0,0)}{(10,4)}
\rput[bl](0,0){\loopb}
\rput[bl](1,0){\loopb}
\rput[bl](2,0){\loopa}
\rput[bl](3,0){\loopa}
\rput[bl](4,0){\loopa}
\rput[bl](5,0){\loopa}
\rput[bl](6,0){\loopa}
\rput[bl](7,0){\loopa}
\rput[bl](8,0){\loopb}
\rput[bl](9,0){\loopb}
\rput[bl](0,1){\loopb}
\rput[bl](1,1){\loopa}
\rput[bl](2,1){\loopa}
\rput[bl](3,1){\loopa}
\rput[bl](4,1){\loopa}
\rput[bl](5,1){\loopa}
\rput[bl](6,1){\loopb}
\rput[bl](7,1){\loopb}
\rput[bl](8,1){\loopb}
\rput[bl](9,1){\loopb}
\rput[bl](0,2){\loopb}
\rput[bl](1,2){\loopb}
\rput[bl](2,2){\loopb}
\rput[bl](3,2){\loopb}
\rput[bl](4,2){\loopb}
\rput[bl](5,2){\loopb}
\rput[bl](6,2){\loopb}
\rput[bl](7,2){\loopa}
\rput[bl](8,2){\loopa}
\rput[bl](9,2){\loopa}
\rput[bl](0,3){\loopa}
\rput[bl](1,3){\loopa}
\rput[bl](2,3){\loopa}
\rput[bl](3,3){\loopa}
\rput[bl](4,3){\loopa}
\rput[bl](5,3){\loopa}
\rput[bl](6,3){\loopa}
\rput[bl](7,3){\loopa}
\rput[bl](8,3){\loopb}
\rput[bl](9,3){\loopa}
\end{pspicture}
\label{typconf}
\ee
\caption{A typical dense polymer configuration on a $10\times 4$ cylindrical lattice. 
The left and right edges are identified to form a cylinder. In this case, the circumference is 
$N=10$ and $M=4$. No local closed loops are formed on the surface of the cylinder in accord with 
the vanishing loop fugacity $\beta=0$. It is, however, possible to allow non-contractible 
loops encircling the cylinder with fugacity $\alpha\ne 0$.}
\end{figure}

An elementary face of the lattice can assume one of two configurations 
with different statistical weights
\psset{unit=.9cm}
\setlength{\unitlength}{.9cm}
\be
\begin{pspicture}[shift=-.6](-.25,-.25)(1.25,1.25)
\facegrid{(0,0)}{(1,1)}
\put(0,0){\loopa}
\end{pspicture}
\quad\mbox{or}\quad
\begin{pspicture}[shift=-.6](-.25,-.25)(1.25,1.25)
\facegrid{(0,0)}{(1,1)}
\put(0,0){\loopb}
\end{pspicture}
\ee
where the arcs represent local segments of polymers. 
The two possible configurations can be combined into a single face operator as
\psset{unit=.9cm}
\setlength{\unitlength}{.9cm}
\be
X(u)\;=\!\!\begin{pspicture}[shift=-.45](-.5,-.1)(1.25,1.1)
\facegrid{(0,0)}{(1,1)}
\psarc[linewidth=.5pt](0,0){.15}{0}{90}
\rput(.5,.5){\small $u$}
\end{pspicture}
=\ \cos u\!\!
\begin{pspicture}[shift=-.45](-.5,-.1)(1.25,1.1)
\facegrid{(0,0)}{(1,1)}
\put(0,0){\loopa}
\end{pspicture}
+\ \sin u\!\!
\begin{pspicture}[shift=-.45](-.5,-.1)(1.25,1.1)
\facegrid{(0,0)}{(1,1)}
\put(0,0){\loopb}
\end{pspicture}
\label{u}
\ee
where $u$ is the {\em spectral parameter} related to spatial anisotropy. The lower left corner 
is marked to fix the orientation of the square.

Since the polymer segments pass uniformly through each face, this is a model of 
{\em dense\/} polymers --- in the continuum scaling limit, a polymer is space-filling and 
has fractal dimension 2. The non-local degrees of freedom correspond to a number of polymers. 
It is often convenient to think of these degrees of freedom as non-local connectivities. 

Critical dense polymers corresponds to the first member ${\cal LM}(1,2)$ of the infinite 
series ${\cal LM}(p,p')$ of logarithmic minimal models~\cite{PRZ}.
Each logarithmic model is characterized by a crossing parameter
$\lambda=\frac{p'-p}{p'}\,\pi$ related to the loop fugacity $\beta$ by
\be
  \beta\;=\;2\cos\lambda\;=\;2\cos\frac{p'-p}{p'}\,\pi,\qquad \mbox{$p,p'$ coprime}
\label{beta}
\ee
It was argued in~\cite{PRZ} that the scaling limits of these integrable lattice
models yield logarithmic CFTs. In the case of critical dense polymers,
\be
  \la\;=\;\frac{\pi}{2},\qquad\quad \beta\ =\ 0
\label{la}
\ee
implying that local contractible loops are not allowed.

\subsection{Cylinder Temperley-Lieb algebra}

For the purposes of this paper, the {\em cylinder} Temperley-Lieb (TL) algebra is a 
diagrammatic algebra built up from elementary faces.
The faces are connected such that the midpoint of an outer edge of a 
face, called a node,
can be linked to a node of
any other (or even the same)
face as long as the total set of links make up a {\em 
non-intersecting} web of connections on the surface of a cylinder. The cylinder TL algebra is equivalent 
to the annular algebra of Jones~\cite{Jones}. 

Two basic local properties of the cylinder TL algebra are the inversion relation
\psset{unit=.7cm}
\setlength{\unitlength}{.7cm}
\be
\begin{pspicture}[shift=-1.1](-.5,0.75)(4,3.25)
\pspolygon[fillstyle=solid,fillcolor=lightlightblue](0,2)(1,1)(2,2)(1,3)(0,2)
\pspolygon[fillstyle=solid,fillcolor=lightlightblue](2,2)(3,1)(4,2)(3,3)(2,2)
\psarc[linewidth=.5pt](0,2){.15}{-45}{45}
\psarc[linewidth=.5pt](2,2){.15}{-45}{45}
\psarc[linecolor=blue,linewidth=1.5pt](2,2){.7}{45}{135}
\psarc[linecolor=blue,linewidth=1.5pt](2,2){.7}{-135}{-45}
\rput(1,2){\small $v$}
\rput(3,2){\small $\!-v$}
\end{pspicture}
  \ \ =\ \cos^2\!v\ \
\begin{pspicture}[shift=-1.1](1,0.75)(4,3.25)
\pspolygon[fillstyle=solid,fillcolor=lightlightblue](1,2)(2,1)(3,2)(2,3)(1,2)
\psarc[linecolor=blue,linewidth=1.5pt](2,1){.7}{45}{135}
\psarc[linecolor=blue,linewidth=1.5pt](2,3){.7}{-135}{-45}
\end{pspicture}
\label{Inv}
\ee
and the Yang-Baxter equation (YBE)~\cite{BaxBook}
\psset{unit=.8cm}
\setlength{\unitlength}{.8cm}
\be
\begin{pspicture}[shift=-1.1](-.5,0.75)(4,3.25)
\facegrid{(2,1)}{(3,3)}
\pspolygon[fillstyle=solid,fillcolor=lightlightblue](0,2)(1,1)(2,2)(1,3)(0,2)
\psarc[linewidth=.5pt](0,2){.15}{-45}{45}
\psline[linecolor=blue,linewidth=1.5pt](1.5,1.5)(2,1.5)
\psline[linecolor=blue,linewidth=1.5pt](1.5,2.5)(2,2.5)
\psarc[linewidth=.5pt](2,1){.15}{0}{90}
\psarc[linewidth=.5pt](2,2){.15}{0}{90}
\rput(2.5,1.5){\small $u$}
\rput(2.5,2.5){\small $v$}
\rput(1,2){\small $u-v$}
\end{pspicture}
  \!\!\! =\
\begin{pspicture}[shift=-1.1](-.5,0.75)(4,3.25)
\facegrid{(0,1)}{(1,3)}
\pspolygon[fillstyle=solid,fillcolor=lightlightblue](1,2)(2,1)(3,2)(2,3)(1,2)
\psarc[linewidth=.5pt](1,2){.15}{-45}{45}
\psline[linecolor=blue,linewidth=1.5pt](1,1.5)(1.5,1.5)
\psline[linecolor=blue,linewidth=1.5pt](1,2.5)(1.5,2.5)
\psarc[linewidth=.5pt](0,1){.15}{0}{90}
\psarc[linewidth=.5pt](0,2){.15}{0}{90}
\rput(0.5,1.5){\small $v$}
\rput(0.5,2.5){\small $u$}
\rput(2,2){\small $u-v$}
\end{pspicture}
\label{YB}
\ee
These are identities for 2- and 3-tangles, respectively, where a $k$-tangle is an
arrangement of faces with $2k$ free nodes. The identities are established
by writing out all the possible configurations, while keeping track of the
associated weights, and collecting them in classes according to
connectivities. The left side of (\ref{Inv}), for example, thus corresponds
to a sum of four terms of which one vanishes since $\beta=0$. The 
remaining three terms fall into the two connectivity classes
\psset{unit=.6cm}
\setlength{\unitlength}{.6cm}
\be
\begin{pspicture}[shift=-1.1](-.5,0.75)(4,3.25)
\pspolygon[fillstyle=solid,fillcolor=lightlightblue](0,2)(1,1)(2,2)(1,3)(0,2)
\pspolygon[fillstyle=solid,fillcolor=lightlightblue](2,2)(3,1)(4,2)(3,3)(2,2)
\psarc[linecolor=blue,linewidth=1.5pt](1,3){.7}{-135}{-45}
\psarc[linecolor=blue,linewidth=1.5pt](2,2){.7}{-135}{135}
\psarc[linecolor=blue,linewidth=1.5pt](1,1){.7}{45}{135}
\psarc[linecolor=blue,linewidth=1.5pt](4,2){.7}{135}{225}
\end{pspicture}
\quad\! = 
\begin{pspicture}[shift=-1.1](-.5,0.75)(4,3.25)
\pspolygon[fillstyle=solid,fillcolor=lightlightblue](0,2)(1,1)(2,2)(1,3)(0,2)
\pspolygon[fillstyle=solid,fillcolor=lightlightblue](2,2)(3,1)(4,2)(3,3)(2,2)
\psarc[linecolor=blue,linewidth=1.5pt](2,2){.7}{45}{315}
\psarc[linecolor=blue,linewidth=1.5pt](0,2){.7}{-45}{45}
\psarc[linecolor=blue,linewidth=1.5pt](3,1){.7}{45}{135}
\psarc[linecolor=blue,linewidth=1.5pt](3,3){.7}{-135}{-45}
\end{pspicture}
\qquad\ \ \mbox{and}\qquad
\begin{pspicture}[shift=-1.1](-.5,0.75)(4,3.25)
\pspolygon[fillstyle=solid,fillcolor=lightlightblue](0,2)(1,1)(2,2)(1,3)(0,2)
\pspolygon[fillstyle=solid,fillcolor=lightlightblue](2,2)(3,1)(4,2)(3,3)(2,2)
\psarc[linecolor=blue,linewidth=1.5pt](2,2){.7}{45}{135}
\psarc[linecolor=blue,linewidth=1.5pt](2,2){.7}{-135}{-45}
\psarc[linecolor=blue,linewidth=1.5pt](1,1){.7}{45}{135}
\psarc[linecolor=blue,linewidth=1.5pt](1,3){.7}{-135}{-45}
\psarc[linecolor=blue,linewidth=1.5pt](3,1){.7}{45}{135}
\psarc[linecolor=blue,linewidth=1.5pt](3,3){.7}{-135}{-45}
\end{pspicture}
\ee
The weights accompanying the two equivalent configurations cancel
since $\cos v\sin(-v)+\sin v\cos(-v)=0$, while the last diagram
comes with the weight $\cos v\cos(-v)$ thereby yielding the identity 
(\ref{Inv}).

Particular elements of the cylinder TL algebra are the {\em shift} or {\em winding}
operator $\Omega$ and its inverse $\Omega^{-1}$ 
\psset{unit=.8cm}
\setlength{\unitlength}{.8cm}
\bea
\Omega\ =\
\begin{pspicture}[shift=-.4](.25,1)(8,2)
\multirput(.5,1)(1,0){8}{
\facegrid{(0,0)}{(1,1)}
\put(0,0){\loopa}
}
\end{pspicture}\\[18pt]
\Omega^{-1}=\,
\begin{pspicture}[shift=-.4](.25,1)(8,1.5)
\multirput(.5,1)(1,0){8}{
\facegrid{(0,0)}{(1,1)}
\put(0,0){\loopb}
}
\end{pspicture}
\label{Omega}
\eea
The multiplication implied in
\be
 \Omega\,\Omega^{-1}\;=\;\Omega^{-1}\Omega\;=\;\Ib
\label{OOI}
\ee
is vertical concatenation, and $\Ib$ 
is the (vertical) identity operator linking every node on the upper horizontal
edge to the node directly below it on the lower horizontal edge.

\subsection{Periodic Temperley-Lieb algebra and its enlargement}

The {\em periodic} (affine) TL algebra~\cite{perTL} of size $N$ is generated by the 
identity $I$ and the generators $e_j$
\be
 {\cal T\!L}(N)\;=\;\langle I, e_0, e_1, \ldots, e_{N-1}\rangle
\ee
subject to the periodicity constraints
\be
 e_j\;\equiv\;e_{j\,\text{mod}\,N},\qquad j\in\mathbb{Z}
\label{eeN}
\ee
and the relations
\bea
 &e_j^2\;=\;\beta e_j,\qquad e_j e_{j\pm 1} e_j\;=\;e_j,\qquad j=0,1,\ldots, N-1&
\label{eebe}
\\[4pt]
 &e_j e_k\;=\;e_k e_j,\qquad j-k\ne 0,\pm 1\,\text{mod}\,N&
\label{eeee}
\eea
where $\beta$ is the fugacity of contractible loops. 
The TL generators $e_j$ are represented diagrammatically by monoids~\cite{Kauffman}.
The face operator at position $j$ is given by
\be
 X_j(u)\;=\;\cos u\,I+\sin u\,e_j
\label{faceop}
\ee
and obtained by rotating the face operator in (\ref{u}) by $45$ degrees in the 
counter-clockwise direction. The periodic TL algebra is infinite dimensional.

For $N$ even, one introduces the combinations
\be
 E\;=\;e_0 e_2 e_4 \cdots e_{N-2},\quad F\;=\;e_1 e_3 e_5\cdots e_{N-1},\quad 
 E^2\;=\;\beta^{\frac{N}{2}} E,\quad F^2\;=\;\beta^{\frac{N}{2}} F
\ee
Letting $\alpha$ denote the fugacity of {\em non-contractible} loops,
these can be `removed' in pairs 
\be
 EFE\;=\;\alpha^2 E,\qquad FEF\;=\;\alpha^2 F
\ee
For $N$ odd, non-contractible loops cannot appear.

The elements $\Omega$ and $\Omega^{-1}$ (\ref{Omega}) of the cylinder TL algebra
are {\em not} elements of the periodic TL algebra. They may be included~\cite{Levy}, though, thereby
enlarging the periodic TL algebra.
Translation on the TL generators is then implemented by conjugation
\be
 e_{j-1}\;=\;\Omega\,e_j\,\Omega^{-1}
\ee
so that the enlarged TL algebra is generated by three independent generators 
\be
 {\cal ET\!L}(N)\;=\;\langle e_0, \Omega, \Omega^{-1}\rangle
\ee
The periodicity constraints (\ref{eeN}) and the relations (\ref{eebe}) and (\ref{eeee})
now read 
\bea
 & \Omega^N e_0\Omega^{-N}\;=\; e_0&
 \\[4pt]
 &e_0^2\;=\;\beta e_0,\qquad e_0\Omega^{\mp1}e_0\Omega^{\pm1}e_0\;=\; e_0&
 \\[4pt]
 &e_0\Omega^j e_0\Omega^{-j}\;=\;\Omega^j e_0\Omega^{-j}e_0,\qquad j=2,\ldots,N-2&
\eea
For $N$ even, we furthermore have
\bea
 &&E\;=\;\big(e_0\Omega^{-2}\big)^{\!\frac{N}{2}}\Omega^N
   \;=\;\Omega^{-N}\big(\Omega^2 e_0\big)^{\!\frac{N}{2}}
 \qquad
\\[4pt]
 &&F\;=\;\Omega^{-1}\big(e_0\Omega^{-2}\big)^{\!\frac{N}{2}}\Omega^{N+1}
  \;=\;\Omega^{-N-1}\big(\Omega^2 e_0\big)^{\!\frac{N}{2}}\Omega
 \\[4pt]
 &&E\;=\;\Omega\,F\,\Omega^{-1},
 \qquad\quad
  F\;=\;\Omega^{-1}\,E\,\Omega
\eea
and
\be
 E\Omega^{\pm1}E\;=\;\al E,\qquad
 F\Omega^{\pm1}F\;=\;\al F
\ee
indicating that non-contractible loops can now be removed one by one.

\section{Cylinder Transfer Matrix}
\label{SecCylTransfer}

\subsection{Single-row transfer matrix}

Having introduced the cylinder TL algebra, we now define diagrammatically
the single-row $N$-tangle
\psset{unit=1cm}
\setlength{\unitlength}{1cm}
\be
  \Tb(u)\ =\!\!\!
\begin{pspicture}[shift=-.6](-.5,.75)(8.5,2.2)
\facegrid{(0,1)}{(8,2)}
\multirput(.5,1)(1,0){8}{\psline[linewidth=1pt](0,0)(0,-.2)}
\multirput(.5,2)(1,0){8}{\psline[linewidth=1pt](0,0)(0,.2)}
\rput(0.5,1.5){\small $u$}
\rput(7.5,1.5){\small $u$}
\rput(2.5,1.5){\small $\dots$}
\rput(5.5,1.5){\small $\dots$}
\psarc[linewidth=.5pt](0,1){.15}{0}{90}
\psarc[linewidth=.5pt](7,1){.15}{0}{90}
\end{pspicture}
\label{TM}
\ee
consisting of $N$ faces where the dependence on $N$ is suppressed. 
The left and right edges are identified in accord with the periodicity of the cylinder. 
As discussed below, $\Tb(u)$ has a natural matrix representation when acting vertically 
from below on a given set of periodic link states.
We thus refer to it as the (single-row) ``{transfer matrix}", even though
it is defined as a cylinder $N$-tangle without reference to any matrix representation.
The shift operator $\Omega$ and its inverse $\Omega^{-1}$ enter naturally as the limits
\be
 \lim_{u\rightarrow0}\Tb(u)\;=\;\Omega,\qquad
 \lim_{u\rightarrow\lambda}\Tb(u)\;=\;\Omega^{-1}
\label{limD}
\ee

Using standard diagrammatic arguments~\cite{BaxBook}, it follows that $\Tb(u)$ 
gives rise to a commuting family of transfer matrices where 
\be
 [\Tb(u),\Tb(v)]\:=\:0
\label{TT0}
\ee 
As in (\ref{OOI}), the implied multiplication in the commutator means 
vertical concatenation of the two $N$-tangles in the cylinder TL algebra.
It follows, in particular, that $[\Tb(u),\Omega^{\pm1}]=0$.

\subsection{Hamiltonian limit}

The Hamiltonian limit of the transfer matrix $\Tb(u)$ is 
defined in the cylinder TL algebra as the $N$-tangle appearing as the first sub-leading 
term in an expansion with respect to $u$. We define $\Hb$ as this
$N$-tangle up to a factor of $\Omega$, that is,
\be
  \Tb(u)\;=\;\Omega\big[\Ib\ -\ u\Hb\ +\ O(u^2)\big]
\label{DIH}
\ee
where $\Ib$ is the vertical identity diagram
\psset{unit=.5cm}
\setlength{\unitlength}{.5cm}
\be
 \Ib\;=\; 
\begin{pspicture}[shift=-1](-.25,.75)(6,3)
\conn{(0,1)}{(6,3)}
\psline[linecolor=blue,linewidth=1.5pt](0.5,1)(0.5,3)
\psline[linecolor=blue,linewidth=1.5pt](1.5,1)(1.5,3)
\psline[linecolor=blue,linewidth=1.5pt](2.5,1)(2.5,3)
\psline[linecolor=blue,linewidth=1.5pt](3.5,1)(3.5,3)
\psline[linecolor=blue,linewidth=1.5pt](5.5,1)(5.5,3)
\rput(4.57,2){\color{blue}$\dots$}
\end{pspicture}
\ee
It follows that
\psset{unit=.5cm}
\setlength{\unitlength}{.5cm}
\be
  -\Hb\ =\
\begin{pspicture}[shift=-1](-.25,.75)(6,3)
\conn{(0,1)}{(6,3)}
\psline[linecolor=blue,linewidth=1.5pt](2.5,1)(2.5,3)
\psline[linecolor=blue,linewidth=1.5pt](3.5,1)(3.5,3)
\psline[linecolor=blue,linewidth=1.5pt](5.5,1)(5.5,3)
\psarc[linecolor=blue,linewidth=1.5pt](1,1){.5}{0}{180}
\psarc[linecolor=blue,linewidth=1.5pt](1,3){.5}{180}{0}
\rput(4.57,2){\color{blue}$\dots$}
\end{pspicture}
\ +
\begin{pspicture}[shift=-1](-.25,.75)(6,3)
\conn{(0,1)}{(6,3)}
\psline[linecolor=blue,linewidth=1.5pt](0.5,1)(0.5,3)
\psline[linecolor=blue,linewidth=1.5pt](3.5,1)(3.5,3)
\psline[linecolor=blue,linewidth=1.5pt](5.5,1)(5.5,3)
\psarc[linecolor=blue,linewidth=1.5pt](2,1){.5}{0}{180}
\psarc[linecolor=blue,linewidth=1.5pt](2,3){.5}{180}{0}
\rput(4.57,2){\color{blue}$\dots$}
\end{pspicture}
\ +\ \dots\ +
\begin{pspicture}[shift=-1](-.25,.75)(6,3)
\conn{(0,1)}{(6,3)}
\psline[linecolor=blue,linewidth=1.5pt](1.5,1)(1.5,3)
\psline[linecolor=blue,linewidth=1.5pt](3.5,1)(3.5,3)
\psline[linecolor=blue,linewidth=1.5pt](4.5,1)(4.5,3)
\psarc[linecolor=blue,linewidth=1.5pt](6,1){.5}{90}{180}
\psarc[linecolor=blue,linewidth=1.5pt](6,3){.5}{180}{270}
\psarc[linecolor=blue,linewidth=1.5pt](0,1){.5}{0}{90}
\psarc[linecolor=blue,linewidth=1.5pt](0,3){.5}{270}{360}
\rput(2.57,2){\color{blue}$\dots$}
\end{pspicture}
\label{H}
\ee
which in terms of the generators of the {\em periodic} TL algebra 
merely corresponds to
\be
  \Hb\;=\;-\sum_{j=0}^{N-1}e_j
\label{Hlin}
\ee

\subsection{Link states}

A matrix representation of $\Tb(u)$ is obtained by acting with $\Tb(u)$ from below 
on a suitable vector space of link states. Suppose there are $N$ nodes arranged 
periodically around the upper horizontal edge of the cylinder. 
For $N$ even, a link state specifies how these $N$ nodes are linked together.
Two nodes can be connected by the front of the cylinder or by the back. We can consider 
these two connections as {\em distinct} or we can choose to {\em identify} the two 
connections and their corresponding link states. In the latter case, we can think of a 
hemi-spherical cap placed on the top of the cylinder so that a connection by the back of the 
cylinder can be continuously deformed to a connection by the front and vice versa. 
The link states are locally planar in the sense that connections are not allowed to cross 
on the extended surface of the cylinder (or capped cylinder).

The enlarged TL algebra is the appropriate algebra when acting on link states with 
{\em distinct connectivities} (DC). In the topology associated with the action 
on link states with {\em identified connectivities} (IC), all loops become contractible. 
This implies that the appropriate algebra in this case is the enlarged
TL algebra with $\alpha=\beta$.
In either case, the enlarged TL algebra is effectively finite when acting on the 
(DC or IC) link states since the corresponding matrix realizations satisfy
\be
 \Omega^N\;=\;(\Omega^{-1})^N\;=\;I
\label{ONI}
\ee
This means that we do not keep track of windings around the cylinder. 
As matrices, the inverse shift operator is given by the Hermitian conjugate of the operator itself
\be
 \Omega^{-1}\;=\;\Omega^\dagger
\label{OOdagger}
\ee
For $N=4$, there are six DC link states 
\psset{unit=.3cm}
\setlength{\unitlength}{.3cm}
\be
\begin{pspicture}(0,0)(3,2)
\psarc[linecolor=purple,linewidth=1.5pt](1.5,0){.5}{0}{180}
\psarc[linecolor=purple,linewidth=1.5pt](1.5,0){1.5}{0}{180}
\end{pspicture}\;\ ,\quad 
\begin{pspicture}(-.5,0)(3.5,2)
\psarc[linecolor=purple,linewidth=1.5pt](-.5,0){.5}{0}{90}
\psarc[linecolor=purple,linewidth=1.5pt](3.5,0){.5}{90}{180}
\psarc[linecolor=purple,linewidth=1.5pt](1.5,0){.5}{0}{180}
\end{pspicture}\;\ ,\quad 
\begin{pspicture}(-.5,0)(3.5,2)
\psarc[linecolor=purple,linewidth=1.5pt](-.5,0){.5}{0}{90}
\psarc[linecolor=purple,linewidth=1.5pt](3.5,0){.5}{90}{180}
\psarc[linecolor=purple,linewidth=1.5pt](-.5,0){1.5}{0}{90}
\psarc[linecolor=purple,linewidth=1.5pt](3.5,0){1.5}{90}{180}
\end{pspicture}\;\ ,\quad
\begin{pspicture}(3,2)
\psarc[linecolor=purple,linewidth=1.5pt](.5,0){.5}{0}{180}
\psarc[linecolor=purple,linewidth=1.5pt](2.5,0){.5}{0}{180}
\end{pspicture}\;\ ,\quad
\begin{pspicture}(-.5,0)(3,2)
\psarc[linecolor=purple,linewidth=1.5pt](2.5,0){.5}{0}{180}
\psarc[linecolor=purple,linewidth=1.5pt](-1.5,0){1.5}{0}{65}
\psarc[linecolor=purple,linewidth=1.5pt](2.5,0){1.5}{65}{180}
\end{pspicture}\;\ ,\quad
\begin{pspicture}(0,0)(3.5,2)
\psarc[linecolor=purple,linewidth=1.5pt](.5,0){.5}{0}{180}
\psarc[linecolor=purple,linewidth=1.5pt](.5,0){1.5}{0}{115}
\psarc[linecolor=purple,linewidth=1.5pt](4.5,0){1.5}{115}{180}
\end{pspicture}
\label{DC}
\ee
and two IC link states 
\be
\begin{pspicture}(0,0)(3,2)
\psarc[linecolor=purple,linewidth=1.5pt](1.5,0){.5}{0}{180}
\psarc[linecolor=purple,linewidth=1.5pt](1.5,0){1.5}{0}{180}
\end{pspicture}\;\ ,\quad
\begin{pspicture}(3,2)
\psarc[linecolor=purple,linewidth=1.5pt](.5,0){.5}{0}{180}
\psarc[linecolor=purple,linewidth=1.5pt](2.5,0){.5}{0}{180}
\end{pspicture}
\label{IC}
\ee
In general, the dimension of the vector space of link states for DC and IC is given by the 
central binomial coefficients and Catalan numbers
\be
  \dim(V_N^\mathrm{DC})\;=\;\bin{2n}n,\qquad
   \dim(V_N^\mathrm{IC})\;=\;\frac{1}{n+1}\,\bin{2n}n,\qquad n\;=\;\frac{N}{2},  \qquad N\ \mbox{even}
\label{dimICDC}
\ee

A node that is not linked to another node gives rise 
to a {\em defect} which may be viewed as a link to the point (above) at infinity. 
For $N$ odd, there is at least one defect. In the presence of defects, 
there is only one way (either by the front or by the back) to connect two nodes, 
so there is no distinction between identified and distinct connectivities.
The dimension of the space of link states with precisely $\ell$ defects is
\be
 \dim(V^{(\ell)}_N)\;=\;\bin{N}{\frac{N-\ell}{2}},\qquad \ell=N\ \mbox{mod 2}
\label{VN}
\ee 
It is the set of DC link states which corresponds to this for $\ell=0$:
$V^{(0)}_N=V_N^\mathrm{DC}$. For $N=3$ and $\ell=1$, there are 3 link states 
\psset{unit=.3cm}
\setlength{\unitlength}{.3cm}
\be
\begin{pspicture}(0,0)(2,2)
\psarc[linecolor=purple,linewidth=1.5pt](1.5,0){.5}{0}{180}
\psline[linecolor=purple,linewidth=1.5pt](0,0)(0,1)
\end{pspicture}\;\ ,\quad
\begin{pspicture}(0,0)(2,2)
\psarc[linecolor=purple,linewidth=1.5pt](.5,0){.5}{0}{180}
\psline[linecolor=purple,linewidth=1.5pt](2,0)(2,1)
\end{pspicture}\;\ ,\quad\
\begin{pspicture}(0,0)(2,2)
\psarc[linecolor=purple,linewidth=1.5pt](-.5,0){.5}{0}{90}
\psarc[linecolor=purple,linewidth=1.5pt](2.5,0){.5}{90}{180}
\psline[linecolor=purple,linewidth=1.5pt](1,0)(1,1)
\end{pspicture}
\label{ODD}
\ee
We note that (\ref{ONI}) and (\ref{OOdagger}) remain valid when acting on link states with defects. 
Also, defects can be annihilated in pairs, but not created, by the action of the cylinder TL algebra.
If we allow for an {\em arbitrary} number of defects of given parity, the number of link states is
\be
 \sum_{\ell\in2\mathbb{N}-1} \bin{N}{\frac{N-\ell}{2}}=2^{N-1},\qquad
 \sum_{\ell\in2\mathbb{N}_0}\bin{N}{\frac{N-\ell}{2}}=2^{N-1}-\frac{1}{2}\bin{N}{\frac{N}{2}}
\ee

\subsection{Augmented link states}

When the direction of transfer is fixed, the cylinder TL algebra reduces to the 
enlarged (periodic) TL algebra. The transfer $N$-tangle then acts naturally on 
link states with $N$ nodes. For the purpose of computer calculations, however,
the transfer matrix is conveniently written in terms of the enlarged TL algebra 
${\cal ET\!L}(N+2)$ acting on a suitable vector space of link states of size $N+2$. Explicitly, 
\be
 \Tb(u)\;=\; e_{N}\prod_{j=-1}^{N-2} X_j(u)\,\Omega
\ee
as shown in Figure~\ref{FigTOmega}, where the face operators are given by (\ref{faceop}).
The link states are all augmented by a spectator half-arc joining the nodes in positions $N$
and $N+1$.
\psset{unit=1cm}
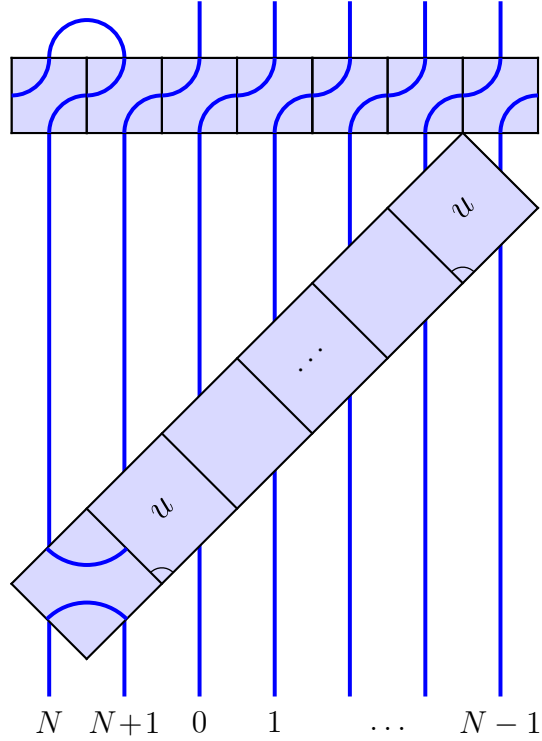
\begin{figure}[htbp]
\begin{center}
\begin{pspicture}(0,.5)(7.5,9.9)
\multiput(0,0)(1,0){5}{\psline[linewidth=1.5pt,linecolor=blue](2.5,9)(2.5,9.75)}
\multiput(0,0)(1,0){6}{\psline[linewidth=1.5pt,linecolor=blue](1.5,.5)(1.5,9)}
\psline[linewidth=1.5pt,linecolor=blue](.5,.5)(.5,1.5)
\psline[linewidth=1.5pt,linecolor=blue](.5,2)(.5,9)
\psarc[linewidth=1.5pt,linecolor=blue](1,9){.5}{0}{180}
\multirput(0,8)(1,0){7}{
\facegrid{(0,0)}{(1,1)}
\put(0,0){\loopa}}
\rput{45}(2,0){
\psset{unit=1.41421cm}
\facegrid{(0,1)}{(6,2)}
\rput(1.5,1.5){\large $u$}
\rput(5.5,1.5){\large $u$}
\rput(3.5,1.5){$\dots$}
\psarc[linewidth=.5pt](1,1){.15}{0}{90}
\psarc[linewidth=.5pt](5,1){.15}{0}{90}}
\rput[t](.5,.3){$N$}
\rput[t](1.5,.3){$N\!+\!1$}
\rput[t](2.5,.3){$0$}
\rput[t](3.5,.3){$1$}
\rput[t](5,.1){$\ldots$}
\rput[t](6.5,.3){$N-1$}
\psarc[linewidth=1.5pt,linecolor=blue](1,1){.75}{45}{135}
\psarc[linewidth=1.5pt,linecolor=blue](1,3){.75}{225}{315}
\end{pspicture}
\end{center}
\caption{Diagrammatic representation of the single-row transfer matrix built from the 
enlarged perodic TL algebra acting on $N+2$ strings.}
\label{FigTOmega}
\end{figure}
\psset{unit=.5cm}
For $N=4$, for example, the six augmented DC link states are given by
\psset{unit=.3cm}
\setlength{\unitlength}{.3cm}
\be
\begin{pspicture}(0,0)(5,2.5)
\psarc[linecolor=purple,linewidth=1.5pt](1.5,0){.5}{0}{180}
\psarc[linecolor=purple,linewidth=1.5pt](1.5,0){1.5}{0}{180}
\psarc[linecolor=purple,linewidth=1.5pt](4.5,0){.5}{0}{180}
\end{pspicture}\;\ ,\quad
\begin{pspicture}(-1,0)(5,2.5)
\psarc[linecolor=purple,linewidth=1.5pt](-1.5,0){1.5}{0}{65}
\psarc[linecolor=purple,linewidth=1.5pt](4.5,0){1.5}{65}{180}
\psarc[linecolor=purple,linewidth=1.5pt](1.5,0){.5}{0}{180}
\psarc[linecolor=purple,linewidth=1.5pt](4.5,0){.5}{0}{180}
\end{pspicture}\;\ ,\quad
\begin{pspicture}(-.5,0)(5.5,2.5)
\psarc[linecolor=purple,linewidth=1.5pt](-1.5,0){1.5}{0}{65}
\psarc[linecolor=purple,linewidth=1.5pt](4.5,0){1.5}{65}{180}
\psarc[linecolor=purple,linewidth=1.5pt](-1.5,0){2.5}{0}{75}
\psarc[linecolor=purple,linewidth=1.5pt](4.5,0){2.5}{75}{180}
\psarc[linecolor=purple,linewidth=1.5pt](4.5,0){.5}{0}{180}
\end{pspicture}\;\ ,\quad
\begin{pspicture}(.5,0)(5,2.5)
\psarc[linecolor=purple,linewidth=1.5pt](.5,0){.5}{0}{180}
\psarc[linecolor=purple,linewidth=1.5pt](2.5,0){.5}{0}{180}
\psarc[linecolor=purple,linewidth=1.5pt](4.5,0){.5}{0}{180}
\end{pspicture}\;\ ,\quad
\begin{pspicture}(-.5,0)(5,2.5)
\psarc[linecolor=purple,linewidth=1.5pt](2.5,0){.5}{0}{180}
\psarc[linecolor=purple,linewidth=1.5pt](-2.5,0){2.5}{0}{50}
\psarc[linecolor=purple,linewidth=1.5pt](3.5,0){2.5}{50}{180}
\psarc[linecolor=purple,linewidth=1.5pt](4.5,0){.5}{0}{180}
\end{pspicture}\;\ ,\quad
\begin{pspicture}(-.5,0)(5,2.5)
\psarc[linecolor=purple,linewidth=1.5pt](.5,0){.5}{0}{180}
\psarc[linecolor=purple,linewidth=1.5pt](0,0){2}{0}{95}
\psarc[linecolor=purple,linewidth=1.5pt](5,0){2}{85}{180}
\psarc[linecolor=purple,linewidth=1.5pt](4.5,0){.5}{0}{180}
\end{pspicture}
\label{DCaug}
\ee
where the nodes are labelled from $j=0$ to $j=N+1$. 
For $N=3$, and an arbitrary odd number of defects, we have four augmented link states
\psset{unit=.3cm}
\setlength{\unitlength}{.3cm}
\be
\begin{pspicture}(0,0)(3,2)
\psarc[linecolor=purple,linewidth=1.5pt](1,0){.5}{0}{180}
\psline[linecolor=purple,linewidth=1.5pt](0,0)(0,1)
\psarc[linecolor=purple,linewidth=1.5pt](2.5,0){.5}{0}{180}
\end{pspicture}\;\ ,\ \ 
\begin{pspicture}(0,0)(3,2)
\psarc[linecolor=purple,linewidth=1.5pt](.5,0){.5}{0}{180}
\psline[linecolor=purple,linewidth=1.5pt](1.5,0)(1.5,1)
\psarc[linecolor=purple,linewidth=1.5pt](2.5,0){.5}{0}{180}
\end{pspicture}\;\ ,\ \
\begin{pspicture}(0,0)(3,2)
\psarc[linecolor=purple,linewidth=1.5pt](-.5,0){1}{0}{90}
\psarc[linecolor=purple,linewidth=1.5pt](2.5,0){1}{90}{180}
\psline[linecolor=purple,linewidth=1.5pt](1,0)(1,1)
\psarc[linecolor=purple,linewidth=1.5pt](2.5,0){.5}{0}{180}
\end{pspicture}\;\ ,\ \
\begin{pspicture}(0,0)(2,2)
\psline[linecolor=purple,linewidth=1.5pt](0,0)(0,1)
\psline[linecolor=purple,linewidth=1.5pt](0.5,0)(0.5,1)
\psline[linecolor=purple,linewidth=1.5pt](1,0)(1,1)
\psarc[linecolor=purple,linewidth=1.5pt](2,0){.5}{0}{180}
\end{pspicture}
\label{ICN3}
\ee
where the nodes are labelled, as before, from $j=0$ to $j=N+1$. 

To further facilitate the implementation of the analysis on a computer, we note that the link states
can be described as sets of pairs of connecting 
arcs and defects. In this language, the set of link states (\ref{ODD}) reads
\be
 \{\{1\},\{2,3\}\},\{\{1,2\},\{3\}\},\{\{2\},\{3,1\}\}
\ee

\section{Inversion Identities}
\label{SecInversion}

Remarkably, the transfer matrix (\ref{TM}) satisfies an inversion
identity in the cylinder TL algebra. This identity is thus independent of the choice of vector space 
of link states eventually acted on to form a matrix representation. First, we describe 
the identity in the general cylinder setting (and prove it in Appendix~\ref{AppInversion}). 
We then characterize it when acting on the various link states introduced above.
The inversion identity is unique to critical dense polymers among the 
logarithmic minimal models~\cite{PRZ}. The analogous inversion identity 
for critical dense polymers on the {\em strip} is discussed in~\cite{PR0610}.
Although the inversion identities on the strip and the cylinder have common features,
their solutions and general properties are very different.

\subsection{Cylinder inversion identity}
\label{SecCylinder}

In preparation for the inversion identity, we introduce the two 3-tangles
\psset{unit=.5cm}
\setlength{\unitlength}{.5cm}
\be
-\;\begin{pspicture}(0,2.1)(1,0.85)
\facegrid{(0,0)}{(1,2)}
\put(0,0){\loopa}
\put(0,1){\loopa}
\end{pspicture}
\qquad\quad \mathrm{and}\qquad \quad
\begin{pspicture}(0,2.1)(1,0.85)
\facegrid{(0,0)}{(1,2)}
\put(0,0){\loopb}
\put(0,1){\loopb}
\end{pspicture}
\label{3m3}
\ee
\\[-.2cm]
as they play important roles as building blocks in the following.
The $N$-tangle $\Jb$, in particular, is defined as 
the sum of the $2^N\!$ possible horizontal combinations of $N$ of these 3-tangles, 
where the left and right edges are identified to respect the cylinder topology.
Due to the minus sign in (\ref{3m3}),
exactly half of the terms in $\Jb$ appear with a minus sign. For small $N$, we thus have
\psset{unit=.5cm}
\setlength{\unitlength}{.5cm}
\bea
 \Jb\big|_{N=1}\!&=&
-\;\begin{pspicture}(0,0.75)(1,2)
\facegrid{(0,0)}{(1,2)}
\put(0,0){\loopa}
\put(0,1){\loopa}
\end{pspicture}
\ +\ 
\begin{pspicture}(0,0.75)(1,2)
\facegrid{(0,0)}{(1,2)}
\put(0,0){\loopb}
\put(0,1){\loopb}
\end{pspicture}
 \ =\ -\Omega^2+\Omega^{-2}
 \\[18pt]
 \Jb\big|_{N=2}\!&=&\Omega^2-\
\begin{pspicture}(0,0.75)(2,2)
\facegrid{(0,0)}{(2,2)}
\put(0,0){\loopa}
\put(0,1){\loopa}
\put(1,0){\loopb}
\put(1,1){\loopb}
\end{pspicture}
\ -\  
\begin{pspicture}(0,0.75)(2,2)
\facegrid{(0,0)}{(2,2)}
\put(0,0){\loopb}
\put(0,1){\loopb}
\put(1,0){\loopa}
\put(1,1){\loopa}
\end{pspicture}
\ +\Omega^{-2}   
 \\[18pt]
 \Jb\big|_{N=3}\!&=&-\Omega^2+\
\begin{pspicture}(0,0.75)(3,2)
\facegrid{(0,0)}{(3,2)}
\put(0,0){\loopa}
\put(0,1){\loopa}
\put(1,0){\loopa}
\put(1,1){\loopa}
\put(2,0){\loopb}
\put(2,1){\loopb}
\end{pspicture}
\ +\  
\begin{pspicture}(0,0.75)(3,2)
\facegrid{(0,0)}{(3,2)}
\put(0,0){\loopa}
\put(0,1){\loopa}
\put(1,0){\loopb}
\put(1,1){\loopb}
\put(2,0){\loopa}
\put(2,1){\loopa}
\end{pspicture}
\ +\  
\begin{pspicture}(0,0.75)(3,2)
\facegrid{(0,0)}{(3,2)}
\put(0,0){\loopb}
\put(0,1){\loopb}
\put(1,0){\loopa}
\put(1,1){\loopa}
\put(2,0){\loopa}
\put(2,1){\loopa}
\end{pspicture}
 \nonumber
 \\[18pt]
 \!&&\hspace{2cm}-\ \ \!
\begin{pspicture}(0,0.75)(3,2)
\facegrid{(0,0)}{(3,2)}
\put(0,0){\loopa}
\put(0,1){\loopa}
\put(1,0){\loopb}
\put(1,1){\loopb}
\put(2,0){\loopb}
\put(2,1){\loopb}
\end{pspicture}
\ -\  
\begin{pspicture}(0,0.75)(3,2)
\facegrid{(0,0)}{(3,2)}
\put(0,0){\loopb}
\put(0,1){\loopb}
\put(1,0){\loopa}
\put(1,1){\loopa}
\put(2,0){\loopb}
\put(2,1){\loopb}
\end{pspicture}
\ -\  
\begin{pspicture}(0,0.75)(3,2)
\facegrid{(0,0)}{(3,2)}
\put(0,0){\loopb}
\put(0,1){\loopb}
\put(1,0){\loopb}
\put(1,1){\loopb}
\put(2,0){\loopa}
\put(2,1){\loopa}
\end{pspicture}
\ +\Omega^{-2}
\eea 
\\[-.1cm]
Although the $N$-tangle $\Jb=\Jb(\alpha)$ depends on the fugacity $\alpha$ of 
non-contractible loops, it is independent of the spectral parameter $u$.
\\[.4cm]
\noindent {\bf Inversion Identity}\ \
{\em The $N$-tangle $\Tb(u)$ defined in {\rm (\ref{TM})} satisfies}
\be
  \Tb(u)\Tb(u+\frac{\pi}{2})\;=\;
   \big(\!\cos^{2N}\!u+(-1)^N\!\sin^{2N}\!u\big)\Ib+(\cos u \sin u)^{N}\Jb
\label{InvCyl}
\ee
The proof of this Inversion Identity is provided in Appendix~\ref{AppInversion}.

It follows from the inversion identity (\ref{InvCyl}) and the commutativity property (\ref{TT0})
that $\Jb$ is a {\em symmetry} of the model in the sense that it commutes with the transfer matrix,
\be
 \big[\Jb,\Tb(u)\big]\;=\;\big[\Jb,\Omega^{\pm1}\big]\;=\;0
\ee
As discussed in Section~\ref{SecBraid}, $\Jb$ is related to the 
so-called braid transfer matrices.

\subsection{Matrix realization of $\Jb$}
\label{SecMatrixJ}

Obtaining a matrix realization of $\Jb$ is greatly simplified by the Drop-Down Lemma below.
To state it, we introduce the {\em arc-part} of a link state as the part remaining when ignoring all
defects. For given number of nodes $N$, a (DC or IC) link state is thus characterized completely 
by its arc-part and its number of defects. 
The arc-part of any of the three link states in (\ref{ODD}) consists of
a single half-arc whose position depends on the original link state.
We also say that the arc-part of a link state is {\em contained} in the arc-part of another
link state if the bigger arc-part can be constructed from the smaller one
by addition of half-arcs. Two identical arc-parts are said to be contained in each other.
\\[.4cm]
\noindent {\bf Drop-Down Lemma}\ \
{\em The action of $\Jb$ on a given input link state results in link states whose arc-part contains
the arc-part of the input link state.}
\\[.4cm]
The proof of this Drop-Down Lemma is provided in Appendix~\ref{AppDrop}.

So far, we have not specified the class of link states which $\Jb$ is acting on. 
The following Sector Lemma concerns the matrix realization of $\Jb$ in a given sector.
\\[.4cm]
\noindent {\bf Sector Lemma}\ \ 
{\em In a given sector defined by a specified number of defects $\ell$, the matrix realization
of $\Jb$ is diagonal and given by}
\be
 \Jb\;=\;
 \begin{cases}(-1)^{\frac{N-\ell}{2}}\big(2+(\al^2-4)\delta_{\ell,0}\big)\Ib,\qquad& 
    N,\ell\ \mathrm{even}
 \\[4pt]
 0, &N,\ell\ \mathrm{odd}
 \end{cases}
\label{SectorLemma}
\ee
\\
The proof of this Sector Lemma is provided in Appendix~\ref{AppSector}.

It is also of interest to examine the action of $\Jb$ on the set of link states with an {\em arbitrary} 
number of defects. After completion of the drop-down process, the remaining part of the
link state consists of defects only and the situation is equivalent to a scenario with 
system size $N_\ell=\ell$ where $\ell$ is the number of defects of the original input link state. 
That is, all the essential data is encoded in the Drop-Down Lemma 
and the action of $\Jb$ on link states with defects only. It is noted that this is true for all sectors
or combinations thereof, in particular for the union of all sectors of the parity of $N$.

As discussed in Appendix~\ref{AppDefects}, the matrix realization of $\Jb$ acting on
the set of link states with an {\em arbitrary odd number} of defects is the {\em zero matrix}.
For $N$ even, on the other hand, the matrix realization of $\Jb$ acting on
the set of link states with an {\em arbitrary even number} of defects is not diagonal, 
not even diagonalizable.
With respect to the number of defects, the matrix is upper block triangular.
The blocks on the diagonal are the same as
the ones obtained by the sector-by-sector analysis above, while the entries
outside these blocks give rise to a non-trivial Jordan decomposition. In particular, 
for DC link states with $\al=2$ and $N=8,10,12,14$, we observe that Jordan blocks of 
rank 2 appear, but not of higher rank, while there are no rank-2 blocks for the smallest 
system sizes $N=2,4,6$. Assuming that no Jordan blocks of rank 3 or higher occur for $N\ge 16$, 
we {\em conjecture} that the minimal polynomial identity satisfied by $\Jb$, 
valid for all even $N$, is
\be
 \big(\Jb^2-4\Ib\big)^2=0,\qquad\quad \al\;=\;2
\label{polJ}
\ee
Such a minimal condition implies the existence of the (diagonalizable) {\em involution} 
\be
 \Rb\;=\;-\frac{1}{16}\big(\Jb^3-12\Jb\big),\qquad\quad \Rb^2\;=\;\Ib
\label{involutionR}
\ee
The eigenvalues are $R=\half J=\pm 1$. 
We will comment on this involution in Section~\ref{SecConclusion},
and refer to Appendix~\ref{AppDefects} for additional details on the matrix realization of $\Jb$
when acting on DC link states with an arbitrary number of defects.

\subsection{Braid limits}
\label{SecBraid}

Let us define the braid limits by
\psset{unit=.9cm}
\setlength{\unitlength}{.9cm}
\bea
b^{\pm}=\lim_{u\to\pm i\infty} \frac{X(u)}{\sin(u+\frac{\pi}{4})}
\;=\;e^{\mp\frac{\pi i}{4}}\!\!\!\!
\begin{pspicture}[shift=-.45](-.5,-.1)(1.25,1.1)
\facegrid{(0,0)}{(1,1)}
\put(0,0){\loopa}
\end{pspicture}
\mbox{}+\,e^{\pm\frac{\pi i}{4}} \!\!\!\!
\begin{pspicture}[shift=-.45](-.5,-.1)(1.25,1.1)
\facegrid{(0,0)}{(1,1)}
\put(0,0){\loopb}
\end{pspicture}
\label{braid}\!\!,\quad
\vec B^{\pm}\;=\;\lim_{u\to\pm i\infty} \frac{\Tb(u)}{\sin^N(u+ 
\frac{\pi}{4})}\quad
\eea
By taking the braid limit of (\ref{InvCyl}),
the matrix $\Jb$ is seen to be simply related to the braid transfer matrices $\Bb^{\pm}$
\be
 (\Bb^{\pm})^2\;=\;2\Ib+(\pm i)^N\Jb
\ee
Using that $\Jb=0$ for $N$ odd, we find
\be
 (\Bb^{\pm})^2\;=\;\begin{cases}
  2\Ib+(-1)^{\frac{N}{2}}\Jb,&\mbox{$N$ even}\\[2pt]
  2\Ib,&\mbox{$N$ odd}
\end{cases},\qquad
 \Jb\;=\;\begin{cases}
  (-1)^{\frac{N}{2}}((\Bb^\pm)^2-2\Ib),&\mbox{$N$ even}\\[2pt]
  0,&\mbox{$N$ odd}
\end{cases}
\ee
Assuming the conjectured minimal polynomial identity (\ref{polJ}) for $\Jb$ 
implies that $\Bb=\Bb^{\pm}$ satisfies
\be
 \Bb^4(\Bb^2-4\vec I)^2\;=\;0,\ \  \mbox{$N$ even};\qquad\quad 
 \Bb^2\;=\;2\Ib,  \ \  \mbox{$N$ odd}
\ee
The eigenvalues of the braid matrices are thus of the form
\be
 B\;=\;2\cos\frac{s\pi}{4},\quad s=0,1,2,3,4
\ee
where
\be
 B\;=\;0,\pm 2,\ \ \mbox{$N$ even};\qquad\quad 
 B\;=\;\pm \sqrt{2},\ \ \mbox{$N$ odd}
\ee

\subsection{Matrix inversion identities}

Once a matrix representation of $\Tb(u)$ has been fixed,
the inversion identity (\ref{InvCyl}) translates into a {\em matrix inversion identity}.
Here, we consider the link states counted in (\ref{VN}), including the DC link states for
$\ell=0$ but not the IC link states. The latter are discussed in Section~\ref{SecIC}.

It follows from the Inversion Identity (\ref{InvCyl}) and the Sector Lemma (\ref{SectorLemma}) 
that the matrix inversion identity for a given sector reads
\be
 \Tb(u)\Tb(u+\frac{\pi}{2})\;=\; 
   \Big(\!\cos^{2N}\!u+\sin^{2N}\!u
     +(-1)^{\frac{N-\ell}{2}}\big(2+(\al^2-4)\delta_{\ell,0}\big)
        \big(\!\cos u \sin u\big)^{\!N}\Big)\Ib
\label{TTeven}
\ee 
for $N$ even, while for $N$ odd, it reads
\be
 \Tb(u)\Tb(u+\frac{\pi}{2})\;=\; \big(\!\cos^{2N}\!u - \sin^{2N}\!u\big) \Ib
\label{TTodd}
\ee

In order to solve the inversion identity (\ref{TTeven})
for the associated eigenvalues ($N$ even), 
we choose to focus on particular values for $\al$ in the following.
For $\al^2=4$, we thus have the matrix inversion identities
\be
 \Tb(u)\Tb(u+\frac{\pi}{2})\;=\; 
   \big(\!\cos^{N}\!u +(-1)^{\frac{N-\ell}{2}} \!\sin^{N}\!u\big)^{\!2}\Ib
\ee
while, for $\al=0$ and $\ell=0$, we have
\be
 \Tb(u)\Tb(u+\frac{\pi}{2})\;=\; 
   \big(\!\cos^{N}\!u -(-1)^{\frac{N}{2}} \!\sin^{N}\!u\big)^{\!2}\Ib
\label{al0}
\ee
We find that
\be
 \al\;=\;2
\label{al2}
\ee
is the most natural value for the fugacity of non-contractible loops.

\subsection{Identified connectivities}
\label{SecIC}

The distinction between DC and IC link states is only meaningful for link states
without defects. In this case, the number of IC link states is smaller than the number of DC
link states, cf. (\ref{dimICDC}), implying that, for given $N$, the corresponding matrix realization
of $\Tb(u)$ or $\Jb$ is of lower dimension in the IC case than in the DC case.
As already mentioned, the appropriate algebra in the IC case is the enlarged
TL algebra with $\alpha=\beta=0$. It follows that the matrix inversion identity takes the
same form as (\ref{al0}), namely
\be
 \Tb(u)\Tb(u+\frac{\pi}{2})\;=\; 
   \big(\!\cos^{N}\!u -(-1)^{\frac{N}{2}} \!\sin^{N}\!u\big)^{\!2}\Ib
\label{IIIC}
\ee
but applies to matrices of smaller dimension than the ones appearing in (\ref{al0}).

\section{Solution on a Finite Cylinder}
\label{SecSolution}

In this section, we solve the inversion identities for the transfer matrix eigenvalues on
finite-size cylinders sector by sector for $\alpha=2$. 
We also discuss the relation, through finite-size
corrections, to the conformal partition functions. The methods build on the previous
works \cite{BaxBook}, \cite{OPW} and \cite{PR0610}.

The key idea is that the eigenvalues $T(u)$ of the transfer matrices in a given sector are
determined, up to an overall constant $\rho$, by the positions $u_j$ of their zeros in the
analyticity strip $-\pi/4\le \Re u<3\pi/4$
\be
 T(u)\;=\;\rho\prod_{j=1}^N \sin(u-u_j)
\ee
These eigenvalues are Laurent polynomials in $z=e^{iu}$.
Given a (right) eigenvector independent of $u$, this follows since each entry of the transfer matrix is of this form.
This argument, which applies to the action of the transfer matrix on a particular eigenvector,
holds even if the commuting transfer matrices are not diagonalizable.
Typically, in the $\ell$ even sectors on the cylinder, we find that the transfer matrices are not diagonalizable as we find Jordan blocks of rank 2.  We have not found Jordan blocks of any higher rank. Moreover,
we find, in a given $\ell$ sector, that all of the eigenvectors (excluding generalized eigenvectors) are independent of $u$.
Degenerate eigenvalues are exactly degenerate as Laurent polynomials so such eigenvalues occur with degeneracy 2.
These observations enable us to numerically obtain all of the eigenvalues as Laurent polynomials in $z=e^{iu}$.

We emphasize that this situation on the cylinder is in contrast to critical dense polymers on the strip in sectors of the extended Kac table
for which the transfer matrices are found empirically~\cite{PR0610} to be simultaneously diagonalizable. It seems that it is not possible to avoid
reducible yet indecomposable representations of rank 2 in the $\ell$ even sectors on the cylinder.

\subsection{Finite-size corrections}

The partition function of critical dense polymers on a periodic lattice of $N$ 
columns and $M$ rows is defined by
\be
  Z_{N,M}\;=\;\mathop{\rm Tr}\Tb(u)^M\;=\;\sum_{n\ge 0} T_{n}(u)^M
   \;=\;\sum_{n\ge 0} e^{-M{\cal E}_{n}(u)}
\label{ZNM}
\ee
Here the sum is over all eigenvalues of $\Tb(u)$, including possible
multiplicities, and ${\cal E}_{n}(u)$ with $n=0,1,2,\ldots$ is the energy associated to the eigenvalue
$T_{n}(u)$. The maximal eigenvalue $T_0(u)$ is labelled by $n=0$. The maximal eigenvalue in the sector with $\ell$ defects is denoted by $T_{0,\ell}(u)$. 
Conformal invariance of the model in the continuum scaling limit dictates 
\cite{BCN,Aff} that the leading finite-size corrections for large $N$ 
are of the form
\bea
\begin{array}{rcl}
{\cal E}_0&=&\disp Nf_{bulk}-\frac{\pi c}{6N}\,\sin \vartheta\\[8pt]
{\cal E}_n-{\cal E}_0&=&\disp 
  \frac{2\pi i}{N}\,[(\Delta+k)e^{-i\vartheta}-(\bar{\Delta}+\kbar)e^{i\vartheta}]\\[8pt]
  &=&\disp \frac{2\pi}{N}\,\big[(\Delta+ \bar{\Delta}+k+\kbar)\sin\vartheta
    +i(\Delta-\bar{\Delta}+k-\kbar)\cos\vartheta\big]
  \end{array}
\label{logLa}\label{FSE}
\eea
Here $f_{bulk}$ is the bulk free energy per face~\cite{PR0610}
\be
  f_{bulk}\;=\;\half\log2-\frac{1}{\pi}\int_{0}^{\pi/2}
   \log\Big(\frac{1}{\sin t}+\sin 2u\Big)dt
\label{fb}
\ee
and $\vartheta=2u$ is the anisotropy angle.
The conformal spectrum is determined by the central charge $c=-2$, the conformal weights 
$\Delta, \bar{\Delta}$ and the excitations or descendants labelled by the non-negative 
integers $k, \kbar$. The conformal weights are given by
\be
 \Delta\;=\;\Delta_t\;=\;\frac{t^2-1}{8},\qquad t\in\half\mathbb{Z}
\ee
where $t$ can be integer or half-integer
\be
\Delta\;=\;\begin{cases}
 \Delta_{r,s}=\Delta_{2r-s}\in\{-\frac{1}{8},0,\frac{3}{8},1,\frac{15}{8},\ldots\},
  &\mbox{$r\in\mathbb{N}, s=1,2$;\quad\  $N$ even}\\[6pt]
\Delta_t\in \{-\frac{3}{32},\frac{5}{32},\frac{21}{32},\frac{45}{32},\frac{77}{32},\frac{117}{32}\ldots\},
  &\mbox{$t\in\mathbb{Z}-\half $;\ \ \ \ \qquad $N$ odd}
\end{cases}
\ee
Here $r,s$ are the Kac labels~\cite{PR0610} and $t=\ell/2$ where $\ell$ is the number of defects. 
The Kac table is shown in Figure~\ref{Kac}.

\begin{figure}[htb]
{\vspace{0in}\psset{unit=.7cm}
{
\small
\begin{center}
\qquad
\begin{pspicture}(0,0)(7,11)
\psframe[linewidth=0pt,fillstyle=solid,fillcolor=lightlightblue](0,0)(7,11)
\multiput(0,0)(0,2){5}{\psframe[linewidth=0pt,fillstyle=solid,fillcolor=midblue](0,1)(7,2)}
\multirput(1,1)(1,0){6}{\pswedge[fillstyle=solid,fillcolor=red,linecolor=red](0,0){.25}{180}{270}}
\multirput(1,2)(1,0){6}{\pswedge[fillstyle=solid,fillcolor=red,linecolor=red](0,0){.25}{180}{270}}
\multirput(1,2)(0,2){5}{\pswedge[fillstyle=solid,fillcolor=red,linecolor=red](0,0){.25}{180}{270}}
\psgrid[gridlabels=0pt,subgriddiv=1]
\rput(.5,10.65){$\vdots$}\rput(1.5,10.65){$\vdots$}\rput(2.5,10.65){$\vdots$}
\rput(3.5,10.65){$\vdots$}\rput(4.5,10.65){$\vdots$}\rput(5.5,10.65){$\vdots$}
\rput(6.5,10.5){$\vvdots$}\rput(.5,9.5){$\frac{63}8$}\rput(1.5,9.5){$\frac{35}8$}
\rput(2.5,9.5){$\frac{15}8$}\rput(3.5,9.5){$\frac{3}8$}\rput(4.5,9.5){$-\frac 18$}
\rput(5.5,9.5){$\frac{3}8$}\rput(6.5,9.5){$\cdots$}
\rput(.5,8.5){$6$}\rput(1.5,8.5){$3$}\rput(2.5,8.5){$1$}\rput(3.5,8.5){$0$}
\rput(4.5,8.5){$0$}\rput(5.5,8.5){$1$}\rput(6.5,8.5){$\cdots$}
\rput(.5,7.5){$\frac{35}8$}\rput(1.5,7.5){$\frac {15}8$}\rput(2.5,7.5){$\frac 38$}
\rput(3.5,7.5){$-\frac{1}8$}\rput(4.5,7.5){$\frac 38$}\rput(5.5,7.5){$\frac{15}8$}
\rput(6.5,7.5){$\cdots$}\rput(.5,6.5){$3$}\rput(1.5,6.5){$1$}\rput(2.5,6.5){$0$}\rput(3.5,6.5){$0$}
\rput(4.5,6.5){$1$}\rput(5.5,6.5){$3$}\rput(6.5,6.5){$\cdots$}
\rput(.5,5.5){$\frac{15}8$}\rput(1.5,5.5){$\frac {3}{8}$}\rput(2.5,5.5){$-\frac 18$}
\rput(3.5,5.5){$\frac{3}{8}$}\rput(4.5,5.5){$\frac {15}8$}\rput(5.5,5.5){$\frac{35}{8}$}
\rput(6.5,5.5){$\cdots$}\rput(.5,4.5){$1$}\rput(1.5,4.5){$0$}\rput(2.5,4.5){$0$}
\rput(3.5,4.5){$1$}\rput(4.5,4.5){$3$}\rput(5.5,4.5){$6$}\rput(6.5,4.5){$\cdots$}
\rput(.5,3.5){$\frac 38$}\rput(1.5,3.5){$-\frac 18$}\rput(2.5,3.5){$\frac 38$}
\rput(3.5,3.5){$\frac{15}8$}\rput(4.5,3.5){$\frac{35}8$}\rput(5.5,3.5){$\frac{63}8$}
\rput(6.5,3.5){$\cdots$}\rput(.5,2.5){$0$}\rput(1.5,2.5){$0$}\rput(2.5,2.5){$1$}\rput(3.5,2.5){$3$}
\rput(4.5,2.5){$6$}\rput(5.5,2.5){$10$}\rput(6.5,2.5){$\cdots$}
\rput(.5,1.5){$-\frac 18$}\rput(1.5,1.5){$\frac 38$}\rput(2.5,1.5){$\frac{15}8$}
\rput(3.5,1.5){$\frac{35}8$}\rput(4.5,1.5){$\frac{63}8$}\rput(5.5,1.5){$\frac{99}8$}
\rput(6.5,1.5){$\cdots$}\rput(.5,.5){$0$}\rput(1.5,.5){$1$}\rput(2.5,.5){$3$}\rput(3.5,.5){$6$}
\rput(4.5,.5){$10$}\rput(5.5,.5){$15$}\rput(6.5,.5){$\cdots$}
{\color{blue}
\rput(.5,-.5){$1$}
\rput(1.5,-.5){$2$}
\rput(2.5,-.5){$3$}
\rput(3.5,-.5){$4$}
\rput(4.5,-.5){$5$}
\rput(5.5,-.5){$6$}
\rput(6.5,-.5){$r$}
\rput(-.5,.5){$1$}
\rput(-.5,1.5){$2$}
\rput(-.5,2.5){$3$}
\rput(-.5,3.5){$4$}
\rput(-.5,4.5){$5$}
\rput(-.5,5.5){$6$}
\rput(-.5,6.5){$7$}
\rput(-.5,7.5){$8$}
\rput(-.5,8.5){$9$}
\rput(-.5,9.5){$10$}
\rput(-.5,10.5){$s$}}
\end{pspicture}
\end{center}}}
\caption{\label{Kac}Kac table of the $\ell$ even sectors of critical dense polymers. 
The relevant rows, $s=1$ and $s=2$, label the Neveu-Schwarz and Ramond sectors, respectively. 
The $(r,s)$ representations indicated with a red quadrant are irreducible representations
with characters $\chi_{r,s}(q)=\mch_{r,s}(q)$.}
\end{figure}

In the scaling limit, the conformal partition functions are sesquilinear forms in characters
\be
 Z(q)\;=\;\sum_{\Delta,\bar\Delta} {\cal N}_{\Delta,\bar\Delta}\,\chit_\Delta(q)\chit_{\bar\Delta}(\qbar)
\ee
where the characters are of the form
\be
 \chit_\Delta(q)\;=\;q^{-c/24}\sum_{k=0}^\infty d_\Delta(k)\,q^{\Delta+k}\;=\;
\begin{cases}
 \displaystyle \mch_{r,s}(q)
  =\frac{q^{-c/24+\Delta_{r,s}}(1-q^{rs})}{\prod_{n=1}^\infty (1-q^n)},&\mbox{$N$ even}\\[12pt]
 \displaystyle \mch_t(q)=\frac{q^{-c/24+\Delta_t}}{\prod_{n=1}^\infty (1-q^n)},&\mbox{$N$ odd}
\end{cases}
\ee
and $d_\Delta(k)$ are the degeneracies at level $k$. The modular nome is
\bea
 q\;=\;\exp(2\pi i\tau),\qquad 
 \tau\;=\;\frac{M}{N}\,\exp[i(\pi-\vartheta)]\;=\;-\delta\,e^{-2iu}\\
 \qbar\;=\;\exp(-2\pi i\bar\tau),\qquad 
 \bar\tau\;=\;\frac{M}{N}\,\exp[-i(\pi-\vartheta)]\;=\;-\delta\,e^{2iu}\\[-14pt] \nonumber
\eea
\be
 |q|^2\;=\;q\qbar\;=\;\exp(-4\pi\delta\sin 2u)
\ee
where $\delta=M/N$ is the aspect ratio and $\Im\tau>0$ in the physical strip 
$0<\Re u<\frac{\pi}{2}$. The characters $\mch_{r,s}(q)$ with $s=1,2$ in the $N$ 
even sectors are the characters of irreducible Kac representations~\cite{PRZ} whereas  
$\mch_t(q)$ in the $N$ odd sectors are the characters of generic (irreducible) Virasoro modules.

\subsection{$\mathbb{Z}_4$ sectors ($N$ odd, $\ell$ odd)}

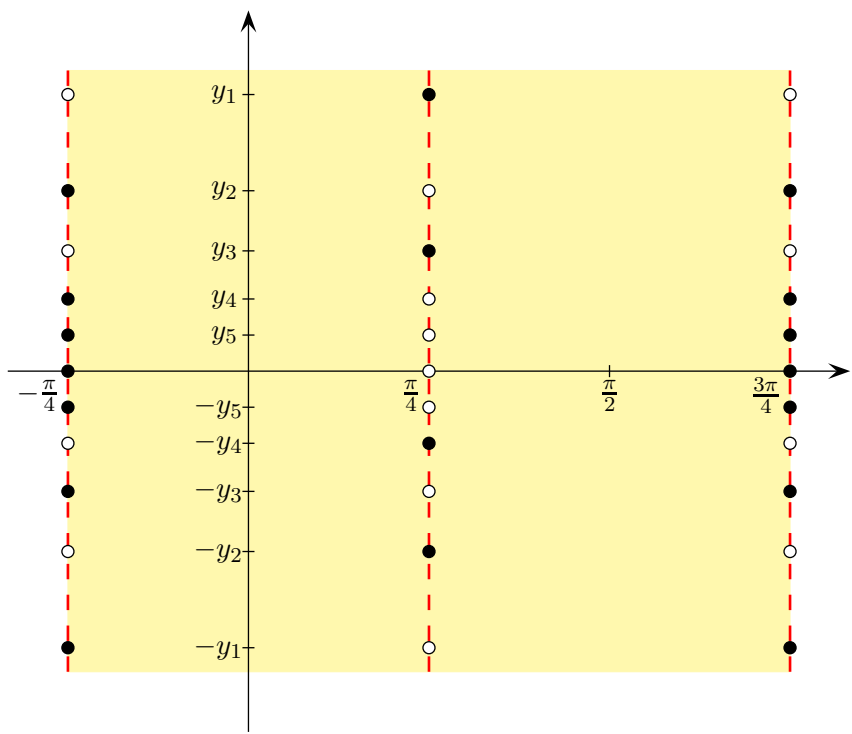
\begin{figure}[htb]
\psset{unit=.8cm}
\setlength{\unitlength}{.8cm}
\begin{center}
\begin{pspicture}(-.25,.5)(14,12)
\psframe[linecolor=yellow!40!white,linewidth=0pt,fillstyle=solid,
   fillcolor=yellow!40!white](1,1)(13,11)
\psline[linecolor=black,linewidth=.5pt,arrowsize=6pt]{->}(4,0)(4,12)
\psline[linecolor=black,linewidth=.5pt,arrowsize=6pt]{->}(0,6)(14,6)
\psline[linecolor=red,linewidth=1pt,linestyle=dashed,dash=.25 .25](1,1)(1,11)
\psline[linecolor=red,linewidth=1pt,linestyle=dashed,dash=.25 .25](7,1)(7,11)
\psline[linecolor=red,linewidth=1pt,linestyle=dashed,dash=.25 .25](13,1)(13,11)
\psline[linecolor=black,linewidth=.5pt](1,5.9)(1,6.1)
\psline[linecolor=black,linewidth=.5pt](7,5.9)(7,6.1)
\psline[linecolor=black,linewidth=.5pt](10,5.9)(10,6.1)
\psline[linecolor=black,linewidth=.5pt](13,5.9)(13,6.1)
\rput(.5,5.6){\small $-\frac{\pi}{4}$}
\rput(6.7,5.6){\small $\frac{\pi}{4}$}
\rput(10,5.6){\small $\frac{\pi}{2}$}
\rput(12.6,5.6){\small $\frac{3\pi}{4}$}
\psline[linecolor=black,linewidth=.5pt](3.9,6.6)(4.1,6.6)
\psline[linecolor=black,linewidth=.5pt](3.9,7.2)(4.1,7.2)
\psline[linecolor=black,linewidth=.5pt](3.9,8.0)(4.1,8.0)
\psline[linecolor=black,linewidth=.5pt](3.9,9.0)(4.1,9.0)
\psline[linecolor=black,linewidth=.5pt](3.9,10.6)(4.1,10.6)
\psline[linecolor=black,linewidth=.5pt](3.9,5.4)(4.1,5.4)
\psline[linecolor=black,linewidth=.5pt](3.9,4.8)(4.1,4.8)
\psline[linecolor=black,linewidth=.5pt](3.9,4.0)(4.1,4.0)
\psline[linecolor=black,linewidth=.5pt](3.9,3.0)(4.1,3.0)
\psline[linecolor=black,linewidth=.5pt](3.9,1.4)(4.1,1.4)
\rput(3.6,6.6){\small $y_5$}
\rput(3.6,7.2){\small $y_4$}
\rput(3.6,8.0){\small $y_3$}
\rput(3.6,9.0){\small $y_2$}
\rput(3.6,10.6){\small $y_1$}
\rput(3.5,5.4){\small $-y_5$}
\rput(3.5,4.8){\small $-y_4$}
\rput(3.5,4.0){\small $-y_3$}
\rput(3.5,3.0){\small $-y_2$}
\rput(3.5,1.4){\small $-y_1$}
\psarc[linecolor=black,linewidth=.5pt,fillstyle=solid,fillcolor=black](1,6.0){.1}{0}{360}
\psarc[linecolor=black,linewidth=.5pt,fillstyle=solid,fillcolor=black](1,6.6){.1}{0}{360}
\psarc[linecolor=black,linewidth=.5pt,fillstyle=solid,fillcolor=black](1,7.2){.1}{0}{360}
\psarc[linecolor=black,linewidth=.5pt,fillstyle=solid,fillcolor=white](1,8.0){.1}{0}{360}
\psarc[linecolor=black,linewidth=.5pt,fillstyle=solid,fillcolor=black](1,9.0){.1}{0}{360}
\psarc[linecolor=black,linewidth=.5pt,fillstyle=solid,fillcolor=white](1,10.6){.1}{0}{360}
\psarc[linecolor=black,linewidth=.5pt,fillstyle=solid,fillcolor=white](7,6.0){.1}{0}{360}
\psarc[linecolor=black,linewidth=.5pt,fillstyle=solid,fillcolor=white](7,6.6){.1}{0}{360}
\psarc[linecolor=black,linewidth=.5pt,fillstyle=solid,fillcolor=white](7,7.2){.1}{0}{360}
\psarc[linecolor=black,linewidth=.5pt,fillstyle=solid,fillcolor=black](7,8.0){.1}{0}{360}
\psarc[linecolor=black,linewidth=.5pt,fillstyle=solid,fillcolor=white](7,9.0){.1}{0}{360}
\psarc[linecolor=black,linewidth=.5pt,fillstyle=solid,fillcolor=black](7,10.6){.1}{0}{360}
\psarc[linecolor=black,linewidth=.5pt,fillstyle=solid,fillcolor=black](13,6.0){.1}{0}{360}
\psarc[linecolor=black,linewidth=.5pt,fillstyle=solid,fillcolor=black](13,6.6){.1}{0}{360}
\psarc[linecolor=black,linewidth=.5pt,fillstyle=solid,fillcolor=black](13,7.2){.1}{0}{360}
\psarc[linecolor=black,linewidth=.5pt,fillstyle=solid,fillcolor=white](13,8.0){.1}{0}{360}
\psarc[linecolor=black,linewidth=.5pt,fillstyle=solid,fillcolor=black](13,9.0){.1}{0}{360}
\psarc[linecolor=black,linewidth=.5pt,fillstyle=solid,fillcolor=white](13,10.6){.1}{0}{360}
\psarc[linecolor=black,linewidth=.5pt,fillstyle=solid,fillcolor=black](1,5.4){.1}{0}{360}
\psarc[linecolor=black,linewidth=.5pt,fillstyle=solid,fillcolor=white](1,4.8){.1}{0}{360}
\psarc[linecolor=black,linewidth=.5pt,fillstyle=solid,fillcolor=black](1,4.0){.1}{0}{360}
\psarc[linecolor=black,linewidth=.5pt,fillstyle=solid,fillcolor=white](1,3.0){.1}{0}{360}
\psarc[linecolor=black,linewidth=.5pt,fillstyle=solid,fillcolor=black](1,1.4){.1}{0}{360}
\psarc[linecolor=black,linewidth=.5pt,fillstyle=solid,fillcolor=white](7,5.4){.1}{0}{360}
\psarc[linecolor=black,linewidth=.5pt,fillstyle=solid,fillcolor=black](7,4.8){.1}{0}{360}
\psarc[linecolor=black,linewidth=.5pt,fillstyle=solid,fillcolor=white](7,4.0){.1}{0}{360}
\psarc[linecolor=black,linewidth=.5pt,fillstyle=solid,fillcolor=black](7,3.0){.1}{0}{360}
\psarc[linecolor=black,linewidth=.5pt,fillstyle=solid,fillcolor=white](7,1.4){.1}{0}{360}
\psarc[linecolor=black,linewidth=.5pt,fillstyle=solid,fillcolor=black](13,5.4){.1}{0}{360}
\psarc[linecolor=black,linewidth=.5pt,fillstyle=solid,fillcolor=white](13,4.8){.1}{0}{360}
\psarc[linecolor=black,linewidth=.5pt,fillstyle=solid,fillcolor=black](13,4.0){.1}{0}{360}
\psarc[linecolor=black,linewidth=.5pt,fillstyle=solid,fillcolor=white](13,3.0){.1}{0}{360}
\psarc[linecolor=black,linewidth=.5pt,fillstyle=solid,fillcolor=black](13,1.4){.1}{0}{360}
\end{pspicture}
\end{center}
\caption{A typical pattern of zeros in the complex $u$-plane for the $\mathbb{Z}_4$ sectors 
($N$ odd, $\ell$ odd). Here, $N=11$ and $\ell=7$. The ordinates of the locations of the zeros $u_j$ are
$y_j=-\half \log \tan{\frac{(2j-1)\pi}{4N}}, j=1,2,\ldots,N$. At each position $j$, there is either 
a 1-string with $\Re u_j=\pi/4$ or a 2-string with $\Re u_j=-\pi/4, 3\pi/4$.
\label{uplaneZ4}}
\end{figure}

In the $\mathbb{Z}_4$ sectors, the sector-by-sector inversion identity for the eigenvalues is
\be
T(u)T(u+\frac{\pi}{2})\;=\;\cos^{2N}\!u-\sin^{2N}\!u
\ee
Factorizing the right side gives
\be
\cos^{2N}\!u- \sin^{2N}\!u
  \;=\;\frac{ e^{-2Niu}}{2^{2N-1}} 
  \prod_{j=1}^{N} {\Big(e^{4i u}+ \tan^{2}{\frac{(2j-1)\pi}{4N}}\Big)}
\ee
Sharing out the zeros to solve the functional equation, gives
\be
T(u)
  \;=\;\epsilon\,\frac{(-i)^{N/2}e^{-Niu}}{2^{N-1/2}}
  \prod_{j=1}^{N} {\Big(e^{2i u}+i\epsilon_j  \tan{\frac{(2j-1)\pi}{4N}}\Big)}
\ee
where $\epsilon^2=\epsilon_j^2=1$. The ordinates of the locations of zeros are
\be
 y_j\;=\;-\half \log \tan{\frac{\half(j-\half)\pi}{N}}, \qquad j=1,2,\ldots,N
\ee
A typical pattern of zeros is shown in Figure~\ref{uplaneZ4}.

\begin{figure}[htbp]
\psset{unit=.6cm}
\setlength{\unitlength}{.6cm}
\begin{center}
\begin{pspicture}[shift=-4](-.25,-1.25)(.5,6.5)
\rput(0,.5){\scriptsize $j=1$}
\rput(0,1.5){\scriptsize $j=2$}
\rput(0,2.5){\scriptsize $j=3$}
\rput(0,3.5){$\vdots$}
\rput(-.2,4.5){\scriptsize $(N\!-\!1)/2$}
\rput(-.2,5.5){\scriptsize $(N\!+\!1)/2$}
\end{pspicture}
\hspace{4pt}
\begin{pspicture}[shift=-3](-.25,-.25)(2,6)
\psframe[linewidth=0pt,fillstyle=solid,fillcolor=yellow!40!white](0,0)(2,6)
\psarc[linecolor=black,linewidth=.5pt,fillstyle=solid,fillcolor=white](1,5.5){.1}{0}{360}
\psarc[linecolor=black,linewidth=.5pt,fillstyle=solid,fillcolor=white](1,4.5){.1}{0}{360}
\psarc[linecolor=black,linewidth=.5pt,fillstyle=solid,fillcolor=white](1,2.5){.1}{0}{360}
\psarc[linecolor= blue,linewidth=.5pt,fillstyle=solid,fillcolor=blue](1,3.5){.1}{0}{360}
\psarc[linecolor=black,linewidth=.5pt,fillstyle=solid,fillcolor=white](1,0.5){.1}{0}{360}
\psarc[linecolor= blue,linewidth=.5pt,fillstyle=solid,fillcolor= blue](1,1.5){.1}{0}{360}
\end{pspicture}
\hspace{.2cm}
\begin{pspicture}[shift=-3](-.25,-.25)(2,5)
\psframe[linewidth=0pt,fillstyle=solid,fillcolor=yellow!40!white](0,0)(2,5)
\psarc[linecolor=black,linewidth=.5pt,fillstyle=solid,fillcolor=white](1,3.5){.1}{0}{360}
\psarc[linecolor= red,linewidth=.5pt,fillstyle=solid,fillcolor=red](1,4.5){.1}{0}{360}
\psarc[linecolor=blue,linewidth=.5pt,fillstyle=solid,fillcolor=blue](1,1.5){.1}{0}{360}
\psarc[linecolor= red,linewidth=.5pt,fillstyle=solid,fillcolor= red](1,2.5){.1}{0}{360}
\psarc[linecolor=red,linewidth=.5pt,fillstyle=solid,fillcolor=red](1,0.5){.1}{0}{360}
\end{pspicture}
\hspace{.2cm}\ \ $\leftrightarrow$ \hspace{.2cm}
\begin{pspicture}[shift=-3](-.25,-.25)(2,6)
\psframe[linewidth=0pt,fillstyle=solid,fillcolor=yellow!40!white](0,0)(2,6)
\psline[linecolor=red,linewidth=1pt,linestyle=dashed,dash=.25 .25](1,0)(1,6)
\psarc[linecolor=black,linewidth=.5pt,fillstyle=solid,fillcolor=white](0.5,5.5){.1}{0}{360}
\psarc[linecolor=black,linewidth=.5pt,fillstyle=solid,fillcolor=white](0.5,4.5){.1}{0}{360}
\psarc[linecolor=black,linewidth=.5pt,fillstyle=solid,fillcolor=white](0.5,2.5){.1}{0}{360}
\psarc[linecolor= blue,linewidth=.5pt,fillstyle=solid,fillcolor= blue](0.5,3.5){.1}{0}{360}
\psarc[linecolor=black,linewidth=.5pt,fillstyle=solid,fillcolor=white](0.5,.5){.1}{0}{360}
\psarc[linecolor= blue,linewidth=.5pt,fillstyle=solid,fillcolor= blue](0.5,1.5){.1}{0}{360}
\psarc[linecolor=black,linewidth=.5pt,fillstyle=solid,fillcolor=white](1.5,3.5){.1}{0}{360}
\psarc[linecolor= red,linewidth=.5pt,fillstyle=solid,fillcolor=red](1.5,4.5){.1}{0}{360}
\psarc[linecolor= blue,linewidth=.5pt,fillstyle=solid,fillcolor= blue](1.5,1.5){.1}{0}{360}
\psarc[linecolor= red,linewidth=.5pt,fillstyle=solid,fillcolor= red](1.5,2.5){.1}{0}{360}
\psarc[linecolor=red,linewidth=.5pt,fillstyle=solid,fillcolor=red](1.5,0.5){.1}{0}{360}
\end{pspicture}
\hspace{.2cm}\ \ $\leftrightarrow$ \hspace{.2cm} 
$q^{\frac{3}{4}+\frac{7}{4}}\qbar^{\frac{1}{4}+\frac{3}{4}+\frac{5}{4}+\frac{9}{4}}
 \;=\;q^{\frac{5}{2}}\qbar^{\frac{9}{2}}$
\label{onetwo}
\end{center}
\caption{\label{pairSingle}The patterns of zeros of $T(u)$ in the $\mathbb{Z}_4$ sectors 
are encoded by introducing pairs of single-column diagrams. The right column corresponds 
to the 1-strings in the lower-half $u$-plane (associated with $\qbar$). 
The left column corresponds to the 1-strings in the upper half-plane (associated with $q$),
including the real axis, but rotated through 180 degrees. Positions occupied by a 1-string 
are indicated by a solid circle and unoccupied positions are indicated by an open circle. 
The 1-string energies are given by $E_j=\half(j-\half)$. Here, $N=11$, $\sigma=2$, 
$\bar\sigma=-2$ and $\ell=2(\sigma+\bar\sigma)+1=1$.}
\end{figure}
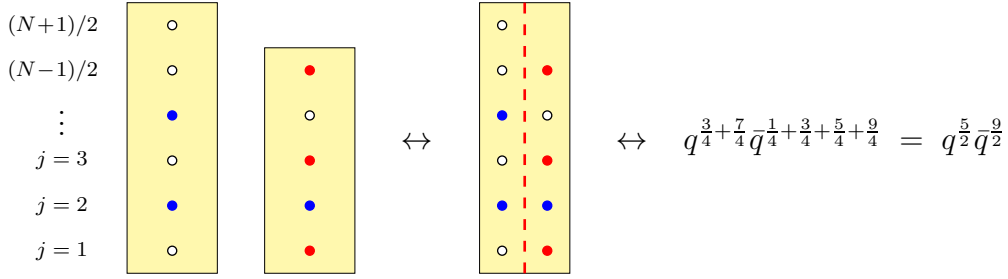

We see that these solutions (eigenvalues) satisfy the crossing symmetry
\be
 \overline{T(\sm{\frac{\pi}{2}}-\bar{u})}\;=\;T(u)
\ee
The choice $\epsilon_j=-1$ for a particular $j$ corresponds to an elementary excitation. 
In principle, up to the overall choice of sign $\epsilon$, there are $2^N$ possible eigenvalues 
allowing for all excitations. However, only $\genfrac{(}{)}{0pt}{}{N}{\frac{N-\ell}{2}}$ of these 
solutions actually occur as eigenvalues and these are determined by selection rules as explained in 
Section~\ref{SecSelection}. For $\ell=1$, the largest eigenvalue $T_{0,1}(u)$ occurs for 
$\epsilon_j=1$ for all $j=1,2,\ldots,N$, that is, there are 2-strings at each position $j$ and 
no 1-strings. The patterns of zeros of $T(u)$ are conveniently encoded  by introducing pairs of 
single-column diagrams as shown in Figure~\ref{pairSingle}. The right column corresponds to 
the 1-strings in the lower-half $u$-plane (associated with $\qbar$). 
The left column corresponds to the 1-strings in the upper half-plane (associated with $q$),
including the real axis, but rotated through 180 degrees. Positions occupied by a 1-string are
indicated by a solid circle and unoccupied positions are indicated by an open circle. 
For $N$ odd and $\ell$ odd, the patterns of zeros of $T_{0,\ell}(u)$ are shown in 
Figure~\ref{groundZ4}.

\psset{unit=.6cm}
\setlength{\unitlength}{.6cm}
\begin{figure}[htbp]
\begin{center}
\begin{pspicture}[shift=-5](-.25,-2.25)(.5,6.5)
\rput(0,.5){\scriptsize $j=1$}
\rput(0,1.5){\scriptsize $j=2$}
\rput(0,2.5){\scriptsize $j=3$}
\rput(0,3.5){$\vdots$}
\rput(-.2,4.5){\scriptsize $(N\!-\!1)/2$}
\rput(-.2,5.5){\scriptsize $(N\!+\!1)/2$}
\end{pspicture}
\hspace{4pt}
\begin{pspicture}[shift=-5](-.25,-2.25)(2,6.5)
\psframe[linewidth=0pt,fillstyle=solid,fillcolor=yellow!40!white](0,0)(2,6)
\psline[linecolor=red,linewidth=1pt,linestyle=dashed,dash=.25 .25](1,0)(1,6)
\psarc[linecolor=black,linewidth=.5pt,fillstyle=solid,fillcolor=white](0.5,5.5){.1}{0}{360}
\psarc[linecolor=black,linewidth=.5pt,fillstyle=solid,fillcolor=white](0.5,4.5){.1}{0}{360}
\psarc[linecolor=black,linewidth=.5pt,fillstyle=solid,fillcolor=white](0.5,3.5){.1}{0}{360}
\psarc[linecolor=black,linewidth=.5pt,fillstyle=solid,fillcolor=white](0.5,2.5){.1}{0}{360}
\psarc[linecolor=black,linewidth=.5pt,fillstyle=solid,fillcolor=white](0.5,1.5){.1}{0}{360}
\psarc[linecolor=black,linewidth=.5pt,fillstyle=solid,fillcolor=white](0.5,0.5){.1}{0}{360}
\psarc[linecolor=black,linewidth=.5pt,fillstyle=solid,fillcolor=white](1.5,4.5){.1}{0}{360}
\psarc[linecolor=black,linewidth=.5pt,fillstyle=solid,fillcolor=white](1.5,3.5){.1}{0}{360}
\psarc[linecolor=black,linewidth=.5pt,fillstyle=solid,fillcolor=white](1.5,2.5){.1}{0}{360}
\psarc[linecolor=black,linewidth=.5pt,fillstyle=solid,fillcolor=white](1.5,1.5){.1}{0}{360}
\psarc[linecolor=black,linewidth=.5pt,fillstyle=solid,fillcolor=white](1.5,0.5){.1}{0}{360}
\rput(1,-.5){\small$\ell=1$}
\rput(1,-1.5){\small$\sigma=\bar\sigma=0$}
\end{pspicture}
\hspace{.6cm}
\begin{pspicture}[shift=-5](-.25,-2.25)(2,6.5)
\psframe[linewidth=0pt,fillstyle=solid,fillcolor=yellow!40!white](0,0)(2,6)
\psline[linecolor=red,linewidth=1pt,linestyle=dashed,dash=.25 .25](1,0)(1,6)
\psarc[linecolor=black,linewidth=.5pt,fillstyle=solid,fillcolor= white](0.5,5.5){.1}{0}{360}
\psarc[linecolor=black,linewidth=.5pt,fillstyle=solid,fillcolor=white](0.5,4.5){.1}{0}{360}
\psarc[linecolor=black,linewidth=.5pt,fillstyle=solid,fillcolor=white](0.5,3.5){.1}{0}{360}
\psarc[linecolor=black,linewidth=.5pt,fillstyle=solid,fillcolor=white](0.5,2.5){.1}{0}{360}
\psarc[linecolor=black,linewidth=.5pt,fillstyle=solid,fillcolor=white](0.5,1.5){.1}{0}{360}
\psarc[linecolor=red,linewidth=.5pt,fillstyle=solid,fillcolor=red](0.5,0.5){.1}{0}{360}
\psarc[linecolor=black,linewidth=.5pt,fillstyle=solid,fillcolor=white](1.5,4.5){.1}{0}{360}
\psarc[linecolor=black,linewidth=.5pt,fillstyle=solid,fillcolor=white](1.5,3.5){.1}{0}{360}
\psarc[linecolor=black,linewidth=.5pt,fillstyle=solid,fillcolor=white](1.5,2.5){.1}{0}{360}
\psarc[linecolor=black,linewidth=.5pt,fillstyle=solid,fillcolor=white](1.5,1.5){.1}{0}{360}
\psarc[linecolor=red,linewidth=.5pt,fillstyle=solid,fillcolor=red](1.5,0.5){.1}{0}{360}
\rput(1,-.5){\small$\ell=3$}
\rput(1,-1.5){\small$\sigma=\bar\sigma=-1$}
\end{pspicture}
\hspace{.6cm}
\begin{pspicture}[shift=-5](-.25,-2.25)(2,6.5)
\psframe[linewidth=0pt,fillstyle=solid,fillcolor=yellow!40!white](0,0)(2,6)
\psline[linecolor=red,linewidth=1pt,linestyle=dashed,dash=.25 .25](1,0)(1,6)
\psarc[linecolor=black,linewidth=.5pt,fillstyle=solid,fillcolor=white](0.5,5.5){.1}{0}{360}
\psarc[linecolor=black,linewidth=.5pt,fillstyle=solid,fillcolor=white](0.5,4.5){.1}{0}{360}
\psarc[linecolor=black,linewidth=.5pt,fillstyle=solid,fillcolor=white](0.5,3.5){.1}{0}{360}
\psarc[linecolor=black,linewidth=.5pt,fillstyle=solid,fillcolor=white](0.5,2.5){.1}{0}{360}
\psarc[linecolor=blue,linewidth=.5pt,fillstyle=solid,fillcolor=blue](0.5,1.5){.1}{0}{360}
\psarc[linecolor=black,linewidth=.5pt,fillstyle=solid,fillcolor=white](0.5,0.5){.1}{0}{360}
\psarc[linecolor=black,linewidth=.5pt,fillstyle=solid,fillcolor=white](1.5,4.5){.1}{0}{360}
\psarc[linecolor=black,linewidth=.5pt,fillstyle=solid,fillcolor=white](1.5,3.5){.1}{0}{360}
\psarc[linecolor=black,linewidth=.5pt,fillstyle=solid,fillcolor=white](1.5,2.5){.1}{0}{360}
\psarc[linecolor=blue,linewidth=.5pt,fillstyle=solid,fillcolor=blue](1.5,1.5){.1}{0}{360}
\psarc[linecolor=black,linewidth=.5pt,fillstyle=solid,fillcolor=white](1.5,0.5){.1}{0}{360}
\rput(1,-.5){\small$\ell=5$}
\rput(1,-1.5){\small$\sigma=\bar\sigma=1$}
\end{pspicture}
\hspace{.6cm}
\begin{pspicture}[shift=-5](-.25,-2.25)(2,6.5)
\psframe[linewidth=0pt,fillstyle=solid,fillcolor=yellow!40!white](0,0)(2,6)
\psline[linecolor=red,linewidth=1pt,linestyle=dashed,dash=.25 .25](1,0)(1,6)
\psarc[linecolor=black,linewidth=.5pt,fillstyle=solid,fillcolor= white](0.5,5.5){.1}{0}{360}
\psarc[linecolor=black,linewidth=.5pt,fillstyle=solid,fillcolor=white](0.5,4.5){.1}{0}{360}
\psarc[linecolor=black,linewidth=.5pt,fillstyle=solid,fillcolor=white](0.5,3.5){.1}{0}{360}
\psarc[linecolor= red,linewidth=.5pt,fillstyle=solid,fillcolor=red](0.5,2.5){.1}{0}{360}
\psarc[linecolor=black,linewidth=.5pt,fillstyle=solid,fillcolor=white](0.5,1.5){.1}{0}{360}
\psarc[linecolor= red,linewidth=.5pt,fillstyle=solid,fillcolor= red](0.5,0.5){.1}{0}{360}
\psarc[linecolor=black,linewidth=.5pt,fillstyle=solid,fillcolor=white](1.5,4.5){.1}{0}{360}
\psarc[linecolor=black,linewidth=.5pt,fillstyle=solid,fillcolor=white](1.5,3.5){.1}{0}{360}
\psarc[linecolor= red,linewidth=.5pt,fillstyle=solid,fillcolor= red](1.5,2.5){.1}{0}{360}
\psarc[linecolor=black,linewidth=.5pt,fillstyle=solid,fillcolor=white](1.5,1.5){.1}{0}{360}
\psarc[linecolor= red,linewidth=.5pt,fillstyle=solid,fillcolor= red](1.5,0.5){.1}{0}{360}
\rput(1,-.5){\small$\ell=7$}
\rput(1,-1.5){\small$\sigma=\bar\sigma=-2$}
\end{pspicture}
\hspace{.6cm}
\begin{pspicture}[shift=-5](-.25,-2.25)(2,6.5)
\psframe[linewidth=0pt,fillstyle=solid,fillcolor=yellow!40!white](0,0)(2,6)
\psline[linecolor=red,linewidth=1pt,linestyle=dashed,dash=.25 .25](1,0)(1,6)
\psarc[linecolor=black,linewidth=.5pt,fillstyle=solid,fillcolor=white](0.5,5.5){.1}{0}{360}
\psarc[linecolor=black,linewidth=.5pt,fillstyle=solid,fillcolor=white](0.5,4.5){.1}{0}{360}
\psarc[linecolor= blue,linewidth=.5pt,fillstyle=solid,fillcolor=blue](0.5,3.5){.1}{0}{360}
\psarc[linecolor=black,linewidth=.5pt,fillstyle=solid,fillcolor=white](0.5,2.5){.1}{0}{360}
\psarc[linecolor= blue,linewidth=.5pt,fillstyle=solid,fillcolor= blue](0.5,1.5){.1}{0}{360}
\psarc[linecolor=black,linewidth=.5pt,fillstyle=solid,fillcolor=white](0.5,0.5){.1}{0}{360}
\psarc[linecolor=black,linewidth=.5pt,fillstyle=solid,fillcolor=white](1.5,4.5){.1}{0}{360}
\psarc[linecolor= blue,linewidth=.5pt,fillstyle=solid,fillcolor= blue](1.5,3.5){.1}{0}{360}
\psarc[linecolor=black,linewidth=.5pt,fillstyle=solid,fillcolor=white](1.5,2.5){.1}{0}{360}
\psarc[linecolor= blue,linewidth=.5pt,fillstyle=solid,fillcolor= blue](1.5,1.5){.1}{0}{360}
\psarc[linecolor=black,linewidth=.5pt,fillstyle=solid,fillcolor=white](1.5,0.5){.1}{0}{360}
\rput(1,-.5){\small$\ell=9$}
\rput(1,-1.5){\small$\sigma=\bar\sigma=2$}
\end{pspicture}
\hspace{.6cm}
\begin{pspicture}[shift=-5](-.25,-2.25)(2,6.5)
\psframe[linewidth=0pt,fillstyle=solid,fillcolor=yellow!40!white](0,0)(2,6)
\psline[linecolor=red,linewidth=1pt,linestyle=dashed,dash=.25 .25](1,0)(1,6)
\psarc[linecolor=black,linewidth=.5pt,fillstyle=solid,fillcolor= white](0.5,5.5){.1}{0}{360}
\psarc[linecolor= red,linewidth=.5pt,fillstyle=solid,fillcolor= red](0.5,4.5){.1}{0}{360}
\psarc[linecolor=black,linewidth=.5pt,fillstyle=solid,fillcolor=white](0.5,3.5){.1}{0}{360}
\psarc[linecolor= red,linewidth=.5pt,fillstyle=solid,fillcolor= red](0.5,2.5){.1}{0}{360}
\psarc[linecolor=black,linewidth=.5pt,fillstyle=solid,fillcolor=white](0.5,1.5){.1}{0}{360}
\psarc[linecolor= red,linewidth=.5pt,fillstyle=solid,fillcolor= red](0.5,0.5){.1}{0}{360}
\psarc[linecolor= red,linewidth=.5pt,fillstyle=solid,fillcolor= red](1.5,4.5){.1}{0}{360}
\psarc[linecolor=black,linewidth=.5pt,fillstyle=solid,fillcolor=white](1.5,3.5){.1}{0}{360}
\psarc[linecolor= red,linewidth=.5pt,fillstyle=solid,fillcolor= red](1.5,2.5){.1}{0}{360}
\psarc[linecolor=black,linewidth=.5pt,fillstyle=solid,fillcolor=white](1.5,1.5){.1}{0}{360}
\psarc[linecolor= red,linewidth=.5pt,fillstyle=solid,fillcolor= red](1.5,0.5){.1}{0}{360}
\rput(1,-.5){\small$\ell=11$}
\rput(1,-1.5){\small$\sigma=\bar\sigma=-3$}
\rput(3,3){$\ldots$}
\end{pspicture}
\end{center}
\caption{Groundstate configurations in the $\mathbb{Z}_4$ sectors with $\ell$ defects 
($N$, $\ell$ odd). The quantum numbers $\sigma, \bar\sigma$ are as shown and the 
conformal weights are $(\Delta,\bar\Delta)=(\Delta_{\ell/2},\Delta_{\ell/2})$.
\label{groundZ4}}
\end{figure}
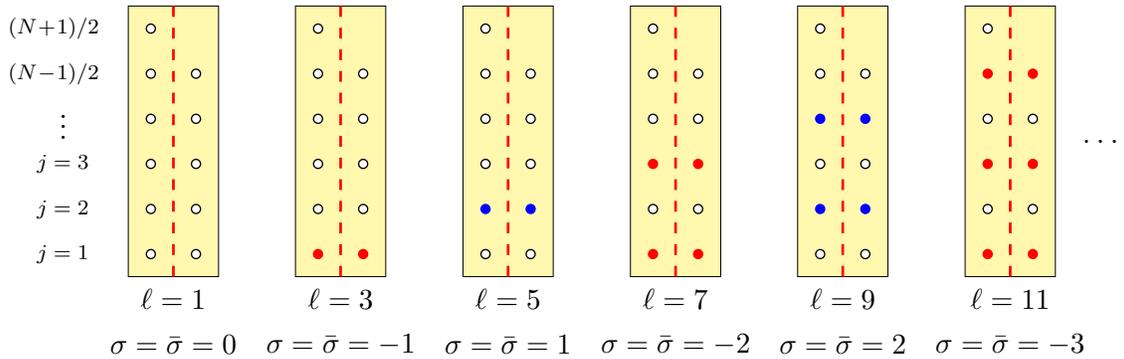

Next, we separate the contribution from zeros in the upper and lower half-planes by keeping 
$\epsilon_j$ for $j=1,2,\ldots,(N+1)/2$ and setting
\be
 \bar\epsilon_j\;=\;\epsilon_{N+1-j}\;=\;\pm 1,\qquad j=1,2,\ldots,\mbox{\small $\frac{N-1}{2}$}
\ee
By convention, we treat the zeros on the real axis labelled by $j=(N+1)/2$ as if they are in 
the upper half-plane. Setting $\epsilon=\mu\,\prod_{j=1}^{(N-1)/2} \bar\epsilon_j$ gives
\be
 T(u)\;=\;\frac{\mu\, (-i)^{N/2}}{2^{N-1/2}e^{Niu}}
  \prod_{j=1}^{\frac{N+1}{2}}{\Big(e^{2i u}+i\epsilon_j  \tan{\frac{(2j-1)\pi}{4N}}\Big)}
  \prod_{j=1}^{\frac{N-1}{2}}{\Big(\bar\epsilon_j e^{2i u}+i \cot{\frac{(2j-1)\pi}{4N}}\Big)}
\label{T(u)Nodd}
\ee
We fix $\mu=+1$ to ensure that $T_{0,1}(0)=1$ consistent with (\ref{limD}). 
Taking the ratio of (\ref{T(u)Nodd}) with precisely one $\epsilon_j=-1$ or $\bar\epsilon_j=-1$ 
to (\ref{T(u)Nodd}) with all $\epsilon_j=\bar\epsilon_j=+1$, and then taking the limit with a 
fixed aspect ratio $\delta=M/N$, gives
\bea
 \lim_{M,N\to\infty}\bigg(\frac{e^{2iu}-i\tan\frac{(2j-1)\pi}{4N}}{e^{2iu}
  +i\tan\frac{(2j-1)\pi}{4N}}\bigg)^{\!M}\!\!
&=&\exp[-(j-\half)\pi i\,\delta\,e^{-2iu}]\;=\;q^{E_j}\\
  \lim_{M,N\to\infty}\bigg(\frac{-e^{2iu}+i\cot\frac{(2j-1)\pi}{4N}}{e^{2iu}
   +i\cot\frac{(2j-1)\pi}{4N}}\bigg)^{\!M}\!\!
 &=&\;\exp[(j-\half)\pi i\,\delta\,e^{2iu}]\;\;=\;\;\qbar^{E_j}
\eea
where 
\be
 E_j\;=\;\half(j-\half),\qquad\mbox{$N$, $\ell$ odd}
\label{Z4Ej}
\ee

It follows that the conformal partition function in the $\mathbb{Z}_4$ sectors with $\ell$ defects is
\be
 Z_\ell(q)\;=\;(q\qbar)^{-c/24+\Delta_{1/2}}\sum_{\epsilon,\bar\epsilon} 
  q^{\sum_{j=1}^{(N+1)/2}\delta(\epsilon_j,-1)E_j}
  \qbar^{\sum_{j=1}^{(N-1)/2}\delta(\bar\epsilon_j,-1)E_j}
\ee
where the sum and the allowed values of $\epsilon_j$ and $\bar\epsilon_j$ are determined 
by selection rules as explained in Section~\ref{SecSelection}. 
Here $\delta(j,k)=\delta_{j,k}$ is the Kronecker delta. 
The prefactor involving the central charge and conformal weights comes from the largest 
eigenvalue $T_{0,\ell}(u)$ with $\ell=1$ and is obtained by applying Euler-Maclaurin as 
explained in Section~\ref{EulerM}.

\subsection{Ramond and Neveu-Schwarz sectors ($N$ even, $\ell$ even)}

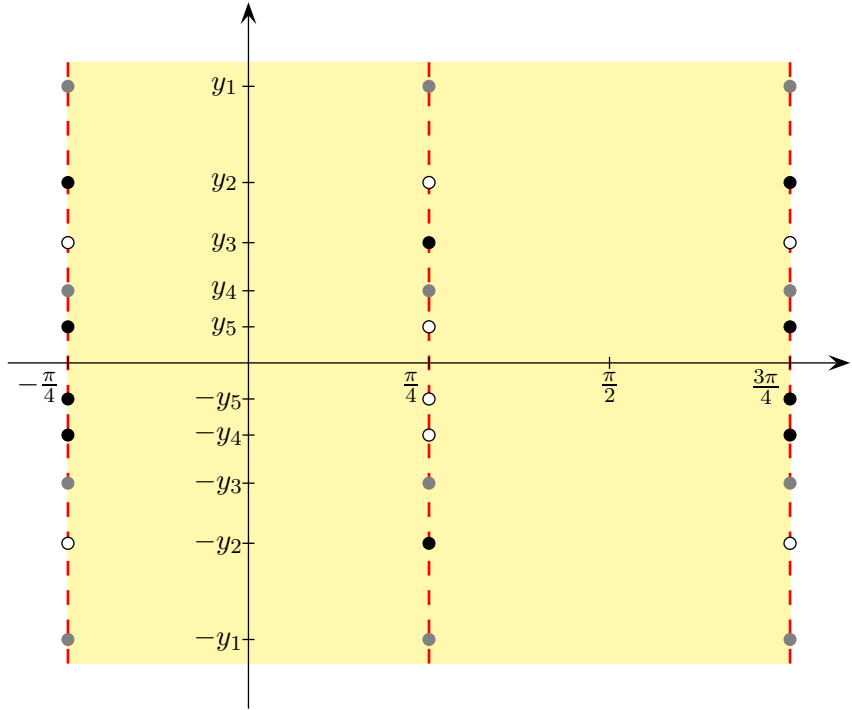
\begin{figure}[htb]
\psset{unit=.8cm}
\setlength{\unitlength}{.8cm}
\begin{center}
\begin{pspicture}[shift=-.45](-.25,.5)(14,12)
\psframe[linecolor=yellow!40!white,linewidth=0pt,fillstyle=solid,
  fillcolor=yellow!40!white](1,1)(13,11)
\psline[linecolor=black,linewidth=.5pt,arrowsize=6pt]{->}(4,.25)(4,12)
\psline[linecolor=black,linewidth=.5pt,arrowsize=6pt]{->}(0,6)(14,6)
\psline[linecolor=red,linewidth=1pt,linestyle=dashed,dash=.25 .25](1,1)(1,11)
\psline[linecolor=red,linewidth=1pt,linestyle=dashed,dash=.25 .25](7,1)(7,11)
\psline[linecolor=red,linewidth=1pt,linestyle=dashed,dash=.25 .25](13,1)(13,11)
\psline[linecolor=black,linewidth=.5pt](1,5.9)(1,6.1)
\psline[linecolor=black,linewidth=.5pt](7,5.9)(7,6.1)
\psline[linecolor=black,linewidth=.5pt](10,5.9)(10,6.1)
\psline[linecolor=black,linewidth=.5pt](13,5.9)(13,6.1)
\rput(.5,5.6){\small $-\frac{\pi}{4}$}
\rput(6.7,5.6){\small $\frac{\pi}{4}$}
\rput(10,5.6){\small $\frac{\pi}{2}$}
\rput(12.6,5.6){\small $\frac{3\pi}{4}$}
\psline[linecolor=black,linewidth=.5pt](3.9,6.6)(4.1,6.6)
\psline[linecolor=black,linewidth=.5pt](3.9,7.2)(4.1,7.2)
\psline[linecolor=black,linewidth=.5pt](3.9,8.0)(4.1,8.0)
\psline[linecolor=black,linewidth=.5pt](3.9,9.0)(4.1,9.0)
\psline[linecolor=black,linewidth=.5pt](3.9,10.6)(4.1,10.6)
\psline[linecolor=black,linewidth=.5pt](3.9,5.4)(4.1,5.4)
\psline[linecolor=black,linewidth=.5pt](3.9,4.8)(4.1,4.8)
\psline[linecolor=black,linewidth=.5pt](3.9,4.0)(4.1,4.0)
\psline[linecolor=black,linewidth=.5pt](3.9,3.0)(4.1,3.0)
\psline[linecolor=black,linewidth=.5pt](3.9,1.4)(4.1,1.4)
\rput(3.6,6.6){\small $y_5$}
\rput(3.6,7.2){\small $y_4$}
\rput(3.6,8.0){\small $y_3$}
\rput(3.6,9.0){\small $y_2$}
\rput(3.6,10.6){\small $y_1$}
\rput(3.5,5.4){\small $-y_5$}
\rput(3.5,4.8){\small $-y_4$}
\rput(3.5,4.0){\small $-y_3$}
\rput(3.5,3.0){\small $-y_2$}
\rput(3.5,1.4){\small $-y_1$}
\psarc[linecolor=black,linewidth=0pt,fillstyle=solid,fillcolor=black](1,6.6){.1}{0}{360}
\psarc[linecolor=gray,linewidth=0pt,fillstyle=solid,fillcolor=gray](1,7.2){.1}{0}{360}
\psarc[linecolor=black,linewidth=.5pt,fillstyle=solid,fillcolor=white](1,8.0){.1}{0}{360}
\psarc[linecolor=black,linewidth=0pt,fillstyle=solid,fillcolor=black](1,9.0){.1}{0}{360}
\psarc[linecolor=gray,linewidth=0pt,fillstyle=solid,fillcolor=gray](1,10.6){.1}{0}{360}
\psarc[linecolor=black,linewidth=.5pt,fillstyle=solid,fillcolor=white](7,6.6){.1}{0}{360}
\psarc[linecolor=gray,linewidth=0pt,fillstyle=solid,fillcolor=gray](7,7.2){.1}{0}{360}
\psarc[linecolor=black,linewidth=0pt,fillstyle=solid,fillcolor=black](7,8.0){.1}{0}{360}
\psarc[linecolor=black,linewidth=.5pt,fillstyle=solid,fillcolor=white](7,9.0){.1}{0}{360}
\psarc[linecolor=gray,linewidth=0pt,fillstyle=solid,fillcolor=gray](7,10.6){.1}{0}{360}
\psarc[linecolor=black,linewidth=0pt,fillstyle=solid,fillcolor=black](13,6.6){.1}{0}{360}
\psarc[linecolor=gray,linewidth=0pt,fillstyle=solid,fillcolor=gray](13,7.2){.1}{0}{360}
\psarc[linecolor=black,linewidth=.5pt,fillstyle=solid,fillcolor=white](13,8.0){.1}{0}{360}
\psarc[linecolor=black,linewidth=0pt,fillstyle=solid,fillcolor=black](13,9.0){.1}{0}{360}
\psarc[linecolor=gray,linewidth=0pt,fillstyle=solid,fillcolor=gray](13,10.6){.1}{0}{360}
\psarc[linecolor=black,linewidth=0pt,fillstyle=solid,fillcolor=black](1,5.4){.1}{0}{360}
\psarc[linecolor=black,linewidth=0pt,fillstyle=solid,fillcolor=black](1,4.8){.1}{0}{360}
\psarc[linecolor=gray,linewidth=0pt,fillstyle=solid,fillcolor=gray](1,4.0){.1}{0}{360}
\psarc[linecolor=black,linewidth=.5pt,fillstyle=solid,fillcolor=white](1,3.0){.1}{0}{360}
\psarc[linecolor=gray,linewidth=0pt,fillstyle=solid,fillcolor=gray](1,1.4){.1}{0}{360}
\psarc[linecolor=black,linewidth=.5pt,fillstyle=solid,fillcolor=white](7,5.4){.1}{0}{360}
\psarc[linecolor=black,linewidth=.5pt,fillstyle=solid,fillcolor=white](7,4.8){.1}{0}{360}
\psarc[linecolor=gray,linewidth=0pt,fillstyle=solid,fillcolor=gray](7,4.0){.1}{0}{360}
\psarc[linecolor=black,linewidth=0pt,fillstyle=solid,fillcolor=black](7,3.0){.1}{0}{360}
\psarc[linecolor=gray,linewidth=0pt,fillstyle=solid,fillcolor=gray](7,1.4){.1}{0}{360}
\psarc[linecolor=black,linewidth=0pt,fillstyle=solid,fillcolor=black](13,5.4){.1}{0}{360}
\psarc[linecolor=black,linewidth=0pt,fillstyle=solid,fillcolor=black](13,4.8){.1}{0}{360}
\psarc[linecolor=gray,linewidth=0pt,fillstyle=solid,fillcolor=gray](13,4.0){.1}{0}{360}
\psarc[linecolor=black,linewidth=.5pt,fillstyle=solid,fillcolor=white](13,3.0){.1}{0}{360}
\psarc[linecolor=gray,linewidth=0pt,fillstyle=solid,fillcolor=gray](13,1.4){.1}{0}{360}
\end{pspicture}
\end{center}
\caption{A typical pattern of zeros in the complex $u$-plane for the $\ell$ even sectors. Here,
$N=12$ and $\ell=2$. The ordinates of the locations of the zeros $u_j$ are
$y_j=-\half \log \tan{\frac{j\pi}{N}}, j=1,2,\ldots,N/2-1$. At each position $j$, there is either 
two 1-strings with $\Re u_j=\pi/4$, two 2-strings with real parts $\Re u_j=-\pi/4, 3\pi/4$ or 
one 1-string and one 2-string. A double zero is indicated by a black circle, a single zero 
by a grey circle and an unoccupied position by an open circle.\label{uplaneR}}
\end{figure}

In the $\ell$ even sectors, the sector-by-sector inversion identity for the eigenvalues is
\be
 T(u)T(u+\frac{\pi}{2})\;=\;\big(\cos^{N}\!u+ (-1)^{(N-\ell)/2}\sin^{N}\!u\big)^2
\ee
Using the identities
\bea
 \cos^N\!u +(-1)^{N/2} \sin^N\!u \!&=&\! \frac{e^{-Niu}}{2^{N-1}} \prod_{j=1}^{N/2} 
  \Big( e^{4iu} + \tan^2 \frac{(2j\!-\!1) \pi}{2N}\Big),\qquad \mbox{$\ell/2$ even}\qquad\mbox{}\\
 \cos^N\!u -(-1)^{N/2} \sin^N\!u \!&=&\!\frac{N e^{(2-N)iu}}{2^{N-1}} \prod_{j=1}^{N/2-1} 
  \Big( e^{4iu} + \tan^2 \frac{j \pi}{N}\Big),\qquad \mbox{$\ell/2$ odd}
\eea
we see that for the two parities of $\ell/2$
\bea
\big(\cos^N\!u\!+\! (-1)^{N/2}\sin^N\!u\big)^2\!\!&=&\!\! \frac{e^{-2Niu}}{2^{2N-2} } 
  \prod_{j=1}^{N/2}\Big( e^{2iu}\! +\! i \epsilon_j \tan \frac{(2j\!-\!1)\pi}{2N}\Big)\!
   \Big( e^{2iu} \!-\! i \epsilon_j \tan \frac{(2j\!-\!1) \pi}{2N}\Big) \nonumber\\
 &&\hspace{-.6in}\mbox{}\times\;\prod_{j=1}^{N/2}\; \Big( e^{2iu} 
   + i \mu_j \tan \frac{(2j\!-\!1)\pi}{2N}\Big)\! \Big( e^{2iu} - i \mu_j \tan \frac{(2j\!-\!1) \pi}{2N}\Big)\\
 \big(\cos^N\!u - (-1)^{N/2}\sin^N\!u\big)^2\!&=&\! \frac{N^2e^{(4-2N)iu}}{2^{2N-2} } 
   \prod_{j=1}^{N/2-1} \Big( e^{2iu} + i \epsilon_j \tan \frac{j\pi}{N}\Big)\!
   \Big( e^{2iu} - i \epsilon_j \tan \frac{j \pi}{N}\Big) \nonumber\\
 &&\hspace{-.6in}\mbox{}\times\prod_{j=1}^{N/2-1} \Big( e^{2iu} + i \mu_j \tan \frac{j\pi}{N}\Big)\! 
   \Big( e^{2iu} - i \mu_j \tan \frac{j \pi}{N}\Big)
\eea
where $\epsilon_j^2 = \mu_j^2=1$. Sharing out the zeros to solve the functional equation, gives
\bea
T(u)=\begin{cases}
\disp \frac{\epsilon(-i)^{\frac{N}{2}}e^{-Niu}}{2^{N-1}} \prod_{j=1}^{N/2} \Big( e^{2iu} 
  + i \epsilon_j \tan \frac{(2j\!-\!1)\pi}{2N}\Big) \Big( e^{2iu} 
  + i \mu_j \tan \frac{(2j\!-\!1) \pi}{2N}\Big),&\mbox{$\frac{\ell}{2}$ even}\\
 \disp\frac{\epsilon(-i)^{\frac{N-2}{2}}Ne^{(2-N)iu}}{2^{N-1}} \prod_{j=1}^{N/2-1} \Big( e^{2iu} 
  + i \epsilon_j \tan \frac{j\pi}{N}\Big) \Big( e^{2iu} 
  + i \mu_j \tan \frac{j\pi}{N}\Big),&\mbox{$\frac{\ell}{2}$ odd}
\end{cases}
\eea
We see that these solutions (eigenvalues) satisfy the 
crossing symmetry
\be
 \overline{T(\sm{\frac{\pi}{2}}-\bar{u})}\;=\;T(u)
\ee
The ordinates of the locations of zeros are
\be
 y_j=\begin{cases}
  \disp-\half \log \tan{\frac{(j\!-\!\half)\pi}{N}}, & \mbox{$\ell/2$ even},\ j=1,2,\ldots,N/2\\[6pt]
  \disp-\half \log \tan{\frac{j\pi}{N}}, & \mbox{$\ell/2$ odd},\  j=1,2,\ldots,N/2-1
\end{cases}
\ee
A typical pattern of zeros is shown in Figure~\ref{uplaneR}.

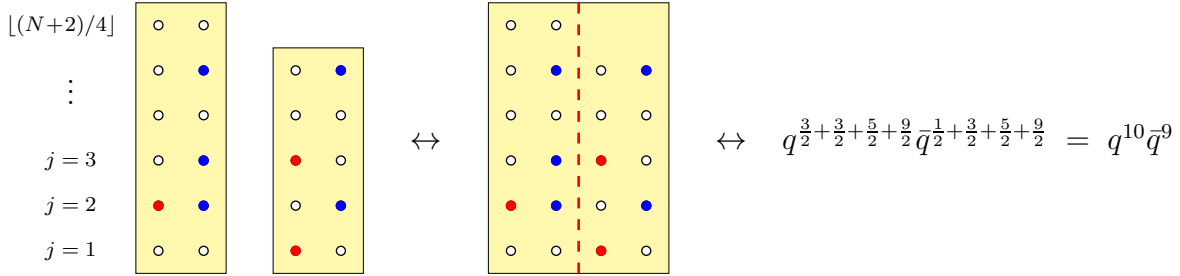
\begin{figure}[p]
\psset{unit=.6cm}
\setlength{\unitlength}{.6cm}
\begin{center}
\begin{pspicture}[shift=-4](-.25,-1.25)(.5,6.5)
\rput(0,.5){\scriptsize $j=1$}
\rput(0,1.5){\scriptsize $j=2$}
\rput(0,2.5){\scriptsize $j=3$}
\rput(0,4.2){$\vdots$}
\rput(-.2,5.5){\scriptsize $\floor{(N\!+\!2)/4}$}
\end{pspicture}
\hspace{4pt}
\begin{pspicture}[shift=-3](-.25,-.25)(2,6)
\psframe[linewidth=0pt,fillstyle=solid,fillcolor=yellow!40!white](0,0)(2,6)
\multirput(0,0)(0,1){6}{\multirput(0,0)(1,0){2}{\psarc[linecolor=black,linewidth=.5pt,fillstyle=solid,
  fillcolor=white](.5,.5){.1}{0}{360}}}
\psarc[linecolor=blue,linewidth=.5pt,fillstyle=solid,fillcolor=blue](1.5,4.5){.1}{0}{360}
\psarc[linecolor=blue,linewidth=.5pt,fillstyle=solid,fillcolor=blue](1.5,2.5){.1}{0}{360}
\psarc[linecolor=blue,linewidth=.5pt,fillstyle=solid,fillcolor=blue](1.5,1.5){.1}{0}{360}
\psarc[linecolor=red,linewidth=.5pt,fillstyle=solid,fillcolor=red](.5,1.5){.1}{0}{360}
\end{pspicture}
\hspace{.2cm}
\begin{pspicture}[shift=-3](-.25,-.25)(2,5)
\psframe[linewidth=0pt,fillstyle=solid,fillcolor=yellow!40!white](0,0)(2,5)
\multirput(0,0)(0,1){5}{\multirput(0,0)(1,0){2}{\psarc[linecolor=black,linewidth=.5pt,fillstyle=solid,
  fillcolor=white](.5,.5){.1}{0}{360}}}
\psarc[linecolor=blue,linewidth=.5pt,fillstyle=solid,fillcolor=blue](1.5,1.5){.1}{0}{360}
\psarc[linecolor=red,linewidth=.5pt,fillstyle=solid,fillcolor=red](.5,.5){.1}{0}{360}
\psarc[linecolor=red,linewidth=.5pt,fillstyle=solid,fillcolor=red](.5,2.5){.1}{0}{360}
\psarc[linecolor=blue,linewidth=.5pt,fillstyle=solid,fillcolor=blue](1.5,4.5){.1}{0}{360}
\end{pspicture}
\hspace{.2cm}\ \ $\leftrightarrow$ \hspace{.2cm}
\begin{pspicture}[shift=-3](-.25,-.25)(4,6)
\psframe[linewidth=0pt,fillstyle=solid,fillcolor=yellow!40!white](0,0)(4,6)
\psline[linecolor=red,linewidth=1pt,linestyle=dashed,dash=.25 .25](2,0)(2,6)
\multirput(0,0)(0,1){5}{\multirput(0,0)(1,0){4}{\psarc[linecolor=black,linewidth=.5pt,fillstyle=solid,
  fillcolor=white](.5,.5){.1}{0}{360}}}
\multirput(0,5)(1,0){2}{\psarc[linecolor=black,linewidth=.5pt,fillstyle=solid,
  fillcolor=white](.5,.5){.1}{0}{360}}
\psarc[linecolor=blue,linewidth=.5pt,fillstyle=solid,fillcolor=blue](1.5,4.5){.1}{0}{360}
\psarc[linecolor=blue,linewidth=.5pt,fillstyle=solid,fillcolor=blue](1.5,2.5){.1}{0}{360}
\psarc[linecolor=blue,linewidth=.5pt,fillstyle=solid,fillcolor=blue](1.5,1.5){.1}{0}{360}
\psarc[linecolor=red,linewidth=.5pt,fillstyle=solid,fillcolor=red](.5,1.5){.1}{0}{360}
\psarc[linecolor=blue,linewidth=.5pt,fillstyle=solid,fillcolor=blue](3.5,1.5){.1}{0}{360}
\psarc[linecolor=blue,linewidth=.5pt,fillstyle=solid,fillcolor=blue](3.5,4.5){.1}{0}{360}
\psarc[linecolor=red,linewidth=.5pt,fillstyle=solid,fillcolor=red](2.5,.5){.1}{0}{360}
\psarc[linecolor=red,linewidth=.5pt,fillstyle=solid,fillcolor=red](2.5,2.5){.1}{0}{360}
\end{pspicture}
\hspace{.2cm}\ \ $\leftrightarrow$ \hspace{.2cm} 
$q^{\frac{3}{2}+\frac{3}{2}+\frac{5}{2}+\frac{9}{2}}\qbar^{\frac{1}{2}+\frac{3}{2}+\frac{5}{2}+\frac{9}{2}}
  \;=\;q^{10}\qbar^{9}$
\end{center}
\vspace{-.2in}
\caption{\label{pairDoubleR} The patterns of zeros of $T(u)$ in the Ramond sectors 
($N$ even, $\ell/2$ even) are encoded by introducing pairs of double-column diagrams. 
The right double-column corresponds to the 1-strings in the lower-half $u$-plane (associated 
with $\qbar$). The left double-column corresponds to the 1-strings in the upper half-plane
(associated with $q$), including the real axis, but rotated through 180 degrees. At each position 
in a double-column diagram, there are 0, 1 or 2 1-strings. Positions occupied by a 1-string 
are indicated by a solid circle and unoccupied positions are indicated by an open circle. 
The 1-string energies are given by $E_j=(j-\half)$. There are $\floor{\frac{N+2}{4}}$ positions 
on the left and $\floor{\frac{N}{4}}$ on the right. Here, $N=22$, $\sigma=2$, $\bar\sigma=0$ and  
$\ell=2(\sigma+\bar\sigma)=4$.}
\end{figure}

The choice $\epsilon_j=-1$ or $\mu_j=-1$ for a particular $j$ corresponds to an elementary
excitation. In principle, up to the overall choice of sign $\epsilon$, there are either $2^N$ 
or $2^{N-2}$ possible eigenvalues allowing for all excitations. 
However, only $\genfrac{(}{)}{0pt}{}{N}{\frac{N-\ell}{2}}$ of these solutions actually occur as eigenvalues and 
these are determined by selection rules as explained in Section~\ref{SecSelection}. For $\ell=0$, 
the largest eigenvalue $T_{0,0}(u)$ occurs for $\epsilon_j=\mu_j=1$ for all $j=1,2,\ldots,N$, 
that is, there are 2-strings at each position $j$ and no 1-strings. 
The patterns of zeros of $T(u)$ are conveniently encoded  by introducing pairs of 
double-column diagrams as shown in Figures~\ref{pairDoubleR} and \ref{pairDoubleNS}. 
The right column corresponds to the 1-strings in the lower-half $u$-plane (associated with 
$\qbar$). The left column corresponds to the 1-strings in the upper half-plane (associated with 
$q$), including the real axis, but rotated through 180 degrees. Positions occupied by a 1-string 
are indicated by a solid circle and unoccupied positions are indicated by an open circle. 
For $N$ even and $\ell/2$ even or odd, the patterns of zeros of $T_{0,\ell}(u)$ are shown in 
Figures~\ref{groundR} and \ref{groundNS}.

\psset{unit=.6cm}
\setlength{\unitlength}{.6cm}
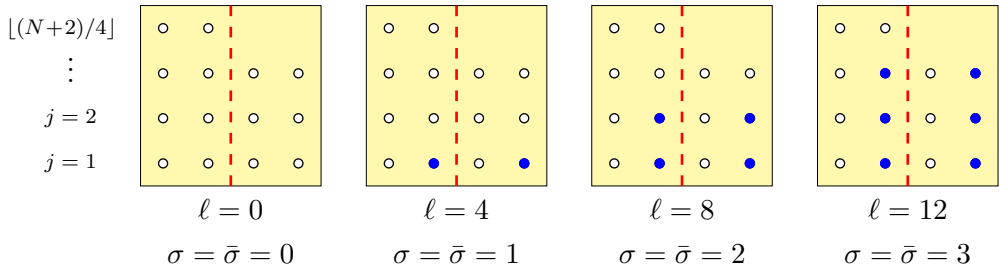
\begin{figure}[p]
\begin{center}
\begin{pspicture}[shift=-5](-.25,-2.25)(.6,5)
\rput(0,.5){\scriptsize $j=1$}
\rput(0,1.5){\scriptsize $j=2$}
\rput(0,2.75){$\vdots$}
\rput(-.2,3.5){\scriptsize $\floor{(N\!+\!2)/4}$}
\end{pspicture}
\hspace{4pt}
\begin{pspicture}[shift=-5](-.25,-2.25)(20,5)
\multirput(0,0)(5,0){4}{\psframe[linewidth=0pt,fillstyle=solid,fillcolor=yellow!40!white](0,0)(4,4)}
\multirput(0,0)(5,0){4}{\psline[linecolor=red,linewidth=1pt,linestyle=dashed,
  dash=.25 .25](2,0)(2,4)}
\multirput(0,0)(5,0){4}{
\multirput(0,0)(0,1){3}{\multirput(0,0)(1,0){4}{\psarc[linecolor=black,linewidth=.5pt,fillstyle=solid,
  fillcolor=white](0.5,0.5){.1}{0}{360}}}}
\multirput(0,0)(5,0){4}{
\multirput(0,3)(1,0){2}{\psarc[linecolor=black,linewidth=.5pt,fillstyle=solid,
  fillcolor=white](0.5,0.5){.1}{0}{360}}}
\rput(2,-.5){\small$\ell=0$}
\rput(2,-1.5){\small$\sigma=\bar\sigma=0$}
\rput(7,-.5){\small$\ell=4$}
\rput(7,-1.5){\small$\sigma=\bar\sigma=1$}
\rput(12,-.5){\small$\ell=8$}
\rput(12,-1.5){\small$\sigma=\bar\sigma=2$}
\rput(17,-.5){\small$\ell=12$}
\rput(17,-1.5){\small$\sigma=\bar\sigma=3$}
\multirput(0,0)(5,0){3}{\multirput(0,0)(2,0){2}{\psarc[linecolor= blue,linewidth=.5pt,fillstyle=solid,
  fillcolor=blue](6.5,.5){.1}{0}{360}}}
\multirput(5,0)(5,0){2}{\multirput(0,0)(2,0){2}{\psarc[linecolor= blue,linewidth=.5pt,fillstyle=solid,
  fillcolor=blue](6.5,1.5){.1}{0}{360}}}
\multirput(10,0)(5,0){1}{\multirput(0,0)(2,0){2}{\psarc[linecolor= blue,linewidth=.5pt,fillstyle=solid,
  fillcolor=blue](6.5,2.5){.1}{0}{360}}}
\end{pspicture}
\vspace{-.2in}
\end{center}
\caption{Groundstate configurations in the Ramond sectors with $\ell$ defects 
($N$, $\ell/2$ even). Here $N=14$, the quantum numbers $\sigma, \bar\sigma$ are as shown 
and the conformal weights are $(\Delta,\bar\Delta)=(\Delta_{\ell/2},\Delta_{\ell/2})$.
\label{groundR}}
\end{figure}

\begin{figure}[p]
\psset{unit=.6cm}
\setlength{\unitlength}{.6cm}
\begin{center}
\begin{pspicture}[shift=-4](-.25,-1.25)(.5,6.5)
\rput(0,.5){\scriptsize $j=1$}
\rput(0,1.5){\scriptsize $j=2$}
\rput(0,2.5){\scriptsize $j=3$}
\rput(0,4.2){$\vdots$}
\rput(-.2,5.5){\scriptsize $\floor{N/4}$}
\end{pspicture}
\hspace{4pt}
\begin{pspicture}[shift=-3](-.25,-.25)(2,6)
\psframe[linewidth=0pt,fillstyle=solid,fillcolor=yellow!40!white](0,0)(2,6)
\multirput(0,0)(0,1){6}{\multirput(0,0)(1,0){2}{\psarc[linecolor=black,linewidth=.5pt,fillstyle=solid,
  fillcolor=white](.5,.5){.1}{0}{360}}}
\psarc[linecolor=blue,linewidth=.5pt,fillstyle=solid,fillcolor=blue](1.5,4.5){.1}{0}{360}
\psarc[linecolor=blue,linewidth=.5pt,fillstyle=solid,fillcolor=blue](1.5,2.5){.1}{0}{360}
\psarc[linecolor=blue,linewidth=.5pt,fillstyle=solid,fillcolor=blue](1.5,1.5){.1}{0}{360}
\psarc[linecolor=red,linewidth=.5pt,fillstyle=solid,fillcolor=red](.5,1.5){.1}{0}{360}
\end{pspicture}
\hspace{.2cm}
\begin{pspicture}[shift=-3](-.25,-.25)(2,5)
\psframe[linewidth=0pt,fillstyle=solid,fillcolor=yellow!40!white](0,0)(2,5)
\multirput(0,0)(0,1){5}{\multirput(0,0)(1,0){2}{\psarc[linecolor=black,linewidth=.5pt,fillstyle=solid,
  fillcolor=white](.5,.5){.1}{0}{360}}}
\psarc[linecolor=red,linewidth=.5pt,fillstyle=solid,fillcolor=red](.5,4.5){.1}{0}{360}
\psarc[linecolor=blue,linewidth=.5pt,fillstyle=solid,fillcolor=blue](1.5,.5){.1}{0}{360}
\psarc[linecolor=red,linewidth=.5pt,fillstyle=solid,fillcolor=red](.5,.5){.1}{0}{360}
\psarc[linecolor=blue,linewidth=.5pt,fillstyle=solid,fillcolor=blue](1.5,3.5){.1}{0}{360}
\end{pspicture}
\hspace{.2cm}\ \ $\leftrightarrow$ \hspace{.2cm}
\begin{pspicture}[shift=-3](-.25,-.25)(4,6)
\psframe[linewidth=0pt,fillstyle=solid,fillcolor=yellow!40!white](0,0)(4,6)
\psline[linecolor=red,linewidth=1pt,linestyle=dashed,dash=.25 .25](2,0)(2,6)
\multirput(0,0)(0,1){5}{\multirput(0,0)(1,0){4}{\psarc[linecolor=black,linewidth=.5pt,fillstyle=solid,
  fillcolor=white](.5,.5){.1}{0}{360}}}
\multirput(0,5)(1,0){2}{\psarc[linecolor=black,linewidth=.5pt,fillstyle=solid,
  fillcolor=white](.5,.5){.1}{0}{360}}
\psarc[linecolor=blue,linewidth=.5pt,fillstyle=solid,fillcolor=blue](1.5,1.5){.1}{0}{360}
\psarc[linecolor=blue,linewidth=.5pt,fillstyle=solid,fillcolor=blue](1.5,2.5){.1}{0}{360}
\psarc[linecolor=blue,linewidth=.5pt,fillstyle=solid,fillcolor=blue](1.5,4.5){.1}{0}{360}
\psarc[linecolor=red,linewidth=.5pt,fillstyle=solid,fillcolor=red](.5,1.5){.1}{0}{360}
\psarc[linecolor=blue,linewidth=.5pt,fillstyle=solid,fillcolor=blue](3.5,3.5){.1}{0}{360}
\psarc[linecolor=blue,linewidth=.5pt,fillstyle=solid,fillcolor=blue](3.5,.5){.1}{0}{360}
\psarc[linecolor=red,linewidth=.5pt,fillstyle=solid,fillcolor=red](2.5,4.5){.1}{0}{360}
\psarc[linecolor=red,linewidth=.5pt,fillstyle=solid,fillcolor=red](2.5,.5){.1}{0}{360}
\end{pspicture}
\hspace{.2cm}\ \ $\leftrightarrow$ \hspace{.2cm} 
$q^{2+2+3+5}\qbar^{1+1+4+5}\;=\;q^{12}\qbar^{11}$
\end{center}
\vspace{-.2in}
\caption{\label{pairDoubleNS} The patterns of zeros of $T(u)$ in the Neveu-Schwarz sectors 
($N$ even, $\ell/2$ odd) are encoded by introducing pairs of double-column diagrams. 
The right double-column corresponds to the 1-strings in the lower-half $u$-plane 
(associated with $\qbar$). The left double-column corresponds to the 1-strings in the 
upper half-plane (associated with $q$), including the real axis, but rotated through 
180 degrees. At each position in a double-column diagram, there are 0, 1 or 2 1-strings. 
Positions occupied by a 1-string are indicated by a solid circle and unoccupied positions 
are indicated by an open circle. The 1-string energies are given by $E_j=j$. There are 
$\floor{\frac{N}{4}}$ positions on the left and $\floor{\frac{N-2}{4}}$ on the right. Here, $N=24$, 
$\sigma=2$, $\bar\sigma=0$ and $\ell=2(\sigma+\bar\sigma+1)=6$.}
\end{figure}
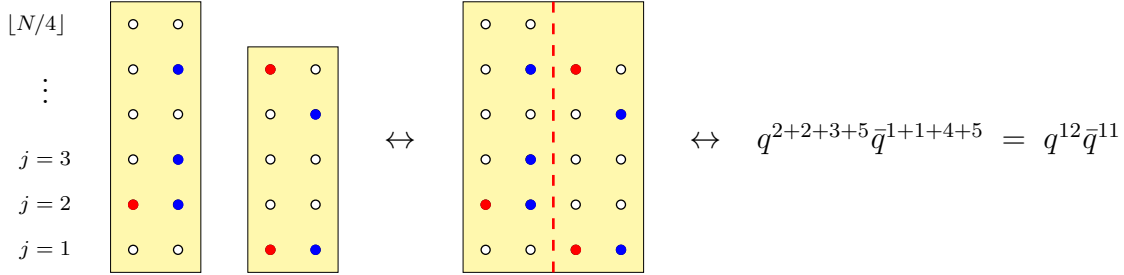

\psset{unit=.6cm}
\setlength{\unitlength}{.6cm}
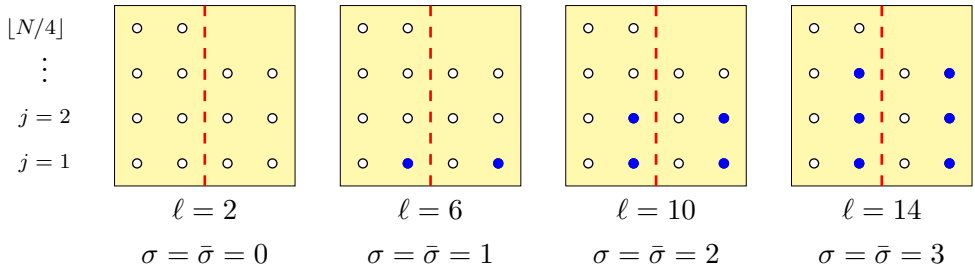
\begin{figure}[p]
\begin{center}
\begin{pspicture}[shift=-5](-.25,-2.25)(.6,5)
\rput(0,.5){\scriptsize $j=1$}
\rput(0,1.5){\scriptsize $j=2$}
\rput(0,2.75){$\vdots$}
\rput(-.2,3.5){\scriptsize $\floor{N/4}$}
\end{pspicture}
\hspace{4pt}
\begin{pspicture}[shift=-5](-.25,-2.25)(20,5)
\multirput(0,0)(5,0){4}{\psframe[linewidth=0pt,fillstyle=solid,fillcolor=yellow!40!white](0,0)(4,4)}
\multirput(0,0)(5,0){4}{\psline[linecolor=red,linewidth=1pt,linestyle=dashed,
  dash=.25 .25](2,0)(2,4)}
\multirput(0,0)(5,0){4}{
\multirput(0,0)(0,1){3}{\multirput(0,0)(1,0){4}{\psarc[linecolor=black,linewidth=.5pt,fillstyle=solid,
  fillcolor=white](0.5,0.5){.1}{0}{360}}}}
\multirput(0,0)(5,0){4}{
\multirput(0,3)(1,0){2}{\psarc[linecolor=black,linewidth=.5pt,fillstyle=solid,
  fillcolor=white](0.5,0.5){.1}{0}{360}}}
\rput(2,-.5){\small$\ell=2$}
\rput(2,-1.5){\small$\sigma=\bar\sigma=0$}
\rput(7,-.5){\small$\ell=6$}
\rput(7,-1.5){\small$\sigma=\bar\sigma=1$}
\rput(12,-.5){\small$\ell=10$}
\rput(12,-1.5){\small$\sigma=\bar\sigma=2$}
\rput(17,-.5){\small$\ell=14$}
\rput(17,-1.5){\small$\sigma=\bar\sigma=3$}
\multirput(0,0)(5,0){3}{\multirput(0,0)(2,0){2}{\psarc[linecolor= blue,linewidth=.5pt,fillstyle=solid,
  fillcolor=blue](6.5,.5){.1}{0}{360}}}
\multirput(5,0)(5,0){2}{\multirput(0,0)(2,0){2}{\psarc[linecolor= blue,linewidth=.5pt,fillstyle=solid,
  fillcolor=blue](6.5,1.5){.1}{0}{360}}}
\multirput(10,0)(5,0){1}{\multirput(0,0)(2,0){2}{\psarc[linecolor= blue,linewidth=.5pt,fillstyle=solid,
  fillcolor=blue](6.5,2.5){.1}{0}{360}}}
\end{pspicture}
\vspace{-.2in}
\end{center}
\caption{Groundstate configurations in the Neveu-Schwarz sectors with $\ell$ defects 
($N$ even, $\ell/2$ odd). Here $N=16$, the quantum numbers $\sigma, \bar\sigma$ are as 
shown and the conformal weights are $(\Delta,\bar\Delta)=(\Delta_{\ell/2},\Delta_{\ell/2})$.
\label{groundNS}}
\end{figure}

Next, we separate the contribution from zeros in the upper and lower half-planes. 
For $\ell/2$ even, we keep $\epsilon_j$ and $\mu_j$ for $j=1,2,\ldots,\floor{(N+2)/4}$ and set
\bea
\bar{\epsilon}_j=\epsilon_{N/2+1-j}=\pm 1,\qquad j=1,2,\ldots,\floor{N/4} \\
\bar{\mu}_j=\mu_{N/2+1-j}=\pm 1,\qquad j=1,2,\ldots,\floor{N/4}
\eea
For $\ell/2$ odd, we keep $\epsilon_j$ and $\mu_j$ for $j=1,2,\ldots,\floor{N/4}$ and set
\bea
\bar{\epsilon}_j=\epsilon_{N/2-j}=\pm 1,\qquad j=1,2,\ldots,\floor{(N-2)/4} \\
\bar{\mu}_j=\mu_{N/2-j}=\pm 1,\qquad j=1,2,\ldots,\floor{(N-2)/4}
\eea
By convention, we treat the zeros on the real axis (labelled by $j=(N+2)/4$ in the case $\ell/2$ 
even and $N=2$ mod 4 and by $j=N/4$ in the case $\ell/2$ odd and $N=0$ mod 4) as if they 
are in the upper half-plane. For $\ell/2$ even, setting 
$\epsilon=\mu\,\prod_{j=1}^{\floor{N/4}} \bar\epsilon_j\,\bar\mu_j$ gives
\bea
T(u)&=&\frac{\mu(-i)^{\frac{N}{2}}e^{-Niu}}{2^{N-1}}
\prod_{j=1}^{\floor{(N+2)/4}}{\Big(e^{2i u}+i\epsilon_j  \tan{\frac{(2j-1)\pi}{2N}}\Big)}
{\Big(\bar{\epsilon}_j e^{2i u}+i  \cot{\frac{(2j-1)\pi}{2N}}\Big)} \nonumber\\
&\times&\prod_{j=1}^{\floor{N/4}}{\Big(e^{2i u}+i\mu_j  \tan{\frac{(2j-1)\pi}{2N}}\Big)}
\Big(\bar{\mu}_j e^{2i u}+i \cot{\frac{(2j-1)\pi}{2N}}\Big)
\label{T(u)even}
\eea
For $\ell/2$ odd, setting
$\epsilon=\mu\,\prod_{j=1}^{\floor{(N-2)/4}} \bar\epsilon_j\,\bar\mu_j$ gives
\bea
T(u)&=&\frac{\mu(-i)^{\frac{N-2}{2}}Ne^{(2-N)iu}}{2^{N-1}}
\prod_{j=1}^{\floor{N/4}}{\Big(e^{2i u}+i\epsilon_j  \tan{\frac{j\pi}{N}}\Big)}
{\Big(\bar{\epsilon}_j e^{2i u}+i  \cot{\frac{j\pi}{N}}\Big)} \nonumber\\
&\times&\prod_{j=1}^{\floor{(N-2)/4}}{\Big(e^{2i u}+i\mu_j  \tan{\frac{j\pi}{N}}\Big)}
\Big(\bar{\mu}_j e^{2i u}+i \cot{\frac{j\pi}{N}}\Big)
\label{T(u)odd}
\eea
Fixing $\mu=+1$ ensures that $T_{0,0}(0)=T_{0,2}(0)=1$ consistent with (\ref{limD}). 
For $\ell/2$ even, taking the ratio of (\ref{T(u)even}) with precisely one $\epsilon_j,\bar\epsilon_j$ 
or $\mu_j,\bar\mu_j=-1$ to (\ref{T(u)even}) with all 
$\epsilon_j=\mu_j=\bar\epsilon_j=\bar\mu_j=+1$ and taking the limit $M,N\to\infty$ 
with a fixed aspect ratio $\delta=M/N$ gives
\bea
 \lim_{M,N\to\infty}\bigg(\frac{e^{2iu}-i\tan\frac{(2j-1)\pi}{2N}}{{e^{2iu}
   +i\tan\frac{(2j-1)\pi}{2N}}}\bigg)^M\!\!
 &=&\exp[-(2j-1)\pi i\,\delta\,e^{-2iu}]\;=\;q^{E_j}\\
 \lim_{M,N\to\infty}\bigg(\frac{-e^{2iu}+i\cot\frac{(2j-1)\pi}{2N}}{{e^{2iu}
  +i\cot\frac{(2j-1)\pi}{2N}}}\bigg)^M\!\!
 &=&\exp[(2j-1)\pi i\,\delta\,e^{2iu}]\;=\;\qbar^{E_j}
\eea
where $E_j=(j-\half)$. Similarly, for $\ell/2$ odd, taking the ratio of (\ref{T(u)odd}) with precisely 
one $\epsilon_j$ or $\mu_j=-1$ to (\ref{T(u)odd}) with all 
$\epsilon_j=\mu_j=\bar\epsilon_j=\bar\mu_j=+1$ and taking the limit $M,N\to\infty$ with a 
fixed aspect ratio $\delta=M/N$ gives
\bea
\lim_{M,N\to\infty}\bigg(\frac{e^{2iu}-i\tan\frac{j\pi}{N}}{e^{2iu}+i\tan\frac{j\pi}{N}}\bigg)^M\!\!
&=&\exp[-2j\pi i\,\delta\,e^{-2iu}]\;=\;q^{E_j}\\
\lim_{M,N\to\infty}\bigg(\frac{-e^{2iu}+i\cot\frac{j\pi}{N}}{e^{2iu}+i\cot\frac{j\pi}{N}}\bigg)^M\!\!
&=&\exp[2j\pi i\,\delta\,e^{2iu}]\;=\;\qbar^{E_j}
\eea
where $E_j=j$.

We conclude that, in the $\ell$ even sectors, 
\be
 E_j\;=\;\begin{cases}
  j-\half,&\mbox{$N$ even, $\ell/2$ even}\\
  j,&\mbox{$N$ even, $\ell/2$ odd}
\end{cases}
\ee
and the conformal partition functions are
\be
Z_\ell(q)\;=\;\begin{cases}
\disp(q\qbar)^{-c/24+\Delta_0}\!\!\!\sum_{\epsilon,\mu,\bar\epsilon,\bar\mu} 
\!\!q^{\sum_{j=1}^{\floor{\frac{N+2}{4}}}[\delta(\epsilon_j,-1)+\delta(\mu_j,-1)] E_j}
\qbar^{\sum_{j=1}^{\floor{\frac{N}{4}}}[\delta(\bar\epsilon_j,-1)+\delta(\bar\mu_j,-1)]E_j},\ 
   \mbox{$\frac{\ell}{2}$ even}\\[8pt]
\disp(q\qbar)^{-c/24+\Delta_1}\!\!\!\sum_{\epsilon,\mu,\bar\epsilon,\bar\mu} 
\!\!q^{\sum_{j=1}^{\floor{\frac{N}{4}}}[\delta(\epsilon_j,-1)+\delta(\mu_j,-1)] E_j}
\qbar^{\sum_{j=1}^{\floor{\frac{N-2}{4}}}[\delta(\bar\epsilon_j,-1)+\delta(\bar\mu_j,-1)]E_j},\ 
  \mbox{$\frac{\ell}{2}$ odd}
\end{cases}
\ee
where the sums and the allowed values of $\epsilon_j,\mu_j,\bar\epsilon_j,\bar\mu_j$ 
are determined by selection rules  as explained in Section~\ref{SecSelection}. 
The prefactor involving the central charge and conformal weights comes from the 
largest eigenvalue $T_{0,\ell}(u)$ with $\ell=0,2$ and is obtained by applying Euler-Maclaurin 
as explained in Section~\ref{EulerM}.

\section{Finite-Size Corrections from Euler-Maclaurin}
\label{EulerM}

In this section, we use the Euler-Maclaurin formula to obtain the finite-size corrections for 
the following three groundstates
\bea
\begin{array}{lll}
\mbox{$T_{0,0}(u)$: Ramond ($N$ even, $\ell=0$)}&
  \disp y_j=-\half\log\tan\frac{(j-\half)\pi}{N}&\mbox{double}\\[8pt]
\mbox{$T_{0,2}(u)$: Neveu-Schwarz ($N$ even, $\ell=2$)}\quad&
  \disp y_j=-\half\log\tan\frac{j\pi}{N}&\mbox{double}\\[8pt]
\mbox{$T_{0,1}(u)$: $\mathbb{Z}_4$ ($N$ odd, $\ell=1$)\quad}&
  \disp y_j=-\half\log\tan\frac{\half(j-\half)\pi}{N}\ \ &\mbox{single}\\
\end{array}
\eea
In each of these cases, there are no 1-strings, only {\em single} or {\em double} 2-strings at the positions 
given by $y_j$. The maximum eigenvalues in these sectors take the following real forms 
involving only the geometric factor $\sin 2u$
\bea
T_{0,0}(u)\!\!\!&=&\!\!\frac{i^{\frac{3N}{2}}e^{-Niu}}{2^{N-1}} \prod_{j=1}^{\frac{N}{2}} 
  \Big( e^{2iu} + i \tan \frac{(2j-1)\pi}{2N}\Big)^2
=\frac{1}{2^{\frac{N}{2}-1}} \prod_{j=1}^{\frac{N}{2}} {\Big(\frac{1}{\sin{\frac{(2j-1)\pi}{N}}} 
  + \sin{2u} \Big)}\qquad\\
T_{0,2}(u)\!\!\!&=&\!\!\frac{Ni^{\frac{3N}{2}+1}e^{(2-N)iu}}{2^{N-1}} 
  \prod_{j=1}^{\frac{N-2}{2}} \Big( e^{2iu} + i \tan \frac{j\pi}{N}\Big)^2
  =\frac{N}{2^{\frac{N}{2}}} \prod_{j=1}^{\frac{N-2}{2}} {\Big(\frac{1}{\sin{\frac{2j\pi}{N}}} 
   + \sin{2u} \Big)}\\
 T_{0,1}(u)\!\!\!&=&\!\!\!\frac{i^{\frac{3N}{2}+2}e^{-Niu}}{2^{N-1/2}}
  \prod_{j=1}^{N} \!{\Big(e^{2i u}+i \tan{\frac{(2j-1)\pi}{4N}}\Big)}
  =2^{\frac{1-N}{2}}\prod_{j=1}^{N}\! {\Big(\frac{1}{\sin{\frac{(2j-1)\pi}{2N}}} 
  + \sin{2u} \Big)^{\!\frac{1}{2}}}
\eea
It is confirmed that these eigenvalues satisfy $T_{0,\ell}(0)=1$ for $\ell=0,1,2$.

The logarithms of the eigenvalues $T_{0,\ell}(u)$ with $\ell=0,1,2$ involve sums
of terms which are singular at both endpoints in the limit $N\rightarrow\infty$.
To remedy this, we introduce~\cite{OPW} the function
\be
  F(t)\ =\ \log\Big[t(\pi-t)\big(\frac{1}{\sin t}+\sin 2u\big)\Big]\ =\ \log 
t+\log(\pi-t)+\log\big(\frac{1}{\sin t}+\sin 2u\big)
\label{F}
\ee
with 
\be
 \lim_{t\to 0^+} F(t)\;=\;\lim_{t\to \pi^-} F(t)\;=\;\log\pi,\qquad -\lim_{t\to 0^+} F'(t)\;=\;\lim_{t\to \pi^-} F'(t)
  \;=\;\frac{1}{\pi}-\sin 2u
\ee
For simplicity, we suppress the $u$ dependence. The endpoint or midpoint Euler-Maclaurin 
formula~\cite{AbramowitzS} can now be applied to approximate the sum by an integral
\bea
 &\disp \sum_{k=0}^M F(a+kh)\;=\;\frac{1}{h}\int_a^bF(t)dt+\frac{1}{2}[F(b)+F(a)]
   +\frac{h}{12}[F'(b)-F'(a)]+O(h^2)&\qquad
\\
 &\disp \sum_{k=0}^{M-1} F\big(a+(k+\half)h\big)\;=\;\frac{1}{h}\int_a^bF(t)dt
   -\frac{h}{24}[F'(b)-F'(a)]+O(h^2)&
\eea
where $b=a+Mh$ and $h$ is small. Due to (\ref{F}),
we also need the asymptotic expansion of the logarithm of the gamma function
\be
  \log\Gamma(y)\;=\;(y-\half)\log y-y+\half\log(2\pi)+\frac{1}{12y}+O(y^{-2}),\qquad y\to \infty
\label{Gamma}
\ee

First, let us consider the Neveu-Schwarz sector with $N$ even and $\ell=2$. 
Setting $t_j=2j\pi/N$, we have
\bea
-\log T_{0,2}(u)\;=\;\mbox{\small$\frac{N}{2}$}\log2-\log N-\sum_{j=1}^{\frac{N}{2}} F(t_j)
  +2\sum_{j=1}^{\frac{N}{2}} \log t_j
\eea
The sum over $F$ can be approximated using the endpoint Euler-Maclaurin formula, 
with $a=0$, $b=\pi$, $h=2\pi/N$, and
the sum over logarithms by the asymptotics of the gamma function. 
This yields finite-size corrections of the form (\ref{FSE}) with $c=-2$ and 
$\Delta=\bar\Delta=\Delta_1=0$. 

Next, consider the Ramond sector with $N=0$ mod 4 and $\ell=0$. 
Setting $t_j=(2j-1)\pi/N$, we have
\bea
-\log T_{0,0}(u)=(\mbox{\small$\frac{N}{2}$}-1)\log 2-
\sum_{k=0}^{\frac{N}{2}-1} F(t_j)+2\sum_{j=1}^{\frac{N}{2}}\log t_j
\eea
The sum over $F$ can be approximated using the midpoint Euler-Maclaurin formula, 
with $a=0$, $b=\pi$, $h=2\pi/N$, and the sum over logarithms by the asymptotics of the 
gamma function. This yields finite-size corrections of the form (\ref{FSE}) with $c=-2$ and 
$\Delta=\bar\Delta=\Delta_0=-1/8$. 

Lastly, consider the $\mathbb{Z}_4$ sector with $N$ odd and $\ell=1$. 
Setting $t_j=(2j-1)\pi/2N$, we have
\be
 -\log T_{0,1}(u)\;=\;\mbox{\small$\frac{N-1}{2}$}\log 2-
  \half\sum_{k=0}^{\frac{N}{2}-1} F(t_j)+\sum_{j=1}^{\frac{N}{2}}\log t_j
\ee
The sum over $F$ can be approximated using the midpoint Euler-Maclaurin formula, 
with $a=0$, $b=\pi$, $h=\pi/N$, and the sum over logarithms by 
the asymptotics of the gamma function. This yields finite-size corrections of the form 
(\ref{FSE}) with $c=-2$ and $\Delta=\bar\Delta=\Delta_{1/2}=-3/32$.

In summary, the conformal finite-size predictions are of the expected form (\ref{logLa}) with
\bea
c=-2,\qquad \Delta=\bar\Delta=\Delta_{\ell/2}=-\frac{1}{8}, -\frac{3}{32}, 0,\qquad \ell=0,1,2
\eea
The Euler-Maclaurin analysis can be extended to the excitations in these sectors as 
in \cite{PR0610} but we do not do this here.  
Instead, we use physical combinatorics to directly obtain the energy levels of the excited states.

\section{Physical Combinatorics and Selection Rules}
\label{SecSelection}

\subsection{$\mathbb{Z}_4$ sectors ($N$ odd, $\ell$ odd)}

The building blocks of the spectra in the $\mathbb{Z}_4$ sectors consist of the $q$-binomials
\bea
\qbin nmq=\qbin n{\floor{n/2}-\sigma}q=q^{-\frac{1}{2}\sigma(\sigma+\frac{1}{2})} 
  \sum_{\genfrac{}{}{0pt}{}{\text{$\sigma$-single}}{\text{columns}}} q^{\sum_j m_j E_j},\qquad \sigma=\floor{n/2}-m
\eea
with $E_j=\half(j-\half)$. The sum is over all single-column diagrams as in Figure~\ref{binomZ4} with a fixed $\sigma$. Here $\sigma$ is a quantum number given by the number of 1-strings 
at even positions $j$ minus the number of 1-strings at odd positions $j$
\bea
\sigma=m_{\text{even}}-m_{\text{odd}}
  =\sum_{k=1}^{\floor{n/2}}m_{2k}-\sum_{k=1}^{\floor{(n+1)/2}}m_{2k-1}
\eea
The number of 1-strings plus 2-strings at any given position is exactly one
\bea
m_j+n_j=1,\qquad j=1,2,\ldots,n
\eea

The single-columns with quantum number $\sigma$ are generated combinatorially by starting 
with the minimum energy configuration of 1-strings for given $\sigma$ as shown in 
Figure~\ref{EsigmaZ4}. Empirically determined selection rules dictate that in a sector with 
$\ell$ defects the quantum numbers of the groundstate satisfy
\be
 \sigma\;=\;\bar\sigma\;=\;\begin{cases}
  (\ell-1)/4,&\mbox{$\ell=1$ mod 4}\\
  -(\ell+1)/4,&\mbox{$\ell=3$ mod 4}
 \end{cases}\qquad 
  \ell=|4\sigma+1|=1,3,5,7,\ldots
\ee
The energy of these groundstates is $E(\sigma)+E(\bar\sigma)=\frac{1}{16}(\ell^2-1)$.

\begin{figure}[htbp]
\setlength{\unitlength}{.8pt}
\psset{unit=.8pt}
\begin{center}
\begin{pspicture}(-50,-20)(280,100)
\thicklines
\multirput(0,0)(40,0){8}{\psframe[linewidth=0pt,fillstyle=solid,
  fillcolor=yellow!40!white](-12,0)(12,90)}
\multiput(0,0)(40,0){8}{\line(0,10){90}}
\multirput(0,0)(40,0){8}{\psline[linewidth=1pt](-14,90)(14,90)}
\multiput(0,0)(40,0){8}{\multiput(0,0)(0,15){7}{\psarc[linecolor=black,linewidth=.5pt,fillstyle=solid,
  fillcolor=white](0,0){3}{0}{360}}}
\multiput(0,0)(0,30){3}{\psarc[linecolor=blue,linewidth=.5pt,fillstyle=solid,
  fillcolor=blue](0,15){3}{0}{360}}
\multiput(40,0)(0,30){2}{\psarc[linecolor=blue,linewidth=.5pt,fillstyle=solid,
  fillcolor=blue](0,15){3}{0}{360}}
\psarc[linecolor=blue,linewidth=.5pt,fillstyle=solid,fillcolor=blue](80,15){3}{0}{360}
\psarc[linecolor=red,linewidth=.5pt,fillstyle=solid,fillcolor=red](160,0){3}{0}{360}
\multiput(200,0)(0,30){2}{\psarc[linecolor=red,linewidth=.5pt,fillstyle=solid,
  fillcolor=red](0,0){3}{0}{360}}
\multiput(240,0)(0,30){3}{\psarc[linecolor=red,linewidth=.5pt,fillstyle=solid,
  fillcolor=red](0,0){3}{0}{360}}
\multiput(280,0)(0,30){4}{\psarc[linecolor=red,linewidth=.5pt,fillstyle=solid,
  fillcolor=red](0,0){3}{0}{360}}
\rput[B](-40,-32){$\sigma$}
\rput[B](0,-32){$3$}
\rput[B](40,-32){$2$}
\rput[B](80,-32){$1$}
\rput[B](120,-32){$0$}
\rput[B](160,-32){$-1$}
\rput[B](200,-32){$-2$}
\rput[B](240,-32){$-3$}
\rput[B](280,-32){$-4$}
\rput[B](-40,-3){\scriptsize $j=1$}
\rput[B](-40,12){\scriptsize $j=2$}
\rput[B](-40,27){\scriptsize $j=3$}
\rput[B](-40,42){\scriptsize $j=4$}
\rput[B](-40,57){\scriptsize $j=5$}
\rput[B](-40,72){\scriptsize $j=6$}
\rput[B](-40,87){\scriptsize $j=7$}
\end{pspicture}
\end{center}
\smallskip
\caption{$\mathbb{Z}_4$ sectors ($N$, $\ell$ odd): Minimal configurations of 
single-columns with energy $E(\sigma)=\half\sigma(\sigma+\half)$.
The quantum number $\sigma=\floor{n/2}-m$ is given by the excess of blue (even $j$) over 
red (odd $j$) 1-strings.  At each empty position $j$, there is a 2-string. This analyticity strip 
is in the upper-half complex $u$-plane rotated by 180 degrees so that position $j=1$ 
(furthest from the real axis) is at the bottom.\label{EsigmaZ4}}
\end{figure}
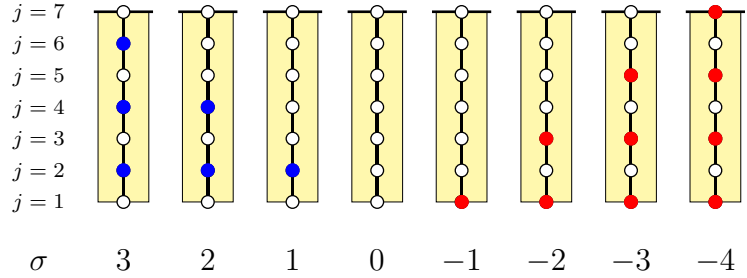
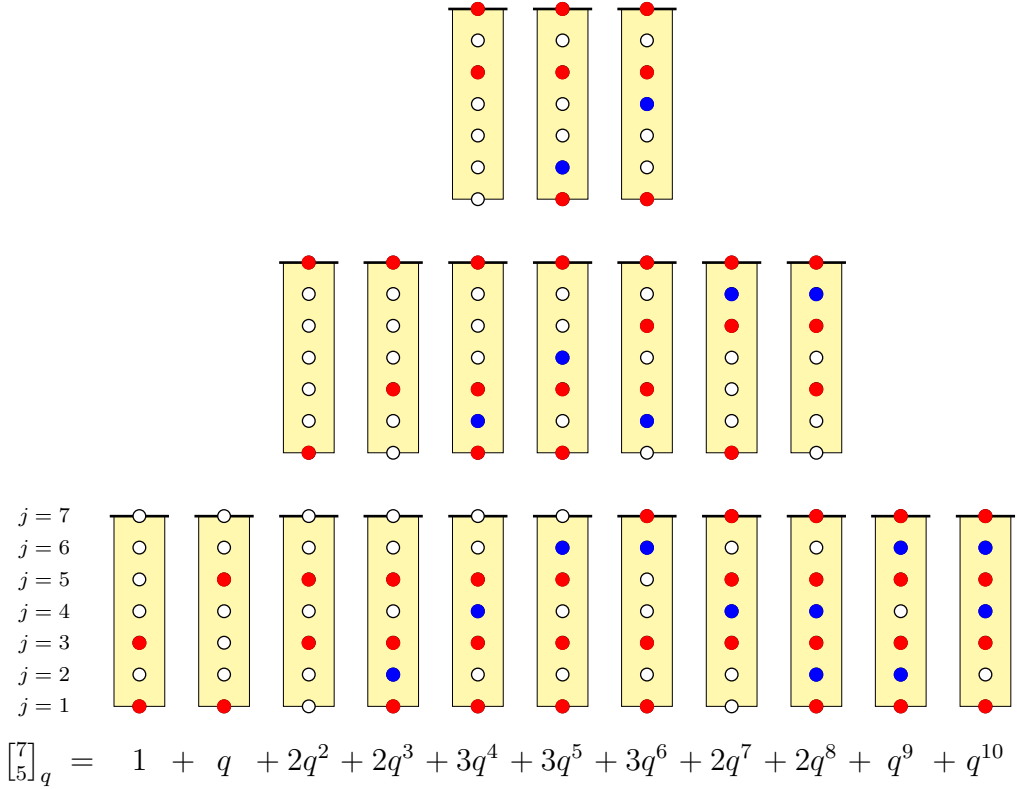
\begin{figure}[htbp]
\setlength{\unitlength}{.8pt}
\psset{unit=.8pt}
\begin{center}
\begin{pspicture}(-50,-30)(400,340)
\thicklines
\multirput(0,0)(40,0){11}{\psframe[linewidth=0pt,fillstyle=solid,
  fillcolor=yellow!40!white](-12,0)(12,90)}
\multirput(0,0)(40,0){11}{\psline[linewidth=1pt](-14,90)(14,90)}
\multirput(80,120)(40,0){7}{\psframe[linewidth=0pt,fillstyle=solid,
  fillcolor=yellow!40!white](-12,0)(12,90)}
\multirput(80,120)(40,0){7}{\psline[linewidth=1pt](-14,90)(14,90)}
\multirput(160,240)(40,0){3}{\psframe[linewidth=0pt,fillstyle=solid,
  fillcolor=yellow!40!white](-12,0)(12,90)}
\multirput(160,240)(40,0){3}{\psline[linewidth=1pt](-14,90)(14,90)}
%
\multiput(0,0)(40,0){11}{\multiput(0,0)(0,15){7}{\psarc[linecolor=black,linewidth=.5pt,fillstyle=solid,
  fillcolor=white](0,0){3}{0}{360}}}
\multiput(80,120)(40,0){7}{\multiput(0,0)(0,15){7}{\psarc[linecolor=black,
  linewidth=.5pt,fillstyle=solid,fillcolor=white](0,0){3}{0}{360}}}
\multiput(160,240)(40,0){3}{\multiput(0,0)(0,15){7}{\psarc[linecolor=black,
  linewidth=.5pt,fillstyle=solid,fillcolor=white](0,0){3}{0}{360}}}
\psarc[linecolor=red,linewidth=.5pt,fillstyle=solid,fillcolor=red](0,0){3}{0}{360}
\psarc[linecolor=red,linewidth=.5pt,fillstyle=solid,fillcolor=red](0,30){3}{0}{360}
\psarc[linecolor=red,linewidth=.5pt,fillstyle=solid,fillcolor=red](40,0){3}{0}{360}
\psarc[linecolor=red,linewidth=.5pt,fillstyle=solid,fillcolor=red](40,60){3}{0}{360}
\psarc[linecolor=red,linewidth=.5pt,fillstyle=solid,fillcolor=red](80,30){3}{0}{360}
\psarc[linecolor=red,linewidth=.5pt,fillstyle=solid,fillcolor=red](80,60){3}{0}{360}
\psarc[linecolor=red,linewidth=.5pt,fillstyle=solid,fillcolor=red](80,120){3}{0}{360}
\psarc[linecolor=red,linewidth=.5pt,fillstyle=solid,fillcolor=red](80,210){3}{0}{360}
\psarc[linecolor=red,linewidth=.5pt,fillstyle=solid,fillcolor=red](120,0){3}{0}{360}
\psarc[linecolor=blue,linewidth=.5pt,fillstyle=solid,fillcolor=blue](120,15){3}{0}{360}
\psarc[linecolor=red,linewidth=.5pt,fillstyle=solid,fillcolor=red](120,30){3}{0}{360}
\psarc[linecolor=red,linewidth=.5pt,fillstyle=solid,fillcolor=red](120,60){3}{0}{360}
\psarc[linecolor=red,linewidth=.5pt,fillstyle=solid,fillcolor=red](120,150){3}{0}{360}
\psarc[linecolor=red,linewidth=.5pt,fillstyle=solid,fillcolor=red](120,210){3}{0}{360}
\psarc[linecolor=red,linewidth=.5pt,fillstyle=solid,fillcolor=red](160,0){3}{0}{360}
\psarc[linecolor=blue,linewidth=.5pt,fillstyle=solid,fillcolor=blue](160,45){3}{0}{360}
\psarc[linecolor=red,linewidth=.5pt,fillstyle=solid,fillcolor=red](160,30){3}{0}{360}
\psarc[linecolor=red,linewidth=.5pt,fillstyle=solid,fillcolor=red](160,60){3}{0}{360}
\psarc[linecolor=red,linewidth=.5pt,fillstyle=solid,fillcolor=red](160,120){3}{0}{360}
\psarc[linecolor=blue,linewidth=.5pt,fillstyle=solid,fillcolor=blue](160,135){3}{0}{360}
\psarc[linecolor=red,linewidth=.5pt,fillstyle=solid,fillcolor=red](160,150){3}{0}{360}
\psarc[linecolor=red,linewidth=.5pt,fillstyle=solid,fillcolor=red](160,210){3}{0}{360}
\psarc[linecolor=red,linewidth=.5pt,fillstyle=solid,fillcolor=red](160,300){3}{0}{360}
\psarc[linecolor=red,linewidth=.5pt,fillstyle=solid,fillcolor=red](160,330){3}{0}{360}
\psarc[linecolor=red,linewidth=.5pt,fillstyle=solid,fillcolor=red](200,0){3}{0}{360}
\psarc[linecolor=red,linewidth=.5pt,fillstyle=solid,fillcolor=red](200,30){3}{0}{360}
\psarc[linecolor=blue,linewidth=.5pt,fillstyle=solid,fillcolor=blue](200,75){3}{0}{360}
\psarc[linecolor=red,linewidth=.5pt,fillstyle=solid,fillcolor=red](200,60){3}{0}{360}
\psarc[linecolor=red,linewidth=.5pt,fillstyle=solid,fillcolor=red](200,120){3}{0}{360}
\psarc[linecolor=blue,linewidth=.5pt,fillstyle=solid,fillcolor=blue](200,165){3}{0}{360}
\psarc[linecolor=red,linewidth=.5pt,fillstyle=solid,fillcolor=red](200,150){3}{0}{360}
\psarc[linecolor=red,linewidth=.5pt,fillstyle=solid,fillcolor=red](200,210){3}{0}{360}
\psarc[linecolor=red,linewidth=.5pt,fillstyle=solid,fillcolor=red](200,240){3}{0}{360}
\psarc[linecolor=blue,linewidth=.5pt,fillstyle=solid,fillcolor=blue](200,255){3}{0}{360}
\psarc[linecolor=red,linewidth=.5pt,fillstyle=solid,fillcolor=red](200,300){3}{0}{360}
\psarc[linecolor=red,linewidth=.5pt,fillstyle=solid,fillcolor=red](200,330){3}{0}{360}
\psarc[linecolor=red,linewidth=.5pt,fillstyle=solid,fillcolor=red](240,0){3}{0}{360}
\psarc[linecolor=red,linewidth=.5pt,fillstyle=solid,fillcolor=red](240,30){3}{0}{360}
\psarc[linecolor=blue,linewidth=.5pt,fillstyle=solid,fillcolor=blue](240,75){3}{0}{360}
\psarc[linecolor=red,linewidth=.5pt,fillstyle=solid,fillcolor=red](240,90){3}{0}{360}
\psarc[linecolor=red,linewidth=.5pt,fillstyle=solid,fillcolor=red](240,150){3}{0}{360}
\psarc[linecolor=blue,linewidth=.5pt,fillstyle=solid,fillcolor=blue](240,135){3}{0}{360}
\psarc[linecolor=red,linewidth=.5pt,fillstyle=solid,fillcolor=red](240,180){3}{0}{360}
\psarc[linecolor=red,linewidth=.5pt,fillstyle=solid,fillcolor=red](240,210){3}{0}{360}
\psarc[linecolor=red,linewidth=.5pt,fillstyle=solid,fillcolor=red](240,240){3}{0}{360}
\psarc[linecolor=blue,linewidth=.5pt,fillstyle=solid,fillcolor=blue](240,285){3}{0}{360}
\psarc[linecolor=red,linewidth=.5pt,fillstyle=solid,fillcolor=red](240,300){3}{0}{360}
\psarc[linecolor=red,linewidth=.5pt,fillstyle=solid,fillcolor=red](240,330){3}{0}{360}
\psarc[linecolor=red,linewidth=.5pt,fillstyle=solid,fillcolor=red](280,30){3}{0}{360}
\psarc[linecolor=blue,linewidth=.5pt,fillstyle=solid,fillcolor=blue](280,45){3}{0}{360}
\psarc[linecolor=red,linewidth=.5pt,fillstyle=solid,fillcolor=red](280,60){3}{0}{360}
\psarc[linecolor=red,linewidth=.5pt,fillstyle=solid,fillcolor=red](280,90){3}{0}{360}
\psarc[linecolor=red,linewidth=.5pt,fillstyle=solid,fillcolor=red](280,120){3}{0}{360}
\psarc[linecolor=red,linewidth=.5pt,fillstyle=solid,fillcolor=red](280,180){3}{0}{360}
\psarc[linecolor=blue,linewidth=.5pt,fillstyle=solid,fillcolor=blue](280,195){3}{0}{360}
\psarc[linecolor=red,linewidth=.5pt,fillstyle=solid,fillcolor=red](280,210){3}{0}{360}
\psarc[linecolor=red,linewidth=.5pt,fillstyle=solid,fillcolor=red](320,0){3}{0}{360}
\psarc[linecolor=blue,linewidth=.5pt,fillstyle=solid,fillcolor=blue](320,15){3}{0}{360}
\psarc[linecolor=red,linewidth=.5pt,fillstyle=solid,fillcolor=red](320,30){3}{0}{360}
\psarc[linecolor=blue,linewidth=.5pt,fillstyle=solid,fillcolor=blue](320,45){3}{0}{360}
\psarc[linecolor=red,linewidth=.5pt,fillstyle=solid,fillcolor=red](320,60){3}{0}{360}
\psarc[linecolor=red,linewidth=.5pt,fillstyle=solid,fillcolor=red](320,90){3}{0}{360}
\psarc[linecolor=red,linewidth=.5pt,fillstyle=solid,fillcolor=red](320,150){3}{0}{360}
\psarc[linecolor=red,linewidth=.5pt,fillstyle=solid,fillcolor=red](320,180){3}{0}{360}
\psarc[linecolor=blue,linewidth=.5pt,fillstyle=solid,fillcolor=blue](320,195){3}{0}{360}
\psarc[linecolor=red,linewidth=.5pt,fillstyle=solid,fillcolor=red](320,210){3}{0}{360}
\psarc[linecolor=red,linewidth=.5pt,fillstyle=solid,fillcolor=red](360,0){3}{0}{360}
\psarc[linecolor=blue,linewidth=.5pt,fillstyle=solid,fillcolor=blue](360,15){3}{0}{360}
\psarc[linecolor=red,linewidth=.5pt,fillstyle=solid,fillcolor=red](360,30){3}{0}{360}
\psarc[linecolor=blue,linewidth=.5pt,fillstyle=solid,fillcolor=blue](360,75){3}{0}{360}
\psarc[linecolor=red,linewidth=.5pt,fillstyle=solid,fillcolor=red](360,60){3}{0}{360}
\psarc[linecolor=red,linewidth=.5pt,fillstyle=solid,fillcolor=red](360,90){3}{0}{360}
\psarc[linecolor=red,linewidth=.5pt,fillstyle=solid,fillcolor=red](400,0){3}{0}{360}
\psarc[linecolor=blue,linewidth=.5pt,fillstyle=solid,fillcolor=blue](400,45){3}{0}{360}
\psarc[linecolor=red,linewidth=.5pt,fillstyle=solid,fillcolor=red](400,30){3}{0}{360}
\psarc[linecolor=blue,linewidth=.5pt,fillstyle=solid,fillcolor=blue](400,75){3}{0}{360}
\psarc[linecolor=red,linewidth=.5pt,fillstyle=solid,fillcolor=red](400,60){3}{0}{360}
\psarc[linecolor=red,linewidth=.5pt,fillstyle=solid,fillcolor=red](400,90){3}{0}{360}
\rput[B](-43,-30){$\qbin75q\;=$}
\multiput(15,-30)(40,0){10}{$+$}
\rput[B](0,-30){$1$}
\rput[B](40,-30){$q$}
\rput[B](80,-30){$2q^2$}
\rput[B](120,-30){$2q^3$}
\rput[B](160,-30){$3q^4$}
\rput[B](200,-30){$3q^5$}
\rput[B](240,-30){$3q^6$}
\rput[B](280,-30){$2q^7$}
\rput[B](320,-30){$2q^8$}
\rput[B](360,-30){$q^9$}
\rput[B](400,-30){$q^{10}$}
\rput[B](-45,-3){\scriptsize $j=1$}
\rput[B](-45,12){\scriptsize $j=2$}
\rput[B](-45,27){\scriptsize $j=3$}
\rput[B](-45,42){\scriptsize $j=4$}
\rput[B](-45,57){\scriptsize $j=5$}
\rput[B](-45,72){\scriptsize $j=6$}
\rput[B](-45,87){\scriptsize $j=7$}
\end{pspicture}
\end{center}
\smallskip
\caption{$\mathbb{Z}_4$ sectors ($N$, $\ell$ odd): Combinatorial enumeration by 
single-columns of the $q$-binomial $\sqbin nmq=\sqbin 75q
=q^{-3/2} \sum q^{\sum_j m_j E_j}$. The excess of blue (even $j$) over red (odd $j$) 1-strings 
is given by the quantum number $\sigma=\floor{n/2}-m=-2$.   The elementary excitation energy 
of a 1-string at position $j$ is $E_j=\half(j-\half)$. The lowest energy configuration has energy 
$E(\sigma)=1/4+5/4=3/2=\half\sigma(\sigma+\half)$. At each empty position $j$, there is a 2-string.
This analyticity strip is in the upper-half complex $u$-plane rotated by 180 degrees so that 
position $j=1$ (furthest from the real axis) is at the bottom. The elementary excitations 
(of energy 1) are generated by either inserting two 1-strings at positions $j=1$ and $j=2$ 
or promoting a 1-string at position $j$ to position $j+2$.  Notice that 
$\sqbin nmq=\sqbin n{n-m}q$ as $q$-polynomials but they have different combinatorial 
interpretations because they have different quantum numbers $\sigma$. In the lower half-plane, 
$q$ is replaced with $\qbar$ and no rotation is required. In this example, $\ell=7$ and the value $\bar\sigma=-2$ of the 
quantum number in the lower half-plane is related to $\sigma=-2$ in the upper half-plane by 
the selection rules $\sigma+\bar\sigma=-(\ell+1)/2$ and 
$\half(\sigma-\bar\sigma)\in\mathbb{Z}$.\label{binomZ4}}
\end{figure}

Excitations, incrementing the energy by one unit, are generated either by inserting a pair 
of 1-strings at positions $j=1$ and $j=2$ or incrementing the position $j$ of a 1-string by 2 
units. The $q$-binomials are illustrated in Figure~\ref{binomZ4}.
Empirically, we find that all the excitations satisfy the selection rules
\bea
\sigma+\bar\sigma=
\begin{cases}
\half(\ell-1),&\mbox{$\ell=1$ mod 4}\\
-\half(\ell+1),&\mbox{$\ell=3$ mod 4}
\end{cases}
\qquad\qquad \half(\sigma-\bar\sigma)\in
\mathbb{Z}
\eea
Using the $q$-binomial building blocks and empirical selection rules, we thus obtain the 
finitized partition functions
\bea
Z_\ell^{(N)}(q)&\!\!=\!\!&
\begin{cases}
\displaystyle(q\qbar)^{-c/24}\sum_{k\in\mathbb{Z}} 
  q^{\Delta_{2k+\ell/2}}\qbin{\sc{\frac{N+1}{2}}}{\sc{\frac{N-\ell}{4}}-k}q  
\qbar^{\Delta_{2k-\ell/2}}\qbin{\sc{\frac{N-1}{2}}}{\sc{\frac{N-\ell}{4}}+k}\qbar\\[12pt]
\displaystyle(q\qbar)^{-c/24}\sum_{k\in\mathbb{Z}} 
q^{\Delta_{2k+\ell/2}}\qbin{\sc{\frac{N+1}{2}}}{\sc{\frac{N+\ell+2}{4}}\!+\!k}q  
\qbar^{\Delta_{2k-\ell/2}}\qbin{\sc{\frac{N-1}{2}}}{\sc{\frac{N+\ell-2}{4}}\!-\!k}\qbar
\end{cases}\mbox{$N\!-\!\ell=0$ mod 4}\qquad\label{Z4Z1}\\[2pt]
Z_\ell^{(N)}(q)&\!\!=\!\!&
\begin{cases}
\displaystyle(q\qbar)^{-c/24}\sum_{k\in\mathbb{Z}} 
  q^{\Delta_{2k+\ell/2}}\qbin{\sc{\frac{N+1}{2}}}{\sc{\frac{N-\ell+2}{4}}\!-\!k}q  
\qbar^{\Delta_{2k-\ell/2}}\qbin{\sc{\frac{N-1}{2}}}{\sc{\frac{N-\ell-2}{4}}\!+\!k}\qbar\\[12pt]
\displaystyle(q\qbar)^{-c/24}\sum_{k\in\mathbb{Z}} 
q^{\Delta_{2k+\ell/2}}\qbin{\sc{\frac{N+1}{2}}}{\sc{\frac{N+\ell}{4}}+k}q  
\qbar^{\Delta_{2k-\ell/2}}\qbin{\sc{\frac{N-1}{2}}}{\sc{\frac{N+\ell}{4}}-k}\qbar
\end{cases}\mbox{$N\!-\!\ell=2$ mod 4}\qquad\label{Z4Z3}
\eea
For given mod 4 parities of $N-\ell$, these expressions are equivalent as partition functions but, 
in each case, the first form is used for the combinatorial interpretation when $\ell=1$ mod 4 
and the second form when $\ell=3$ mod 4.
We further observe that
\bea
\sum_{\ell\in2\mathbb{N}-1}^{\ell\le N}\!\!\!Z_\ell^{(N)}(q)=\half(q\qbar)^{-\frac{c}{24}-\frac{3}{32}}\!\bigg[\!
\prod_{n=1}^{\frac{N+1}{2}} (1\!+\!q^{\frac{2n-1}{4}})\!\prod_{n=1}^{\frac{N-1}{2}} (1\!+\!\qbar^{\frac{2n-1}{4}})
+\!\prod_{n=1}^{\frac{N+1}{2}} (1\!-\!q^{\frac{2n-1}{4}})\!\prod_{n=1}^{\frac{N-1}{2}} (1\!-\!\qbar^{\frac{2n-1}{4}})\!\bigg]
\eea

\subsection{Ramond sectors ($N$ even, $\ell/2$ even)}

The building blocks of the spectra in the Ramond sectors consist of the $q$-binomials
\bea
\qbin nmq=\qbin n{\floor{n/2}-\sigma}q=q^{-\frac{1}{2}\sigma^2} 
\sum_{\genfrac{}{}{0pt}{}{\text{$\sigma$-double}}{\text{columns}}} q^{\sum_j m_j E_j},\qquad \sigma=\floor{n/2}-m
\eea
with $E_j=j-\half$. The sum is over all double-column diagrams as in Figure~\ref{binomR1} with a fixed $\sigma$. Here $\sigma$ is a quantum number given by the number of 1-strings in 
the right column minus the number of 1-strings in the left column
\bea
\sigma=m_{\text{right}}-m_{\text{left}}
\eea
The number of 1-strings plus 2-strings at any given position is exactly two
\bea
m_j+n_j=2,\qquad j=1,2,\ldots,n
\eea

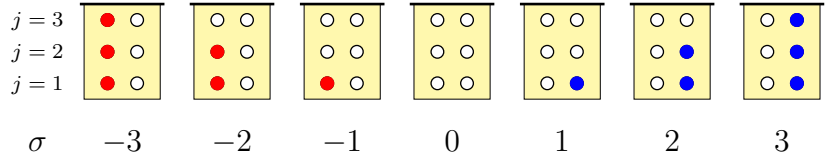
\begin{figure}[p]
\setlength{\unitlength}{.8pt}
\psset{unit=.8pt}
\begin{center}
\begin{pspicture}(-50,-20)(335,50)
\thicklines
\multirput(0,0)(52,0){7}{\psframe[linewidth=0pt,fillstyle=solid,
  fillcolor=yellow!40!white](-18,-7.5)(18,37.5)}
\multirput(0,0)(52,0){7}{\psline[linewidth=1pt](-20,37.5)(20,37.5)}
\multiput(-7,0)(52,0){7}{\multiput(0,0)(0,15){3}{\psarc[linecolor=black,
  linewidth=.5pt,fillstyle=solid,fillcolor=white](0,0){3}{0}{360}}}
\multiput(7,0)(52,0){7}{\multiput(0,0)(0,15){3}{\psarc[linecolor=black,
  linewidth=.5pt,fillstyle=solid,fillcolor=white](0,0){3}{0}{360}}}
\multiput(-7,0)(0,15){3}{\psarc[linecolor=red,linewidth=.5pt,fillstyle=solid,
  fillcolor=red](0,0){3}{0}{360}}
\multiput(45,0)(0,15){2}{\psarc[linecolor=red,linewidth=.5pt,fillstyle=solid,
  fillcolor=red](0,0){3}{0}{360}}
\multiput(97,0)(0,15){1}{\psarc[linecolor=red,linewidth=.5pt,fillstyle=solid,
  fillcolor=red](0,0){3}{0}{360}}
\multiput(215,0)(0,15){1}{\psarc[linecolor=blue,linewidth=.5pt,fillstyle=solid,
  fillcolor=blue](0,0){3}{0}{360}}
\multiput(267,0)(0,15){2}{\psarc[linecolor=blue,linewidth=.5pt,fillstyle=solid,
  fillcolor=blue](0,0){3}{0}{360}}
\multiput(319,0)(0,15){3}{\psarc[linecolor=blue,linewidth=.5pt,fillstyle=solid,
  fillcolor=blue](0,0){3}{0}{360}}
\rput[B](-40,-32){$\sigma$}
\rput[B](0,-32){$-3$}
\rput[B](52,-32){$-2$}
\rput[B](104,-32){$-1$}
\rput[B](156,-32){$0$}
\rput[B](208,-32){$1$}
\rput[B](260,-32){$2$}
\rput[B](312,-32){$3$}
\rput[B](-40,-3){\scriptsize $j=1$}
\rput[B](-40,12){\scriptsize $j=2$}
\rput[B](-40,27){\scriptsize $j=3$}
\end{pspicture}
\end{center}
\smallskip
\caption{Ramond sectors ($N$ even, $\ell/2$ even): Minimal configurations of double-columns 
with energy $E(\sigma)=\half\sigma^2$. The quantum number $\sigma=\floor{n/2}-m$ 
is given by the excess of blue (right) over red (left) 1-strings.  At each position $j$, the number 
of 1-strings $m_j$ plus the number of 2-strings $n_j$ is 2. This analyticity strip is in the 
upper-half complex $u$-plane rotated by 180 degrees so that position $j=1$ (furthest from the 
real axis) is at the bottom.\label{EsigmaR1}}
\end{figure}

\begin{figure}[p]
\setlength{\unitlength}{.7pt}
\psset{unit=.7pt}
\begin{center}
\begin{pspicture}(-60,-30)(450,170)
\thicklines
\multirput(0,0)(52,0){9}{\psframe[linewidth=0pt,fillstyle=solid,
  fillcolor=yellow!40!white](-18,-7.5)(18,37.5)}
\multirput(0,0)(52,0){9}{\psline[linewidth=1pt](-20,37.5)(20,37.5)}
\multirput(104,60)(52,0){5}{\psframe[linewidth=0pt,fillstyle=solid,
  fillcolor=yellow!40!white](-18,-7.5)(18,37.5)}
\multirput(104,60)(52,0){5}{\psline[linewidth=1pt](-20,37.5)(20,37.5)}
\multirput(208,120)(52,0){1}{\psframe[linewidth=0pt,fillstyle=solid,
  fillcolor=yellow!40!white](-18,-7.5)(18,37.5)}
\multirput(208,120)(52,0){1}{\psline[linewidth=1pt](-20,37.5)(20,37.5)}
%
\multiput(-7,0)(52,0){9}{\multiput(0,0)(0,15){3}{\psarc[linecolor=black,
  linewidth=.5pt,fillstyle=solid,fillcolor=white](0,0){3}{0}{360}}}
\multiput(7,0)(52,0){9}{\multiput(0,0)(0,15){3}{\psarc[linecolor=black,
  linewidth=.5pt,fillstyle=solid,fillcolor=white](0,0){3}{0}{360}}}
\multiput(97,60)(52,0){5}{\multiput(0,0)(0,15){3}{\psarc[linecolor=black,
  linewidth=.5pt,fillstyle=solid,fillcolor=white](0,0){3}{0}{360}}}
\multiput(111,60)(52,0){5}{\multiput(0,0)(0,15){3}{\psarc[linecolor=black,
  linewidth=.5pt,fillstyle=solid,fillcolor=white](0,0){3}{0}{360}}}
\multiput(201,120)(52,0){1}{\multiput(0,0)(0,15){3}{\psarc[linecolor=black,
  linewidth=.5pt,fillstyle=solid,fillcolor=white](0,0){3}{0}{360}}}
\multiput(215,120)(52,0){1}{\multiput(0,0)(0,15){3}{\psarc[linecolor=black,
  linewidth=.5pt,fillstyle=solid,fillcolor=white](0,0){3}{0}{360}}}
\psarc[linecolor=blue,linewidth=.5pt,fillstyle=solid,fillcolor=blue](7,0){3}{0}{360}
\psarc[linecolor=blue,linewidth=.5pt,fillstyle=solid,fillcolor=blue](59,15){3}{0}{360}
\psarc[linecolor=red,linewidth=.5pt,fillstyle=solid,fillcolor=red](97,0){3}{0}{360}
\psarc[linecolor=blue,linewidth=.5pt,fillstyle=solid,fillcolor=blue](111,0){3}{0}{360}
\psarc[linecolor=blue,linewidth=.5pt,fillstyle=solid,fillcolor=blue](111,15){3}{0}{360}
\psarc[linecolor=blue,linewidth=.5pt,fillstyle=solid,fillcolor=blue](111,90){3}{0}{360}
\psarc[linecolor=red,linewidth=.5pt,fillstyle=solid,fillcolor=red](149,0){3}{0}{360}
\psarc[linecolor=red,linewidth=.5pt,fillstyle=solid,fillcolor=red](149,75){3}{0}{360}
\psarc[linecolor=blue,linewidth=.5pt,fillstyle=solid,fillcolor=blue](163,0){3}{0}{360}
\psarc[linecolor=blue,linewidth=.5pt,fillstyle=solid,fillcolor=blue](163,30){3}{0}{360}
\psarc[linecolor=blue,linewidth=.5pt,fillstyle=solid,fillcolor=blue](163,60){3}{0}{360}
\psarc[linecolor=blue,linewidth=.5pt,fillstyle=solid,fillcolor=blue](163,75){3}{0}{360}
\psarc[linecolor=red,linewidth=.5pt,fillstyle=solid,fillcolor=red](201,15){3}{0}{360}
\psarc[linecolor=red,linewidth=.5pt,fillstyle=solid,fillcolor=red](201,60){3}{0}{360}
\psarc[linecolor=red,linewidth=.5pt,fillstyle=solid,fillcolor=red](201,150){3}{0}{360}
\psarc[linecolor=blue,linewidth=.5pt,fillstyle=solid,fillcolor=blue](215,0){3}{0}{360}
\psarc[linecolor=blue,linewidth=.5pt,fillstyle=solid,fillcolor=blue](215,30){3}{0}{360}
\psarc[linecolor=blue,linewidth=.5pt,fillstyle=solid,fillcolor=blue](215,75){3}{0}{360}
\psarc[linecolor=blue,linewidth=.5pt,fillstyle=solid,fillcolor=blue](215,90){3}{0}{360}
\psarc[linecolor=blue,linewidth=.5pt,fillstyle=solid,fillcolor=blue](215,120){3}{0}{360}
\psarc[linecolor=blue,linewidth=.5pt,fillstyle=solid,fillcolor=blue](215,135){3}{0}{360}
\psarc[linecolor=red,linewidth=.5pt,fillstyle=solid,fillcolor=red](253,15){3}{0}{360}
\psarc[linecolor=red,linewidth=.5pt,fillstyle=solid,fillcolor=red](253,90){3}{0}{360}
\psarc[linecolor=blue,linewidth=.5pt,fillstyle=solid,fillcolor=blue](267,15){3}{0}{360}
\psarc[linecolor=blue,linewidth=.5pt,fillstyle=solid,fillcolor=blue](267,30){3}{0}{360}
\psarc[linecolor=blue,linewidth=.5pt,fillstyle=solid,fillcolor=blue](267,60){3}{0}{360}
\psarc[linecolor=blue,linewidth=.5pt,fillstyle=solid,fillcolor=blue](267,90){3}{0}{360}
\psarc[linecolor=red,linewidth=.5pt,fillstyle=solid,fillcolor=red](305,0){3}{0}{360}
\psarc[linecolor=red,linewidth=.5pt,fillstyle=solid,fillcolor=red](305,15){3}{0}{360}
\psarc[linecolor=red,linewidth=.5pt,fillstyle=solid,fillcolor=red](305,90){3}{0}{360}
\psarc[linecolor=blue,linewidth=.5pt,fillstyle=solid,fillcolor=blue](319,0){3}{0}{360}
\psarc[linecolor=blue,linewidth=.5pt,fillstyle=solid,fillcolor=blue](319,15){3}{0}{360}
\psarc[linecolor=blue,linewidth=.5pt,fillstyle=solid,fillcolor=blue](319,30){3}{0}{360}
\psarc[linecolor=blue,linewidth=.5pt,fillstyle=solid,fillcolor=blue](319,75){3}{0}{360}
\psarc[linecolor=blue,linewidth=.5pt,fillstyle=solid,fillcolor=blue](319,90){3}{0}{360}
\psarc[linecolor=red,linewidth=.5pt,fillstyle=solid,fillcolor=red](357,0){3}{0}{360}
\psarc[linecolor=red,linewidth=.5pt,fillstyle=solid,fillcolor=red](357,30){3}{0}{360}
\psarc[linecolor=blue,linewidth=.5pt,fillstyle=solid,fillcolor=blue](371,0){3}{0}{360}
\psarc[linecolor=blue,linewidth=.5pt,fillstyle=solid,fillcolor=blue](371,15){3}{0}{360}
\psarc[linecolor=blue,linewidth=.5pt,fillstyle=solid,fillcolor=blue](371,30){3}{0}{360}
\psarc[linecolor=red,linewidth=.5pt,fillstyle=solid,fillcolor=red](409,15){3}{0}{360}
\psarc[linecolor=red,linewidth=.5pt,fillstyle=solid,fillcolor=red](409,30){3}{0}{360}
\psarc[linecolor=blue,linewidth=.5pt,fillstyle=solid,fillcolor=blue](423,0){3}{0}{360}
\psarc[linecolor=blue,linewidth=.5pt,fillstyle=solid,fillcolor=blue](423,15){3}{0}{360}
\psarc[linecolor=blue,linewidth=.5pt,fillstyle=solid,fillcolor=blue](423,30){3}{0}{360}
\rput[B](-43,-32){$\qbin62q\;=$}
\multiput(21,-32)(52,0){8}{$+$}
\rput[B](0,-32){$1$}
\rput[B](52,-32){$q$}
\rput[B](104,-32){$2q^2$}
\rput[B](156,-32){$2q^3$}
\rput[B](208,-32){$3q^4$}
\rput[B](260,-32){$2q^5$}
\rput[B](312,-32){$2q^6$}
\rput[B](364,-32){$q^7$}
\rput[B](416,-32){$q^8$}
\rput[B](-45,-3){\scriptsize $j=1$}
\rput[B](-45,12){\scriptsize $j=2$}
\rput[B](-45,27){\scriptsize $j=3$}
\end{pspicture}
\end{center}
\smallskip
\caption{Ramond sectors ($N$ even, $\ell/2$ even): Combinatorial enumeration by 
double-columns of the $q$-binomial $\sqbin nmq=\sqbin 62q
=q^{-1/2} \sum q^{\sum_j m_j E_j}$.  The excess of blue (right) over red (left) 1-strings is given 
by the quantum number $\sigma=\floor{n/2}-m=1$.  The elementary excitation energy of a 
1-string at position $j$ is $E_j=j-\half$. The lowest energy configuration has energy 
$E(\sigma)=\half\sigma^2=\half$. At each position $j$, there are $m_j$ 1-strings and 
$n_j=2-m_j$ 2-strings. This analyticity strip is in the upper-half complex $u$-plane rotated by 
180 degrees so that position $j=1$ (furthest from the real axis) is at the bottom. The elementary 
excitations (of energy 1) are generated by either inserting a left-right pair of 1-strings at position 
$j=1$ or promoting a 1-string at position $j$ to position $j+1$.  Notice that 
$\sqbin nmq=\sqbin n{n-m}q$ as $q$-polynomials but they have different combinatorial 
interpretations because they have different quantum numbers $\sigma$. In the lower half-plane, 
$q$ is replaced with $\qbar$ and no rotation is required. The value $\bar\sigma$ of the quantum 
number in the lower half-plane is related to $\sigma$ in the upper half-plane by the selection 
rules $\sigma+\bar\sigma=\ell/2$ and $\half(\sigma-\bar\sigma)\in\mathbb{Z}$.
\label{binomR1}}
\end{figure}
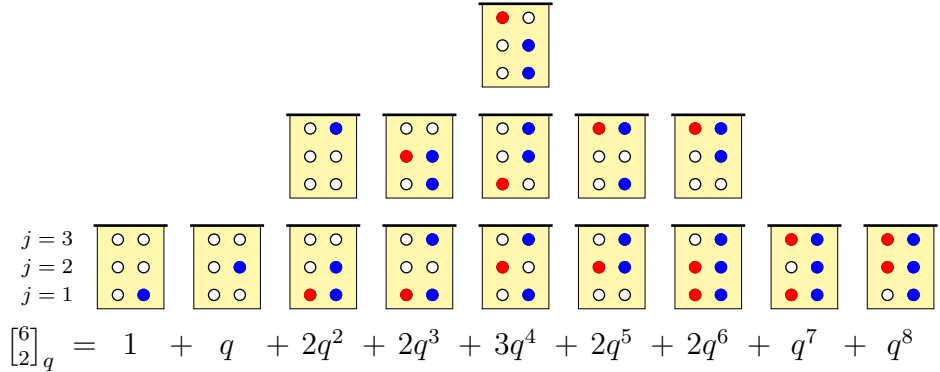

The double-columns with quantum number $\sigma$ are generated combinatorially by 
starting with the minimum energy configuration of 1-strings for given $\sigma$ as shown 
in Figure~\ref{EsigmaR1}. Empirically determined selection rules dictate that in a sector 
with $\ell$ defects the quantum numbers of the groundstate satisfy
\be
 \sigma\;=\;\bar\sigma\;=\;\ell/4,\qquad \ell=0,4,8,\ldots
\ee
The energy of these groundstates is $E(\sigma)+E(\bar\sigma)=\frac{1}{16}\,\ell^2$.

Excitations, incrementing the energy by one unit, are generated either by inserting a left-right 
pair of 1-strings at position $j=1$ or incrementing the position $j$ of a 1-string by 1 unit. 
The $q$-binomials are illustrated in Figure~\ref{binomR1}. 
Empirically, we find that all the excitations satisfy the selection rules
\be
 \sigma+\bar\sigma\;=\;\ell/2,\qquad \half(\sigma-\bar\sigma)\in\mathbb{Z}
\ee
Using the $q$-binomial building blocks and empirical selection rules, we thus obtain 
the finitized partition functions
\bea
Z_\ell^{(N)}(q)=(q\qbar)^{-c/24}\sum_{k\in\mathbb{Z}} 
q^{\Delta_{2k+\ell/2}}\qbin{\sc{2\floor{\frac{N+2}{4}}}}{\sc{\floor{\frac{N+2-\ell}{4}}}\!-\!k}q  
\qbar^{\Delta_{2k-\ell/2}}\qbin{\sc{2\floor{\frac{N}{4}}}}{\sc{\floor{\frac{N-\ell}{4}}}\!+\!k}\qbar,
\ \  \mbox{$N$ even, $\frac{\ell}{2}$ even}\qquad\label{RZ0}
\eea
We further observe that
\bea
Z_0^{(N)}+2\sum_{\ell\in4\mathbb{N}}^{\ell\le N}Z_\ell^{(N)}(q)&=&\half(q\qbar)^{-\frac{c}{24}-\frac{1}{8}}\bigg[
\prod_{n=1}^{\floor{\frac{N+2}{4}}} (1+q^{n-\frac{1}{2}})^2\prod_{n=1}^{\floor{\frac{N}{4}}} (1+\qbar^{n-\frac{1}{2}})^2\qquad\qquad\nonumber\\
&&\hspace{.4in}\mbox{}+\prod_{n=1}^{\floor{\frac{N+2}{4}}} (1-q^{n-\frac{1}{2}})^2\prod_{n=1}^{\floor{\frac{N}{4}}} (1-\qbar^{n-\frac{1}{2}})^2\bigg]
\eea

\subsection{Neveu-Schwarz sectors ($N$ even, $\ell/2$ odd)}

The building blocks of the spectra in the Neveu-Schwarz sectors consist of the $q$-binomials
\bea
\qbin nmq=\qbin n{\floor{n/2}-\sigma}q=q^{-\frac{1}{2}\sigma(\sigma+1)} 
  \sum_{\genfrac{}{}{0pt}{}{\text{$\sigma$-double}}{\text{columns}}} q^{\sum_j m_j E_j},\qquad \sigma=\floor{n/2}-m
\eea
with $E_j=j$. The sum is over all double-column diagrams as in Figure~\ref{binomR2} with fixed $\sigma$. In these sectors, the number of 1-strings in the right column minus the 
number of 1-strings in the left column is related to the quantum number $\sigma$ by
\bea
 m_{\text{right}}-m_{\text{left}}=\mbox{$\sigma\ $ or $\ \sigma+1$}
\eea
The number of 1-strings plus 2-strings at any given position is exactly two
\be
 m_j+n_j\;=\;2,\qquad j=1,2,\ldots,n
\ee

The double-columns with quantum number $\sigma$ are generated combinatorially by 
starting with the minimum energy configuration of 1-strings for given $\sigma$ as shown in 
Figure~\ref{EsigmaR2}. For these {\em minimum} energy configurations
\bea
 m_{\text{right}}-m_{\text{left}}\;=\;\sigma_{\text{min}}\;=\;\begin{cases}
\sigma,&\mbox{$\sigma\ge 0$}\\
\sigma+1,&\mbox{$\sigma<0$}
\end{cases}\label{sigmamin}
\eea
Empirically determined selection rules dictate that in a sector with 
$\ell$ defects the quantum numbers of the groundstate satisfy
\be
 \sigma\;=\;\bar\sigma\;=\;(\ell-2)/4,\qquad \ell=2,6,10,\ldots
\ee
The energy of these groundstates is $E(\sigma)+E(\bar\sigma)=\frac{1}{16}(\ell^2-4)$.

\begin{figure}[p]
\setlength{\unitlength}{.8pt}
\psset{unit=.8pt}
\begin{center}
\begin{pspicture}(-50,-20)(387,50)
\thicklines
\multirput(0,0)(52,0){8}{\psframe[linewidth=0pt,fillstyle=solid,
  fillcolor=yellow!40!white](-18,-7.5)(18,37.5)}
\multirput(0,0)(52,0){8}{\psline[linewidth=1pt](-20,37.5)(20,37.5)}
\multiput(-7,0)(52,0){8}{\multiput(0,0)(0,15){3}{\psarc[linecolor=black,
  linewidth=.5pt,fillstyle=solid,fillcolor=white](0,0){3}{0}{360}}}
\multiput(7,0)(52,0){8}{\multiput(0,0)(0,15){3}{\psarc[linecolor=black,
  linewidth=.5pt,fillstyle=solid,fillcolor=white](0,0){3}{0}{360}}}
\multiput(-7,0)(0,15){3}{\psarc[linecolor=red,linewidth=.5pt,fillstyle=solid,
  fillcolor=red](0,0){3}{0}{360}}
\multiput(45,0)(0,15){2}{\psarc[linecolor=red,linewidth=.5pt,fillstyle=solid,
  fillcolor=red](0,0){3}{0}{360}}
\multiput(97,0)(0,15){1}{\psarc[linecolor=red,linewidth=.5pt,fillstyle=solid,
  fillcolor=red](0,0){3}{0}{360}}
\multiput(267,0)(0,15){1}{\psarc[linecolor=blue,linewidth=.5pt,fillstyle=solid,
  fillcolor=blue](0,0){3}{0}{360}}
\multiput(319,0)(0,15){2}{\psarc[linecolor=blue,linewidth=.5pt,fillstyle=solid,
  fillcolor=blue](0,0){3}{0}{360}}
\multiput(371,0)(0,15){3}{\psarc[linecolor=blue,linewidth=.5pt,fillstyle=solid,
  fillcolor=blue](0,0){3}{0}{360}}
%
\rput[B](2,50){$\qbin77q$}
\rput[B](54,50){$\qbin76q$}
\rput[B](106,50){$\qbin75q$}
\rput[B](158,50){$\qbin74q$}
\rput[B](210,50){$\qbin73q$}
\rput[B](262,50){$\qbin72q$}
\rput[B](314,50){$\qbin71q$}
\rput[B](366,50){$\qbin70q$}
\rput[B](-40,-25){$\sigma$}
\rput[B](0,-25){$-4$}
\rput[B](52,-25){$-3$}
\rput[B](104,-25){$-2$}
\rput[B](156,-25){$-1$}
\rput[B](208,-25){$0$}
\rput[B](260,-25){$1$}
\rput[B](312,-25){$2$}
\rput[B](364,-25){$3$}
\rput[B](-40,-42){$\sigma_{\text{min}}$}
\rput[B](0,-42){$-3$}
\rput[B](52,-42){$-2$}
\rput[B](104,-42){$-1$}
\rput[B](156,-42){$0$}
\rput[B](208,-42){$0$}
\rput[B](260,-42){$1$}
\rput[B](312,-42){$2$}
\rput[B](364,-42){$3$}
\rput[B](-40,-3){\scriptsize $j=1$}
\rput[B](-40,12){\scriptsize $j=2$}
\rput[B](-40,27){\scriptsize $j=3$}
\end{pspicture}
\end{center}
\smallskip
\caption{Neveu-Schwarz sectors ($N$ even, $\ell/2$ odd): Minimal configurations of 
double-columns within the binomials $\qbin nmq=\qbin7mq$. The energy is $E(\sigma)=\frac{1}{2}\sigma(\sigma+1)$ where
the quantum number is $\sigma=\floor{n/2}-m$. The excess of blue (right) over red (left) 
1-strings in these minimal configurations is $\sigma_{\text{min}}$ as given in (\ref{sigmamin}).  At each position $j$, 
the number of 1-strings $m_j$ plus the number of 2-strings $n_j$ is 2. This analyticity strip is in 
the upper-half complex $u$-plane rotated by 180 degrees so that position $j=1$ 
(furthest from the real axis) is at the bottom.
\label{EsigmaR2}}
\end{figure}
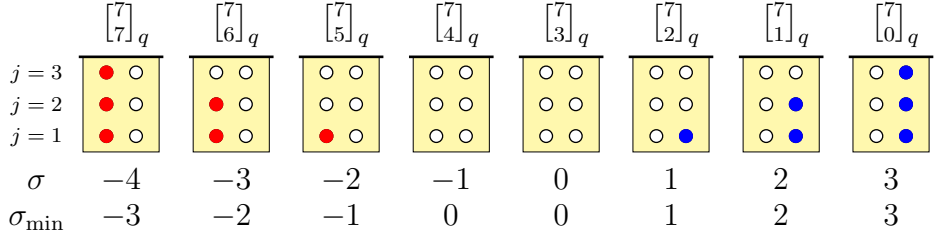

\begin{figure}[p]
\setlength{\unitlength}{.7pt}
\psset{unit=.7pt}
\begin{center}
\begin{pspicture}(-60,-30)(545,170)
\thicklines
\multirput(0,0)(52,0){11}{\psframe[linewidth=0pt,fillstyle=solid,
  fillcolor=yellow!40!white](-18,-7.5)(18,37.5)}
\multirput(0,0)(52,0){11}{\psline[linewidth=1pt](-20,37.5)(20,37.5)}
\multirput(104,60)(52,0){7}{\psframe[linewidth=0pt,fillstyle=solid,
  fillcolor=yellow!40!white](-18,-7.5)(18,37.5)}
\multirput(104,60)(52,0){7}{\psline[linewidth=1pt](-20,37.5)(20,37.5)}
\multirput(208,120)(52,0){3}{\psframe[linewidth=0pt,fillstyle=solid,
  fillcolor=yellow!40!white](-18,-7.5)(18,37.5)}
\multirput(208,120)(52,0){3}{\psline[linewidth=1pt](-20,37.5)(20,37.5)}
%
\multiput(-7,0)(52,0){11}{\multiput(0,0)(0,15){3}{\psarc[linecolor=black,
  linewidth=.5pt,fillstyle=solid,fillcolor=white](0,0){3}{0}{360}}}
\multiput(7,0)(52,0){11}{\multiput(0,0)(0,15){3}{\psarc[linecolor=black,
  linewidth=.5pt,fillstyle=solid,fillcolor=white](0,0){3}{0}{360}}}
\multiput(97,60)(52,0){7}{\multiput(0,0)(0,15){3}{\psarc[linecolor=black,
  linewidth=.5pt,fillstyle=solid,fillcolor=white](0,0){3}{0}{360}}}
\multiput(111,60)(52,0){7}{\multiput(0,0)(0,15){3}{\psarc[linecolor=black,
  linewidth=.5pt,fillstyle=solid,fillcolor=white](0,0){3}{0}{360}}}
\multiput(201,120)(52,0){3}{\multiput(0,0)(0,15){3}{\psarc[linecolor=black,
  linewidth=.5pt,fillstyle=solid,fillcolor=white](0,0){3}{0}{360}}}
\multiput(215,120)(52,0){3}{\multiput(0,0)(0,15){3}{\psarc[linecolor=black,
  linewidth=.5pt,fillstyle=solid,fillcolor=white](0,0){3}{0}{360}}}
\psarc[linecolor=blue,linewidth=.5pt,fillstyle=solid,fillcolor=blue](7,0){3}{0}{360}
\psarc[linecolor=blue,linewidth=.5pt,fillstyle=solid,fillcolor=blue](59,15){3}{0}{360}
\psarc[linecolor=blue,linewidth=.5pt,fillstyle=solid,fillcolor=blue](111,0){3}{0}{360}
\psarc[linecolor=blue,linewidth=.5pt,fillstyle=solid,fillcolor=blue](111,15){3}{0}{360}
\psarc[linecolor=blue,linewidth=.5pt,fillstyle=solid,fillcolor=blue](111,90){3}{0}{360}
\psarc[linecolor=red,linewidth=.5pt,fillstyle=solid,fillcolor=red](149,60){3}{0}{360}
\psarc[linecolor=blue,linewidth=.5pt,fillstyle=solid,fillcolor=blue](163,0){3}{0}{360}
\psarc[linecolor=blue,linewidth=.5pt,fillstyle=solid,fillcolor=blue](163,30){3}{0}{360}
\psarc[linecolor=blue,linewidth=.5pt,fillstyle=solid,fillcolor=blue](163,60){3}{0}{360}
\psarc[linecolor=blue,linewidth=.5pt,fillstyle=solid,fillcolor=blue](163,75){3}{0}{360}
\psarc[linecolor=red,linewidth=.5pt,fillstyle=solid,fillcolor=red](201,0){3}{0}{360}
\psarc[linecolor=red,linewidth=.5pt,fillstyle=solid,fillcolor=red](201,135){3}{0}{360}
\psarc[linecolor=blue,linewidth=.5pt,fillstyle=solid,fillcolor=blue](215,0){3}{0}{360}
\psarc[linecolor=blue,linewidth=.5pt,fillstyle=solid,fillcolor=blue](215,30){3}{0}{360}
\psarc[linecolor=blue,linewidth=.5pt,fillstyle=solid,fillcolor=blue](215,75){3}{0}{360}
\psarc[linecolor=blue,linewidth=.5pt,fillstyle=solid,fillcolor=blue](215,90){3}{0}{360}
\psarc[linecolor=blue,linewidth=.5pt,fillstyle=solid,fillcolor=blue](215,120){3}{0}{360}
\psarc[linecolor=blue,linewidth=.5pt,fillstyle=solid,fillcolor=blue](215,135){3}{0}{360}
\psarc[linecolor=red,linewidth=.5pt,fillstyle=solid,fillcolor=red](253,0){3}{0}{360}
\psarc[linecolor=red,linewidth=.5pt,fillstyle=solid,fillcolor=red](253,75){3}{0}{360}
\psarc[linecolor=red,linewidth=.5pt,fillstyle=solid,fillcolor=red](253,150){3}{0}{360}
\psarc[linecolor=blue,linewidth=.5pt,fillstyle=solid,fillcolor=blue](267,15){3}{0}{360}
\psarc[linecolor=blue,linewidth=.5pt,fillstyle=solid,fillcolor=blue](267,30){3}{0}{360}
\psarc[linecolor=blue,linewidth=.5pt,fillstyle=solid,fillcolor=blue](267,60){3}{0}{360}
\psarc[linecolor=blue,linewidth=.5pt,fillstyle=solid,fillcolor=blue](267,90){3}{0}{360}
\psarc[linecolor=blue,linewidth=.5pt,fillstyle=solid,fillcolor=blue](267,120){3}{0}{360}
\psarc[linecolor=blue,linewidth=.5pt,fillstyle=solid,fillcolor=blue](267,135){3}{0}{360}
\psarc[linecolor=red,linewidth=.5pt,fillstyle=solid,fillcolor=red](305,15){3}{0}{360}
\psarc[linecolor=red,linewidth=.5pt,fillstyle=solid,fillcolor=red](305,90){3}{0}{360}
\psarc[linecolor=red,linewidth=.5pt,fillstyle=solid,fillcolor=red](305,120){3}{0}{360}
\psarc[linecolor=blue,linewidth=.5pt,fillstyle=solid,fillcolor=blue](319,15){3}{0}{360}
\psarc[linecolor=blue,linewidth=.5pt,fillstyle=solid,fillcolor=blue](319,30){3}{0}{360}
\psarc[linecolor=blue,linewidth=.5pt,fillstyle=solid,fillcolor=blue](319,60){3}{0}{360}
\psarc[linecolor=blue,linewidth=.5pt,fillstyle=solid,fillcolor=blue](319,90){3}{0}{360}
\psarc[linecolor=blue,linewidth=.5pt,fillstyle=solid,fillcolor=blue](319,120){3}{0}{360}
\psarc[linecolor=blue,linewidth=.5pt,fillstyle=solid,fillcolor=blue](319,135){3}{0}{360}
\psarc[linecolor=blue,linewidth=.5pt,fillstyle=solid,fillcolor=blue](319,150){3}{0}{360}
\psarc[linecolor=red,linewidth=.5pt,fillstyle=solid,fillcolor=red](357,30){3}{0}{360}
\psarc[linecolor=red,linewidth=.5pt,fillstyle=solid,fillcolor=red](357,75){3}{0}{360}
\psarc[linecolor=blue,linewidth=.5pt,fillstyle=solid,fillcolor=blue](371,15){3}{0}{360}
\psarc[linecolor=blue,linewidth=.5pt,fillstyle=solid,fillcolor=blue](371,30){3}{0}{360}
\psarc[linecolor=blue,linewidth=.5pt,fillstyle=solid,fillcolor=blue](371,60){3}{0}{360}
\psarc[linecolor=blue,linewidth=.5pt,fillstyle=solid,fillcolor=blue](371,75){3}{0}{360}
\psarc[linecolor=blue,linewidth=.5pt,fillstyle=solid,fillcolor=blue](371,90){3}{0}{360}
\psarc[linecolor=red,linewidth=.5pt,fillstyle=solid,fillcolor=red](409,30){3}{0}{360}
\psarc[linecolor=red,linewidth=.5pt,fillstyle=solid,fillcolor=red](409,60){3}{0}{360}
\psarc[linecolor=red,linewidth=.5pt,fillstyle=solid,fillcolor=red](409,75){3}{0}{360}
\psarc[linecolor=blue,linewidth=.5pt,fillstyle=solid,fillcolor=blue](423,0){3}{0}{360}
\psarc[linecolor=blue,linewidth=.5pt,fillstyle=solid,fillcolor=blue](423,15){3}{0}{360}
\psarc[linecolor=blue,linewidth=.5pt,fillstyle=solid,fillcolor=blue](423,30){3}{0}{360}
\psarc[linecolor=blue,linewidth=.5pt,fillstyle=solid,fillcolor=blue](423,60){3}{0}{360}
\psarc[linecolor=blue,linewidth=.5pt,fillstyle=solid,fillcolor=blue](423,75){3}{0}{360}
\psarc[linecolor=blue,linewidth=.5pt,fillstyle=solid,fillcolor=blue](423,90){3}{0}{360}
\psarc[linecolor=red,linewidth=.5pt,fillstyle=solid,fillcolor=red](461,0){3}{0}{360}
\psarc[linecolor=red,linewidth=.5pt,fillstyle=solid,fillcolor=red](461,30){3}{0}{360}
\psarc[linecolor=blue,linewidth=.5pt,fillstyle=solid,fillcolor=blue](475,0){3}{0}{360}
\psarc[linecolor=blue,linewidth=.5pt,fillstyle=solid,fillcolor=blue](475,15){3}{0}{360}
\psarc[linecolor=blue,linewidth=.5pt,fillstyle=solid,fillcolor=blue](475,30){3}{0}{360}
\psarc[linecolor=red,linewidth=.5pt,fillstyle=solid,fillcolor=red](513,15){3}{0}{360}
\psarc[linecolor=red,linewidth=.5pt,fillstyle=solid,fillcolor=red](513,30){3}{0}{360}
\psarc[linecolor=blue,linewidth=.5pt,fillstyle=solid,fillcolor=blue](527,0){3}{0}{360}
\psarc[linecolor=blue,linewidth=.5pt,fillstyle=solid,fillcolor=blue](527,15){3}{0}{360}
\psarc[linecolor=blue,linewidth=.5pt,fillstyle=solid,fillcolor=blue](527,30){3}{0}{360}
\rput[B](-43,-32){$\qbin72q\;=$}
\multiput(21,-32)(52,0){10}{$+$}
\rput[B](0,-32){$1$}
\rput[B](52,-32){$q$}
\rput[B](104,-32){$2q^2$}
\rput[B](156,-32){$2q^3$}
\rput[B](208,-32){$3q^4$}
\rput[B](260,-32){$3q^5$}
\rput[B](312,-32){$3q^6$}
\rput[B](364,-32){$2q^7$}
\rput[B](416,-32){$2q^8$}
\rput[B](468,-32){$q^9$}
\rput[B](520,-32){$q^{10}$}
\rput[B](-45,-3){\scriptsize $j=1$}
\rput[B](-45,12){\scriptsize $j=2$}
\rput[B](-45,27){\scriptsize $j=3$}
\end{pspicture}
\end{center}
\smallskip
\caption{Neveu-Schwarz sectors ($N$ even, $\ell/2$ odd): Combinatorial enumeration by 
double-columns of the $q$-binomial $\sqbin nmq=\sqbin 72q
=q^{-1} \sum q^{\sum_j m_j E_j}$.  The number of positions is \mbox{$(n-1)/2=3$}. The quantum number is $\sigma=\floor{n/2}-m=1$.  
The excess of blue (right) over red (left) 1-strings is $\sigma=1$ or $\sigma+1=2$. The 
elementary excitation energy of a 1-string at position $j$ is $E_j=j$. The lowest energy 
configuration has energy $E(\sigma)=\frac{1}{2}\sigma(\sigma+1)=1$. 
At each position $j$, there are $m_j$ 1-strings and $n_j=2-m_j$ 2-strings. This analyticity strip 
is in the upper-half complex $u$-plane rotated by 180 degrees so that position $j=1$ 
(furthest from the real axis) is at the bottom. The elementary excitations (of energy 1) are 
generated by either inserting a left or right 1-string at position $j=1$ or promoting a 1-string 
at position $j$ to position $j+1$.  Notice that $\sqbin nmq=\sqbin n{n-m}q$ as $q$-polynomials 
but they have different combinatorial interpretations because they have different quantum 
numbers $\sigma$ and $\sigma'=-\sigma-1$. In calculating $\sigma'$, we have used the fact that $n$ in (\ref{RZ2}) is odd. 
In the lower half-plane, $q$ is replaced with $\qbar$ and no rotation is 
required. The value $\bar\sigma$ of the quantum number in the lower half-plane is related to 
$\sigma$ in the upper half-plane by the selection rules $\sigma+\bar\sigma=(\ell-2)/2$ and 
$\half(\sigma-\bar\sigma)\in\mathbb{Z}$.
\label{binomR2}}
\end{figure}

Excitations, incrementing the energy by one unit, are generated either by inserting a left or right 
1-string at position $j=1$ or incrementing the position $j$ of a 1-string by 1 unit. 
The $q$-binomials are illustrated in Figure~\ref{binomR2}. 
Empirically, we find that all the excitations are consistent with the selection rules
\be
 \sigma+\bar\sigma\;=\;(\ell-2)/2,\qquad \half(\sigma-\bar\sigma)\in\mathbb{Z}
\ee
Using the $q$-binomial building blocks and empirical selection rules, we thus obtain the 
finitized partition functions
\bea
Z_\ell^{(N)}(q)=(q\qbar)^{-c/24}\sum_{k\in\mathbb{Z}} 
q^{\Delta_{2k+\ell/2}}\qbin{\sc{2\floor{\frac{N}{4}}}+1}{\sc{\floor{\frac{N+2-\ell}{4}}}\!-\!k}q  
\qbar^{\Delta_{2k-\ell/2}}\qbin{\sc{2\floor{\frac{N+2}{4}}}-1}{\sc{\floor{\frac{N-\ell}{4}}}\!+\!k}\qbar,
\ \  \mbox{$N$ even, $\frac{\ell}{2}$ odd}\qquad
\label{RZ2}
\eea
We further observe that
\bea
\sum_{\ell\in 4\mathbb{N}-2}^{\ell\le N}Z_\ell^{(N)}(q)&=&(q\qbar)^{-\frac{c}{24}}\prod_{n=1}^{\floor{\frac{N}{4}}} (1+q^n)^2\,\prod_{n=1}^{\floor{\frac{N-2}{4}}} (1+\qbar^n)^2
\eea

\goodbreak
\subsection{Finitized characters in Ramond and Neveu-Schwarz sectors}

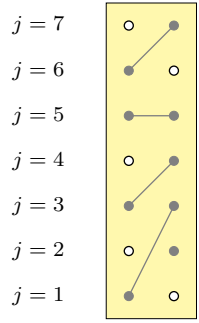
\begin{figure}[htbp]
\psset{unit=.6cm}
\setlength{\unitlength}{.6cm}
\begin{center}
\begin{pspicture}(-.25,0)(2,7)
\psframe[linewidth=0pt,fillstyle=solid,fillcolor=yellow!40!white](0,0)(2,7)
\psarc[linecolor=black,linewidth=.5pt,fillstyle=solid,fillcolor=white](0.5,6.5){.1}{0}{360}
\psarc[linecolor=gray,linewidth=0pt,fillstyle=solid,fillcolor=gray](0.5,5.5){.1}{0}{360}
\psarc[linecolor=gray,linewidth=0pt,fillstyle=solid,fillcolor=gray](0.5,4.5){.1}{0}{360}
\psarc[linecolor=black,linewidth=.5pt,fillstyle=solid,fillcolor=white](0.5,3.5){.1}{0}{360}
\psarc[linecolor=gray,linewidth=0pt,fillstyle=solid,fillcolor=gray](0.5,2.5){.1}{0}{360}
\psarc[linecolor=black,linewidth=.5pt,fillstyle=solid,fillcolor=white](0.5,1.5){.1}{0}{360}
\psarc[linecolor=gray,linewidth=0pt,fillstyle=solid,fillcolor=gray](0.5,0.5){.1}{0}{360}
\psarc[linecolor=gray,linewidth=0pt,fillstyle=solid,fillcolor=gray](1.5,6.5){.1}{0}{360}
\psarc[linecolor=black,linewidth=.5pt,fillstyle=solid,fillcolor=white](1.5,5.5){.1}{0}{360}
\psarc[linecolor=gray,linewidth=0pt,fillstyle=solid,fillcolor=gray](1.5,4.5){.1}{0}{360}
\psarc[linecolor=gray,linewidth=0pt,fillstyle=solid,fillcolor=gray](1.5,3.5){.1}{0}{360}
\psarc[linecolor=gray,linewidth=0pt,fillstyle=solid,fillcolor=gray](1.5,2.5){.1}{0}{360}
\psarc[linecolor=gray,linewidth=0pt,fillstyle=solid,fillcolor=gray](1.5,1.5){.1}{0}{360}
\psarc[linecolor=black,linewidth=.5pt,fillstyle=solid,fillcolor=white](1.5,0.5){.1}{0}{360}
\psline[linecolor=gray,linewidth=.5pt](0.5,5.5)(1.5,6.5)
\psline[linecolor=gray,linewidth=.5pt](0.5,4.5)(1.5,4.5)
\psline[linecolor=gray,linewidth=.5pt](0.5,2.5)(1.5,3.5)
\psline[linecolor=gray,linewidth=.5pt](0.5,0.5)(1.5,2.5)
\rput(-1.5,.5){\scriptsize $j=1$}
\rput(-1.5,1.5){\scriptsize $j=2$}
\rput(-1.5,2.5){\scriptsize $j=3$}
\rput(-1.5,3.5){\scriptsize $j=4$}
\rput(-1.5,4.5){\scriptsize $j=5$}
\rput(-1.5,5.5){\scriptsize $j=6$}
\rput(-1.5,6.5){\scriptsize $j=7$}
\end{pspicture}
\end{center}
\caption{\label{doubdom}An admissible double-column diagram with $m=4$ and $n=5$ 
satisfying dominance $L=\{6,5,3,1\}\preceq \{7,5,4,3,2\}=R$. The energy is $E=15+21=36$ 
and the associated monomial is $q^{36}$.}
\end{figure}

In the Ramond and Neveu-Schwarz sectors with $N$ even, the $q$-binomials do not give 
the appropriate characters. In fact, the relevant finitized irreducible
characters are given~\cite{PR0610,PR_phys} by the generalized $q$-Catalan numbers
\bea
\mch^{(n)}_{r,1}(q)\!\!\!&=&\!\!\! q^{-\frac{c}{24}} \sum_{m=0}^{\frac{n-2r}{2}} 
   \sbinlr{\frac{n-2}{2}}{m,m\!+\!r\!-\!1}_{\!q}
=q^{-\frac{c}{24}+\D_{r,1}} \frac{1-q^r}{1-q^{n/2}} \gauss{n}{\frac{n}{2}-r}\!\!,\ \ \mbox{$n$ even}\\
\mch^{(n)}_{r,2}(q) \!\!\!&=&\!\!\! q^{-\frac{c}{24}-\frac{4r-3}{8}}\!\!\sum_{m=0}^{\frac{n-2r+1}{2}} \!\!
 q^{-m}  \sbinlr{\frac{n-1}{2}}{m,m\!+\!r\!-\!1}_{\!q}
=q^{-\frac{c}{24}+\D_{r,2}} \frac{1-q^{2r}}{1-q^{n+1}} \gauss{n+1}{\frac{n+1}{2}-r}\!\!,\ 
  \mbox{$n$ odd}\ \ \qquad
\eea
Here
\be
  \sbinlr{M}{m,n}_{\!q}\;=\;q^{\hf m(m+1)+\hf n(n+1)}
   \bigg(\gauss{M}{m}\gauss{M}{n}-q^{n-m+1}\gauss{M}{n+1}
    \gauss{M}{m-1}\bigg)
\label{skewbin}
\ee
are generalized versions~\cite{PR0610} of the $q$-Narayana numbers~\cite{Narayana} and the generalized Catalan 
numbers are
\bea
 \mch^{(n)}_{r,1}(1)\!\!\!&=&\!\!\!\bin{n-2}{\frac{n-2}{2}-r+1}-\bin{n-2}{\frac{n-2}{2}-r-1},
  \quad\mbox{$r\ge 1$, $n$ even}\\
 \mch^{(n)}_{r,2}(1)\!\!\!&=&\!\!\!\bin{n-1}{\frac{n-1}{2}-r+1}-\bin{n-1}{\frac{n-1}{2}-r-1},
   \quad \mbox{$r\ge 1$, $n$ odd}
\eea
Despite the minus sign, (\ref{skewbin}) is actually a polynomial with only non-negative coefficients. Combinatorially, 
the generalized $q$-Narayana numbers are generated by double-column diagrams subject 
to admissibility (dominance) as shown in Figure~\ref{doubdom}.

A {\em double-column configuration} of height $M$ consists of a pair of
single-column configurations of height $M$. Suppose that there are $m$ occupied heights in 
the left column and $n$ occupied heights in the right column with $0\le m\le n\le M$. 
Let $L=\{L_1,L_2,\ldots,L_m\}$ be the occupied heights $j$ in descending order of the left 
column and $R=\{R_1,R_2,\ldots,R_n\}$ be the occupied heights $j$ in descending order of 
the right column. We say the double-column diagram is admissible or satisfies dominance if 
$L\ \preceq\ R$ with the partial order
\be
  L\ \preceq\ R\qquad\Longleftrightarrow\qquad L_k\ \leq\ R_k,\quad j=1,2,\ldots,m
\label{order}
\ee
Geometrically, one draws line segments between the occupied sites of greatest height
in the two columns, then between the occupied sites of second-to-greatest
height, and so on. Admissible double-column configurations are now characterized
by {\em not} involving line segments with a {\em strictly negative slope}.

We associate the monomial $q^E$ to each double-column configuration where 
\be
 E\;=\;\sum_{k=1}^mL_k+\sum_{k=1}^nR_k
\label{wLR}
\ee 
is the energy or weight of a double-column configuration. 
The generalized $q$-Narayana number 
\be
 \sbinlr{M}{m,n}_{\!q}\;=
    \sum_{\genfrac{}{}{0pt}{}{\text{admissible}}{\text{double-columns}}}\!\!\! q^E
\ee
is the sum of the monomials associated to the admissible double-column diagrams of height 
$M$ and occupancy $m$ and $n$ in the left and right columns, respectively.

The $q$-binomials decompose in terms of finitized irreducible characters
\bea
 q^{-\frac{c}{24}+\D_{r,1}} \gauss{2n+1}{n-r-1} 
  &=& \sum_{t=r}^{n+1} \mch_{t,1}^{(2n+2)}{(q)}\label{decomp1}\\
 q^{-\frac{c}{24}+\D_{r,2}} \gauss{2n}{n-r+1} 
  &=& \sum_{t = r, \text{ by 2}}^{n\text{ or }n+1} \mch_{t,2}^{(2n+1)}{(q)}
\label{decomp2}
\eea
Substituting the decompositions (\ref{decomp1}) and (\ref{decomp2}) into 
(\ref{RZ0}) and (\ref{RZ2}), we obtain the finitized partition functions
\bea
Z_\ell^{(N)}(q)&=& \begin{cases}
\disp\sum_{r=1}^{\floor{\frac{N+6}{4}}} \sum_{\rbar=1}^{\floor{\frac{N+4}{4}}} 
  Z_{\ell,r,\rbar} \ \mch_{r,2}^{(2\floor{\frac{N+2}{4}}+1)}{(q)} \ 
  \mch_{\rbar,2}^{(2\floor{\frac{N}{4}}+1)}{(\qbar)},
  &\mbox{$\frac{\ell}{2}$ even}
\\[16pt]
\disp\sum_{r=1}^{\floor{\frac{N+4}{4}}} \sum_{\rbar=1}^{\floor{\frac{N+2}{4}}} 
Z_{\ell,r,\rbar} \ \mch_{r,1}^{(2\floor{\frac{N}{4}}+2)}{(q)} \ 
  \mch_{\rbar,1}^{(2\floor{\frac{N+2}{4}})}{(\qbar)},
  &\mbox{$\frac{\ell}{2}$ odd}
\label{finZR}
\end{cases}
\eea
where the coefficients $Z_{\ell,r,\rbar}$ are given by
\be
Z_{\ell,r,\rbar}\;=\;\begin{cases}
\frac{1}{4}(1+(-1)^{r + \rbar}) \big[\max{\big(\frac{\ell}{2},r+\rbar\big)} 
  - \max{\big(\frac{\ell}{2},|r-\rbar| \big)} \big], &\mbox{$\frac{\ell}{2}$ even}
\\[8pt]
\max{\big(\frac{\ell}{2},r+\rbar\big)} - \max{\big(\frac{\ell}{2},|r-\rbar| \big)}, 
  &\mbox{$\frac{\ell}{2}$ odd} 
\end{cases}
\ee

\section{Conformal Partition Functions}
\label{SecConformal}

As $N \to \infty$, the finitized characters become conformal characters. From (\ref{Z4Z1}),
(\ref{Z4Z3}) and (\ref{finZR}), the conformal partition functions are given by the sesquilinear forms
\be
 Z_\ell(q)\;=\;\begin{cases}
 \disp\sum_{k=-\infty}^\infty \mch_{2k+\ell/2}(q)\mch_{2k-\ell/2}(\qbar),\ &
  \mbox{$\mathbb{Z}_4$ ($\ell$ odd)}\\[14pt]
 \disp\sum_{r,\rbar=1}^\infty Z_{\ell,r,\rbar} \ \mch_{r,2} (q) \ \mch_{\rbar,2} (\qbar),&
  \mbox{Ramond ($\frac{\ell}{2}$ even)}\\[14pt]
 \disp\sum_{r,\rbar=1}^\infty Z_{\ell,r,\rbar} \ \mch_{r,1} (q) \ \mch_{\rbar,1} (\qbar),&
   \mbox{Neveu-Schwarz ($\frac{\ell}{2}$ odd)}
\end{cases}
\ee
Summing over sectors with either $\ell/2$ even or $\ell/2$ odd, we find 
\be
\sum_\ell Z_{\ell,r,\rbar}\;=\;\begin{cases}
 Z_{r,\rbar}=\frac{1}{4}\big[1 + (-1)^{r + \rbar}\big] 
  \big[\binom{r+1}{2} + \binom{\rbar+1}{2} - \binom{|r-\rbar|+1}{2} \big], &
    \mbox{$\frac{\ell}{2}$ even}\\[8pt] 
  r\,\rbar, &\mbox{$\frac{\ell}{2}$ odd}
\end{cases}
\ee
and hence
\be
 Z(q)\;=\;\begin{cases}
  \disp\sum_{\ell\in 2\mathbb{N}-1}\sum_{k=-\infty}^\infty \mch_{2k+\ell/2}(q)\mch_{2k-\ell/2}(\qbar),
  &\ \mbox{$\mathbb{Z}_4$ ($\ell$ odd)}\\[14pt]
  \disp\sum_{r,\rbar=1}^\infty Z_{r,\rbar} \ \mch_{r,2} (q) \ \mch_{\rbar,2} (\qbar),
  &\ \mbox{Ramond ($\frac{\ell}{2}$ even)}\\[14pt]
\disp\sum_{r,\rbar=1}^\infty r\,\rbar \ \mch_{r,1} (q) \ \mch_{\rbar,1} (\qbar),
  &\ \mbox{Neveu-Schwarz ($\frac{\ell}{2}$ odd)}
\end{cases}
\ee

Let us define the theta functions
\be
 \vartheta_{s,p}(q)\;=\;\sum_{\lambda\in\mathbb{Z}+\frac{s}{2p}} q^{p\lambda^2}
\ee
It follows from (\ref{Z4Z1}) and (\ref{Z4Z3}), after summing over the $\mathbb{Z}_4$ sectors, that 
\be
 \sum_{\ell\in 2\mathbb{N}-1} Z_\ell(q)
   \;=\;\frac{|\vartheta_{\frac{1}{2},2}(q)|^2+|\vartheta_{\frac{3}{2},2}(q)|^2}{|\eta(q)|^2}
\ee
where the Dedekind eta function is
\be
 \eta(q)\;=\;q^{1/24}\prod_{n=1}^\infty (1-q^n)
\ee
Similarly, from (\ref{RZ0}) and (\ref{RZ2}), we note that summing over  the $\ell$ even sectors 
with suitable mutiplicities yields
\bea
\begin{array}{rcl}
\disp Z_0(q)\!+\!2\sum_{\ell\in 4\mathbb{N}} Z_\ell(q)\!\!\!&=&\!\!\!
  \disp{\frac{|\vartheta_{0,2}(q)|^2+|\vartheta_{2,2}(q)|^2}{|\eta(q)|^2}}
=|\hat\chi_{-1/8}(q)|^2+|\hat\chi_{3/8}(q)|^2\\
\disp2\!\!\sum_{\ell\in 4\mathbb{N}-2} Z_\ell(q)\!\!\!&=&\!\!\!
  \disp{\frac{|\vartheta_{1,2}(q)|^2+|\vartheta_{3,2}(q)|^2}{|\eta(q)|^2}}
  =\frac{2|\vartheta_{1,2}(q)|^2}{|\eta(q)|^2}
=2|\hat\chi_0(q)+\hat\chi_1(q)|^2\qquad\\
\disp Z_0(q)\!+\!2\sum_{\ell\in 2\mathbb{N}} Z_\ell(q)\!\!\!&=&\!\!\!
  \disp{\frac{1}{|\eta(q)|^2}} \sum_{s=0}^3|\vartheta_{s,2}(q)|^2
=|\hat\chi_{-1/8}(q)|^2+|\hat\chi_{3/8}(q)|^2+2|\hat\chi_0(q)\!+\!\hat\chi_1(q)|^2
    \qquad\hspace{-.7in}\mbox{}
\end{array}
\label{ZZZ}
\eea
where the ${\cal W}$-irreducible characters are
\be
\begin{array}{rclrcl}
\hat\chi_{-1/8}(q)&=&\disp{\frac{1}{\eta(q)}}\,\vartheta_{0,2}(q),\qquad&
\hat\chi_0(q)&=& \disp{\frac{1}{2\eta(q)}}[\vartheta_{1,2}(q)+\eta(q)^3]\\[10pt]
\hat\chi_{3/8}(q)&=& \disp{\frac{1}{\eta(q)}}\,\vartheta_{2,2}(q),\qquad&
\hat\chi_1(q)&=& \disp{\frac{1}{2\eta(q)}}[\vartheta_{1,2}(q)-\eta(q)^3]
\end{array}
\ee
In the last equation of (\ref{ZZZ}), we have produced a modular invariant which we believe is the correct modular invariant partition
function for critical dense polymers on the torus.
However, we stress that this partition function is not obtained by the naive trace
summing over all even $N$ sectors since we have changed the multiplicity of all of the 
$\ell\ne 0$ sectors by a factor of 2.
We note that the naive trace does not yield a modular invariant.

Although we do not give details, the methods of this paper also apply to the case of Identified Connectivities (IC) 
so we include the results for completeness. 
In this case there are no defects, $E_j=j$ and the finitized partition function is
\bea
Z_{IC}^{(N)}(q)=\sum_{r=1}^{\floor{\frac{N}{4}}} \mch_{r,1}^{(2\floor{\frac{N}{4}})}(q)\,\mch_{r,1}^{(2\floor{\frac{N-2}{4}})}(\qbar)
\eea
It follows that the conformal partition function is
\bea
Z_{IC}(q)=\sum_{r=1}^\infty |\mch_{r,1}(q)|^2
\eea
Further results could be obtained for the case of Distinguished Connectivities (DC) with $\ell$ defects and $\alpha=0$ or $\alpha=-2$. The cases $\alpha\ne 0,\pm 2$ are also integrable but can not be solved in the $\ell=0$ sector by the simple factorization techniques used in this paper.

\goodbreak
\section{Conclusion}
\label{SecConclusion}

In this paper, we have solved a model of critical dense polymers exactly on 
arbitrary finite-size cylinders. This topology allows for non-contractible loops with 
fugacity $\alpha$ that wind around the cylinder or for an arbitrary number $\ell$ of defects
that propagate along the full length of the cylinder. 
We have set up commuting transfer matrices that satisfy a functional equation in the form of 
an inversion identity. 
For even $N$ with loop fugacity $\alpha=2$, this involves a non-diagonalizable braid operator 
$\Jb$ and an involution ${\Rb}=-(\vec J^3-12\vec J)/16=(-1)^{\svec F}$ with eigenvalues 
$R=(-1)^{\ell/2}$. This is reminiscent of supersymmetry with a pair of defects interpreted as 
a fermion and $\Fb$ as a fermi number operator. 
However, we have no direct evidence of fermions or supersymmetry in our model. 
Nevertheless, since the number of defects $\ell$ is a quantum number that separates the 
theory into various sectors, we call these sectors Ramond ($\ell/2$ even), 
Neveu-Schwarz ($\ell/2$ odd) and $\mathbb{Z}_4$ ($\ell$ odd), respectively. 
We conjecture that, when acting on link states with an arbitrary even number of defects, the braid operator $\Rb$ and transfer matrices $\Tb(u)$ are of rank 2 and 
have no rank-3 or higher Jordan blocks for arbitrary even $N$. 
The existence of non-trivial Jordan blocks is indicative that the theory on a long cylinder or torus is logarithmic in the sense that it gives rise to indecomposable representations of the Virasoro generator $L_0$.

For  $\alpha=2$ and arbitrary $N$, the inversion identity is solved exactly sector by sector for 
the eigenvalues of the transfer matrices in finite geometry. We thus obtain finitized conformal 
partition functions as sesquilinear forms in appropriate finitized characters.
The eigenvalues are classified by the physical combinatorics 
of the patterns of zeros in the complex spectral-parameter plane.
This yields selection rules for the physically relevant solutions to the inversion identity. 
The finite-size corrections follow from Euler-Maclaurin formulas.
In the scaling limit, we obtain the conformal partition functions as sesquilinear forms and 
confirm~\cite{Saleur87,Duplantier,SaleurSUSY} the central charge $c=-2$ and conformal
weights $\Delta,\bar\Delta=\Delta_t=(t^2-1)/8$. Here $t=\ell/2$ and $t=2r-s\in\mathbb{N}$ in the 
$\ell$ even sectors with Kac labels $r=1,2,3,\ldots; s=1,2$ while $t\in\mathbb{Z}-\half$ in the 
$\ell$ odd sectors. Strikingly, the $\ell/2$ odd sectors exhibit a ${\cal W}$-extended 
symmetry but the $\ell/2$ even sectors do not. Moreover, the naive trace summing over all 
even $N$ sectors does not yield a modular invariant.

Recently, we considered~\cite{PRR0803} critical dense polymers on the strip with 
boundary conditions that, in the continuum scaling limit, respect a ${\cal W}$-extended 
symmetry. Our analysis enabled us to identify this model 
with the $c_{1,2}$ triplet model or symplectic fermions on the strip. 
It appears that no such simple identification can be made between these models on a long 
cylinder or torus after taking a naive trace. 
To resolve this seeming paradox, it may be necessary to modify the lattice model on a cylinder to 
properly restore the fermion structure and supersymmetry, or to introduce a modified trace implementing the proper torus geometry.  
Only then can we hope to have a proper interpretation of the fermi number operator $\Fb$ and 
find agreement, in the $N$ even sectors, 
with the conformal partition functions obtained by Saleur~\cite{SaleurSUSY}. 
We intend to return to this question in a later publication.

\subsection*{Acknowledgments}

This work is supported by the Australian Research Council.

\appendix

\section{Properties of $\Jb$}
\label{AppJ}

For $N$ even, $\Jb$ contains a non-contractible loop in exactly two diagrams
\be
 \Jb\;=\;\Jb_{\!0}+\Jb_{\!1}
\ee
where $\Jb_{\!0}$ refers to the part without non-contractible loops, while
\psset{unit=.5cm}
\setlength{\unitlength}{.5cm}
\be
 (-1)^{\frac{N}{2}}\Jb_{\!1}\; =\ \;\!
\begin{pspicture}(0,0.8)(8,2)
\facegrid{(0,0)}{(8,2)}
\put(0,0){\loopa}
\put(0,1){\loopa}
\put(1,0){\loopb}
\put(1,1){\loopb}
\put(2,0){\loopa}
\put(2,1){\loopa}
\put(3,0){\loopb}
\put(3,1){\loopb}
\put(4,0){\loopa}
\put(4,1){\loopa}
\put(5,0){\loopb}
\put(5,1){\loopb}
\put(6,0){\loopa}
\put(6,1){\loopa}
\put(7,0){\loopb}
\put(7,1){\loopb}
\end{pspicture}
\ +\ 
\begin{pspicture}(0,0.8)(8,2)
\facegrid{(0,0)}{(8,2)}
\put(0,0){\loopb}
\put(0,1){\loopb}
\put(1,0){\loopa}
\put(1,1){\loopa}
\put(2,0){\loopb}
\put(2,1){\loopb}
\put(3,0){\loopa}
\put(3,1){\loopa}
\put(4,0){\loopb}
\put(4,1){\loopb}
\put(5,0){\loopa}
\put(5,1){\loopa}
\put(6,0){\loopb}
\put(6,1){\loopb}
\put(7,0){\loopa}
\put(7,1){\loopa}
\end{pspicture}
\label{R1}
\ee
\\
(here illustrated for $N=8$). In the enlarged TL algebra, this is just $\al\;\!\Omega\big(E+F\big)$.

In order to describe $\Jb$, and not just $\Jb_{\!1}$, in terms of the enlarged TL algebra, 
we introduce the `periodically ordered subsets' $\bar\sigma_n$ defined as follows.
For $n<N$, let $\{\sigma_1,\ldots,\sigma_n\}$ be a set of distinct elements of
$\{0,\ldots,N-1\}$, and let $\bar\sigma_n$ be an ordering of $\{\sigma_1,\ldots,\sigma_n\}$ 
for which $\sigma_{j+1}\neq\sigma_j-1$ and $\sigma_{1}\neq\sigma_n-1$.
Two such orderings are equivalent if the corresponding ordered products of
periodic TL generators match
\be
 \bar\sigma_n\;\sim\;\bar\sigma'_n\quad\Leftrightarrow\quad
 e_{\sigma_1}\ldots e_{\sigma_n}\;=\;e_{\sigma'_1}\ldots e_{\sigma'_n}
\ee
We can now write $\Jb$ in terms of the enlarged TL algebra as
\bea
 \Jb&=&(-1)^N\Omega^2+\sum_{n=1}^{N-1}(-1)^{N-n}
   \sum_{[\bar\sigma_n]}\big(e_{\sigma_1}\ldots
  e_{\sigma_n}\Omega\big)^{\!2}+\Omega^{-2}\nn
 &=&(-1)^N\Omega^2+\sum_{n=1}^{N-1}(-1)^{N-n}\sum_{[\bar\sigma_n]}
    \big(\Omega\;\! e_{\sigma_1}\ldots e_{\sigma_n}\big)^{\!2}+\Omega^{-2}\nn
 &=&(-1)^N\Omega^2+\sum_{n=1}^{N-1}(-1)^{N-n}\sum_{[\bar\sigma_n]}\big(e_{\sigma_n}\ldots
  e_{\sigma_1}\Omega^{-1}\big)^{\!2}+\Omega^{-2}\nn
 &=&(-1)^N\Omega^2+\sum_{n=1}^{N-1}(-1)^{N-n}\sum_{[\bar\sigma_n]}
  \big(\Omega^{-1}e_{\sigma_n}\ldots e_{\sigma_1}\big)^{\!2}+\Omega^{-2}
\eea
where the sum $\sum_{[\bar\sigma_n]}$ is over the set of equivalence classes of 
periodically ordered subsets of cardinality $n$.
$\bar\sigma_n$ is thus a representative of the given equivalence class.

\subsection{Proof of Inversion Identity}
\label{AppInversion}

\noindent {\bf Proof of Inversion Identity} (Section~\ref{SecCylinder})\ \ 
Recalling that $\la=\frac{\pi}{2}$,
the left side of (\ref{InvCyl}) corresponds diagrammatically to
\psset{unit=.7cm}
\setlength{\unitlength}{.7cm}
\be
  \Tb(u)\Tb(u+\frac{\pi}{2})\;=\ \;\!
\begin{pspicture}(0,0.8)(8,2.5)
\facegrid{(0,0)}{(8,2)}
\rput(.5,.5){\scriptsize $u$}
\rput(7.5,.5){\scriptsize $u$}
\rput(.5,1.5){\scriptsize $u\!\!+\!\!\la$}
\rput(7.5,1.5){\scriptsize $u\!\!+\!\!\la$}
\psarc[linewidth=.5pt](0,0){.15}{0}{90}
\psarc[linewidth=.5pt](7,0){.15}{0}{90}
\psarc[linewidth=.5pt](0,1){.15}{0}{90}
\psarc[linewidth=.5pt](7,1){.15}{0}{90}
\rput(2.5,0.5){\scriptsize $\dots$}
\rput(5.5,0.5){\scriptsize $\dots$}
\rput(2.5,1.5){\scriptsize $\dots$}
\rput(5.5,1.5){\scriptsize $\dots$}
\end{pspicture}
\label{TMTM}
\ee
\\[.2cm]
with the left and right (vertical) edges identified.
First, we examine the consequences of having a half-arc connecting the two left nodes
(or two right nodes) of a 3-tangle (two-column) appearing in (\ref{TMTM}). 
Expanding in terms of connectivities, we find
\psset{unit=.5cm}
\setlength{\unitlength}{.5cm}
\bea
&\!\!\begin{pspicture}(0,0.75)(1,2.3)
\leftarc{(0,1)}
\facegrid{(0,0)}{(1,2)}
\rput(.5,1.5){\tiny $u\!\!+\!\!\la$}
\rput(.5,.5){\tiny $u$}
\psarc[linewidth=.5pt](0,0){.15}{0}{90}
\psarc[linewidth=.5pt](0,1){.15}{0}{90}
\end{pspicture}
\;=\; \!\!-\sin u\cos u\ \ \ 
\begin{pspicture}(0,0.75)(1,2.3)
\leftarc{(0,1)}
\facegrid{(0,0)}{(1,2)}
\put(0,0){\loopa}
\put(0,1){\loopa}
\end{pspicture}
\;\!+\!\cos u\sin u\ \ \ 
\begin{pspicture}(0,0.75)(1,2.3)
\leftarc{(0,1)}
\facegrid{(0,0)}{(1,2)}
\put(0,0){\loopb}
\put(0,1){\loopb}
\end{pspicture}
\;\!-\sin^2 \!u\ \ \ 
\begin{pspicture}(0,0.75)(1,2.3)
\leftarc{(0,1)}
\facegrid{(0,0)}{(1,2)}
\put(0,0){\loopb}
\put(0,1){\loopa}
\end{pspicture}
\;\!+0\ \ \ 
\begin{pspicture}(0,0.75)(1,2.3)
\leftarc{(0,1)}
\facegrid{(0,0)}{(1,2)}
\put(0,0){\loopa}
\put(0,1){\loopb}
\end{pspicture}
\;=\;\!\!-\sin^2 \!u\ \ \ 
\begin{pspicture}(0,0.75)(1,2.3)
\leftarc{(0,1)}
\facegrid{(0,0)}{(1,2)}
\put(0,0){\loopb}
\put(0,1){\loopa}
\end{pspicture}
&
\label{right}
\\[18pt]
&\!\!\!\!\!\!\!
\begin{pspicture}(0,0.75)(1,2.3)
\rightarc{(1,1)}
\facegrid{(0,0)}{(1,2)}
\rput(.5,1.5){\tiny $u\!\!+\!\!\la$}
\rput(.5,.5){\tiny $u$}
\psarc[linewidth=.5pt](0,0){.15}{0}{90}
\psarc[linewidth=.5pt](0,1){.15}{0}{90}
\end{pspicture}
\ \ =\; \!\!-\sin u\cos u\ \;\!
\begin{pspicture}(0,0.75)(1,2.3)
\rightarc{(1,1)}
\facegrid{(0,0)}{(1,2)}
\put(0,0){\loopa}
\put(0,1){\loopa}
\end{pspicture}
\ \ +\!\cos u\sin u\ \;\!
\begin{pspicture}(0,0.75)(1,2.3)
\rightarc{(1,1)}
\facegrid{(0,0)}{(1,2)}
\put(0,0){\loopb}
\put(0,1){\loopb}
\end{pspicture}
\ \ +0\ 
\begin{pspicture}(0,0.75)(1,2.3)
\rightarc{(1,1)}
\facegrid{(0,0)}{(1,2)}
\put(0,0){\loopb}
\put(0,1){\loopa}
\end{pspicture}
\ \ +\cos^2 \!u\ \;\!
\begin{pspicture}(0,0.75)(1,2.3)
\rightarc{(1,1)}
\facegrid{(0,0)}{(1,2)}
\put(0,0){\loopa}
\put(0,1){\loopb}
\end{pspicture}
 \ \ =\,\cos^2 \!u\ \;\!
\begin{pspicture}(0,0.75)(1,2.3)
\rightarc{(1,1)}
\facegrid{(0,0)}{(1,2)}
\put(0,0){\loopa}
\put(0,1){\loopb}
\end{pspicture}
\label{left}
\eea
\\
and observe that such a half-arc will `propagate' and ultimately lead to a vertical
identity diagram $\Ib$. This is possible for a right- or left-moving half-arc. Taking the
coefficients $-\sin^2u$ (in the right-moving scenario (\ref{right})) and $\cos^2u$ 
(in the left-moving scenario (\ref{left})) into consideration,
this immediately gives the term proportional to $\Ib$ in (\ref{InvCyl}).
Having accounted for all situations with a half-arc, we are left with the two weighted 
connectivities
\psset{unit=.5cm}
\setlength{\unitlength}{.5cm}
\be
-\cos u\sin u\
\begin{pspicture}(0,0.8)(1,2.3)
\facegrid{(0,0)}{(1,2)}
\put(0,0){\loopa}
\put(0,1){\loopa}
\end{pspicture}
\qquad\quad \mathrm{and}\qquad \quad
\cos u\sin u\
\begin{pspicture}(0,0.8)(1,2.3)
\facegrid{(0,0)}{(1,2)}
\put(0,0){\loopb}
\put(0,1){\loopb}
\end{pspicture}
\ee
\\
for each of the $N$ 3-tangles in (\ref{TMTM}). Adding up all possible combinations of
these connectivities readily produces the term proportional to $\Jb$ in (\ref{InvCyl}).
\quad$\Box$

\subsection{Proof of Drop-Down Lemma}
\label{AppDrop}

\noindent {\bf Proof of Drop-Down Lemma} (Section~\ref{SecMatrixJ})\ \ 
Since a non-trivial arc-part only appears for 
$N\geq2$, we assume $N\geq2$ and consider a segment of $\Jb$ consisting of two 
3-tangles (\ref{3m3}) with a half-arc linking the two nodes on the upper edge
\psset{unit=.7cm}
\setlength{\unitlength}{.7cm}
\be
\begin{pspicture}(0,3)(2,0.85)
\facegrid{(0,0)}{(2,2)}
\psarc[linewidth=1.5pt,linecolor=blue](1,2){.5}{0}{180}
\end{pspicture}
\ \;=\;\ 
\begin{pspicture}(0,3)(2,0.85)
\facegrid{(0,0)}{(2,2)}
\put(0,0){\loopa}
\put(0,1){\loopa}
\put(1,0){\loopa}
\put(1,1){\loopa}
\psarc[linewidth=1.5pt,linecolor=blue](1,2){.5}{0}{180}
\end{pspicture}
\ -\ 
\begin{pspicture}(0,3)(2,0.85)
\facegrid{(0,0)}{(2,2)}
\put(0,0){\loopa}
\put(0,1){\loopa}
\put(1,0){\loopb}
\put(1,1){\loopb}
\psarc[linewidth=1.5pt,linecolor=blue](1,2){.5}{0}{180}
\end{pspicture}
\ -\ 
\begin{pspicture}(0,3)(2,0.85)
\facegrid{(0,0)}{(2,2)}
\put(0,0){\loopb}
\put(0,1){\loopb}
\put(1,0){\loopa}
\put(1,1){\loopa}
\psarc[linewidth=1.5pt,linecolor=blue](1,2){.5}{0}{180}
\end{pspicture}
\ +\ 
\begin{pspicture}(0,3)(2,0.85)
\facegrid{(0,0)}{(2,2)}
\put(0,0){\loopb}
\put(0,1){\loopb}
\put(1,0){\loopb}
\put(1,1){\loopb}
\psarc[linewidth=1.5pt,linecolor=blue](1,2){.5}{0}{180}
\end{pspicture}
\label{J2}
\ee
\\
The third diagram of the right side vanishes due to the contractible loop.
For $N=2$, the identification of the left and right vertical edges gives back a half-arc
multiplied by $2-\al^2$, and the proof is complete. For $N>2$, we include the 3-tangle 
to the left or right, respectively, of the first and fourth diagrams
\psset{unit=.7cm}
\setlength{\unitlength}{.7cm}
\bea
\begin{pspicture}(0,3)(3,0.85)
\facegrid{(0,0)}{(3,2)}
\put(1,0){\loopa}
\put(1,1){\loopa}
\put(2,0){\loopa}
\put(2,1){\loopa}
\psarc[linewidth=1.5pt,linecolor=blue](2,2){.5}{0}{180}
\end{pspicture}
\ \;=\ \;
\begin{pspicture}(0,3)(3,0.85)
\facegrid{(0,0)}{(3,2)}
\put(0,0){\loopb}
\put(0,1){\loopb}
\put(1,0){\loopa}
\put(1,1){\loopa}
\put(2,0){\loopa}
\put(2,1){\loopa}
\psarc[linewidth=1.5pt,linecolor=blue](2,2){.5}{0}{180}
\end{pspicture}
\ \;-\ \;
\begin{pspicture}(0,3)(3,0.85)
\facegrid{(0,0)}{(3,2)}
\put(0,0){\loopa}
\put(0,1){\loopa}
\put(1,0){\loopa}
\put(1,1){\loopa}
\put(2,0){\loopa}
\put(2,1){\loopa}
\psarc[linewidth=1.5pt,linecolor=blue](2,2){.5}{0}{180}
\end{pspicture}
&=&\!0\\[.4cm]
\begin{pspicture}(0,3)(3,0.85)
\facegrid{(0,0)}{(3,2)}
\put(0,0){\loopb}
\put(0,1){\loopb}
\put(1,0){\loopb}
\put(1,1){\loopb}
\psarc[linewidth=1.5pt,linecolor=blue](1,2){.5}{0}{180}
\end{pspicture}
\ \;=\ \;
\begin{pspicture}(0,3)(3,0.85)
\facegrid{(0,0)}{(3,2)}
\put(0,0){\loopb}
\put(0,1){\loopb}
\put(1,0){\loopb}
\put(1,1){\loopb}
\put(2,0){\loopb}
\put(2,1){\loopb}
\psarc[linewidth=1.5pt,linecolor=blue](1,2){.5}{0}{180}
\end{pspicture}
\ \;-\ \;
\begin{pspicture}(0,3)(3,0.85)
\facegrid{(0,0)}{(3,2)}
\put(0,0){\loopb}
\put(0,1){\loopb}
\put(1,0){\loopb}
\put(1,1){\loopb}
\put(2,0){\loopa}
\put(2,1){\loopa}
\psarc[linewidth=1.5pt,linecolor=blue](1,2){.5}{0}{180}
\end{pspicture}
&=&\!0
\eea
\\
It follows that, for $N>2$, we simply have
\psset{unit=.7cm}
\setlength{\unitlength}{.7cm}
\be
\begin{pspicture}(0,3)(2,0.85)
\facegrid{(0,0)}{(2,2)}
\psarc[linewidth=1.5pt,linecolor=blue](1,2){.5}{0}{180}
\end{pspicture}
\ \;=\; 
-\ 
\begin{pspicture}(0,3)(2,0.85)
\facegrid{(0,0)}{(2,2)}
\put(0,0){\loopa}
\put(0,1){\loopa}
\put(1,0){\loopb}
\put(1,1){\loopb}
\psarc[linewidth=1.5pt,linecolor=blue](1,2){.5}{0}{180}
\end{pspicture}
\label{2m2}
\ee
\\
implying that a single half-arc drops down leaving behind a couple of horizontal links
(indicated by wavy links in (\ref{2m2})) between the two vertical edges. 
For given arc-part of the input link state, this procedure is repeated starting with the 
innermost half-arc(s) and working out, as illustrated here
\psset{unit=.5cm}
\setlength{\unitlength}{.5cm}
\bea
&\begin{pspicture}(0,5.5)(12,0.8)
\facegrid{(0,0)}{(12,2)}
\psarc[linewidth=1.5pt,linecolor=blue](3,2){.5}{0}{180}
\psarc[linewidth=1.5pt,linecolor=blue](6,2){.5}{0}{180}
\psarc[linewidth=1.5pt,linecolor=blue](10,2){.5}{0}{180}
\psarc[linewidth=1.5pt,linecolor=blue](3,2){1.5}{0}{180}
\psarc[linewidth=1.5pt,linecolor=blue](4,2){3.5}{0}{180}
\end{pspicture}
\ \;=\; 
-\ 
\begin{pspicture}(0,5.5)(12,0.8)
\facegrid{(0,0)}{(12,2)}
\put(2,0){\loopa}
\put(2,1){\loopa}
\put(3,0){\loopb}
\put(3,1){\loopb}
\put(5,0){\loopa}
\put(5,1){\loopa}
\put(6,0){\loopb}
\put(6,1){\loopb}
\put(9,0){\loopa}
\put(9,1){\loopa}
\put(10,0){\loopb}
\put(10,1){\loopb}
\psarc[linewidth=1.5pt,linecolor=blue](3,2){.5}{0}{180}
\psarc[linewidth=1.5pt,linecolor=blue](6,2){.5}{0}{180}
\psarc[linewidth=1.5pt,linecolor=blue](10,2){.5}{0}{180}
\psarc[linewidth=1.5pt,linecolor=blue](3,2){1.5}{0}{180}
\psarc[linewidth=1.5pt,linecolor=blue](4,2){3.5}{0}{180}
\end{pspicture}
\nonumber
\\[0.8cm]
=&\begin{pspicture}(0,5.5)(12,0.8)
\facegrid{(0,0)}{(12,2)}
\put(1,0){\loopa}
\put(1,1){\loopa}
\put(2,0){\loopa}
\put(2,1){\loopa}
\put(3,0){\loopb}
\put(3,1){\loopb}
\put(4,0){\loopb}
\put(4,1){\loopb}
\put(5,0){\loopa}
\put(5,1){\loopa}
\put(6,0){\loopb}
\put(6,1){\loopb}
\put(9,0){\loopa}
\put(9,1){\loopa}
\put(10,0){\loopb}
\put(10,1){\loopb}
\psarc[linewidth=1.5pt,linecolor=blue](3,2){.5}{0}{180}
\psarc[linewidth=1.5pt,linecolor=blue](6,2){.5}{0}{180}
\psarc[linewidth=1.5pt,linecolor=blue](10,2){.5}{0}{180}
\psarc[linewidth=1.5pt,linecolor=blue](3,2){1.5}{0}{180}
\psarc[linewidth=1.5pt,linecolor=blue](4,2){3.5}{0}{180}
\end{pspicture}
\ \;=\; 
-\ 
\begin{pspicture}(0,5.5)(12,0.8)
\facegrid{(0,0)}{(12,2)}
\put(0,0){\loopa}
\put(0,1){\loopa}
\put(1,0){\loopa}
\put(1,1){\loopa}
\put(2,0){\loopa}
\put(2,1){\loopa}
\put(3,0){\loopb}
\put(3,1){\loopb}
\put(4,0){\loopb}
\put(4,1){\loopb}
\put(5,0){\loopa}
\put(5,1){\loopa}
\put(6,0){\loopb}
\put(6,1){\loopb}
\put(7,0){\loopb}
\put(7,1){\loopb}
\put(9,0){\loopa}
\put(9,1){\loopa}
\put(10,0){\loopb}
\put(10,1){\loopb}
\psarc[linewidth=1.5pt,linecolor=blue](3,2){.5}{0}{180}
\psarc[linewidth=1.5pt,linecolor=blue](6,2){.5}{0}{180}
\psarc[linewidth=1.5pt,linecolor=blue](10,2){.5}{0}{180}
\psarc[linewidth=1.5pt,linecolor=blue](3,2){1.5}{0}{180}
\psarc[linewidth=1.5pt,linecolor=blue](4,2){3.5}{0}{180}
\end{pspicture}
\label{12}
\eea
\\
in an example for $N=12$.\quad$\Box$

\subsection{Proof of Sector Lemma}
\label{AppSector}

\noindent {\bf Proof of Sector Lemma} (Section~\ref{SecMatrixJ})\ \ 
The lemma is readily verified for $N=1$. 
For $N=2$, there are two link states with $\ell=0$ and one link state with $\ell=2$. 
$\Jb_0=\Omega^2+\Omega^{-2}$ acts as twice the identity. $\Jb_1$ maps the link state with 
$\ell=2$ into minus the sum of the link states with 
$\ell=0$ (which is outside the sector and therefore set to zero), while it maps both link states 
with $\ell=0$ into themselves multiplied by $-\al^2$. This completes the proof for $N=2$.
For $N>2$ and $\ell>0$, we get a factor of $(-1)^{\frac{N-\ell}{2}}$ from the drop-down process.
Within the given sector, the $\ell$ defects are mapped into $\ell$ defects. This requires
the 3-tangles not affected by the drop-down process (as the 9th and 12th 3-tangles 
in the last diagram of (\ref{12})) to be identical thereby giving rise to an additional
factor of $((-1)^\ell+1)$. Since this factor is 2 for $N,\ell$ even and vanishes for $N,\ell$ odd,
we immediately recognize (\ref{SectorLemma}).
Finally, for $N>2$ and $\ell=0$, we perform the drop-down process for all but the final, outmost
half-arc. This gives a factor of $(-1)^{\frac{N-2}{2}}$ and we are left with a scenario
as in (\ref{J2}) giving rise to an additional factor of $2-\al^2$. \quad$\Box$

\subsection{Matrix $\Jb$ for an arbitrary number of defects}
\label{AppDefects}

Here, we continue the description at the end of Section~\ref{SecMatrixJ} of the matrix 
realization of $\Jb$ appearing in the Inversion Identity (\ref{TMTM}) for given $N$
when an arbitrary number of defects is allowed.

To simplify the description, let us introduce the following algebraic abbreviations
\psset{unit=.5cm}
\setlength{\unitlength}{.5cm}
\be
u\;=\;
-\;\begin{pspicture}(0,2.1)(1,0.85)
\facegrid{(0,0)}{(1,2)}
\put(0,0){\loopa}
\put(0,1){\loopa}
\end{pspicture}
\quad , \qquad \quad
d\;=\
\begin{pspicture}(0,2.1)(1,0.85)
\facegrid{(0,0)}{(1,2)}
\put(0,0){\loopb}
\put(0,1){\loopb}
\end{pspicture}
\ee
\\[-.2cm]
allowing us to write a horizontal concatenation of such 3-tangles simply as products
of $u$'s and $d$'s. For example, ignoring the half-arc linking the nodes on the upper edge 
in (\ref{J2}), the sum of diagrams on the right side of (\ref{J2}) reads $uu+ud+du+dd$.

For a link state $A$, we let $\ell(A)$ denote
the number of defects, $arc(A)$ the arc-part, and $h(A)$ the maximal height of the arcs, where the
height of an arc is the maximal number of points of intersection of arcs 
(on or below the arc itself) with a vertical line through the horizontal line segment
between the two end-nodes of the arc.
In the first diagram in (\ref{12}), there are thus three arcs of height 1, one arc of height 2, and
one arc of height 3. 

Letting $A$ and $A'$ denote an input and an output link state, respectively, the corresponding
entry of $\Jb$ is denoted by $\Jb_{\!A',A}$. 
We immediately see that $\Jb_{\!A',A}=0$ if $\ell(A)<\ell(A')$ since defects cannot be
created by the action of $\Jb$. Furthermore, the blocks with $\ell(A)=\ell(A')$
appearing on the diagonal of $\Jb$ are themselves diagonal and given by the Sector Lemma
(\ref{SectorLemma}). To complete the description of the matrix realization of $\Jb$,
we therefore turn to the cases where $\ell(A)>\ell(A')$. The Drop-Down Lemma
implies that $arc(A)\subset arc(A')$, which we will assume in the following. 
As discussed in Section~\ref{SecMatrixJ}, the data remaining to
characterize the matrix $\Jb$ can be extracted by examining the input link state $A_\ell$, 
for which $\ell(A_\ell)=\ell$ and $arc(A_\ell)=\emptyset$, for each
of the corresponding system sizes $N_\ell=\ell$ where $\ell=N$ mod 2.

We let $A_\ell'$ denote an output state resulting when $\Jb$ acts on $A_\ell$. 
Since $A_\ell$ consists of defects only, all arcs in $A_\ell'$ are created by $\Jb$. 
This yields three cases distinguished by the height of $A_\ell'$ which can take on the 
values $h(A_\ell')=0,1,2$ since $\Jb_{\!A_\ell',A_\ell}=0$ if $h(A_\ell')>2$.

For $h(A_\ell')=0$, we see that $A_\ell'=A_\ell$ and $\Jb_{\!A_\ell',A_\ell}=(-1)^\ell+1$.

For $h(A_\ell')=1$, we initially assume that $\ell(A_\ell')=0$. 
This presupposes $\ell$ even, and we readily see that 
$\Jb_{\!A_\ell',A_\ell}=(-1)^{\frac{\ell}{2}}\al$. For given $A_\ell'$ with $\ell(A_\ell')>0$,
on the other hand, the contributing terms in the expansion of $\Jb$ have a specified distribution 
of 4-tangles $ud$
\be
 \ldots g_{k_n}^{(n)}(ud)g_1^{(1)}\ldots g_{k_1}^{(1)}(ud)g_1^{(2)}\ldots 
   g_{k_2}^{(2)}(ud)g_1^{(3)}\ldots
\label{gudg}
\ee
where every $g$ is a $u$ or a $d$, and where
\be
 \ell\;=\;\ell'+2n,\qquad\quad \sum_{j=1}^n k_j\;=\;\ell'
\ee
Since $\ell(A_\ell')>0$, there is at least one segment between two neighbouring $ud$'s 
(which can be the same $ud$ due to the cylinder topology) in which a $u$ is not followed 
by a $d$, that is, $n>0$. With reference to (\ref{gudg}), let us assume that
this segment is the one indicated by $g_1^{(1)}\ldots g_{k_1}^{(1)}$.
If $g_{k_1}^{(1)}=u$, we see that $g_1^{(2)}=u$ ($g_1^{(1)}=u$ for $n=1$) 
in order to prevent $h(A_\ell')=2$.
In order to preserve the specified distribution of $(ud)$'s, every $g_i^{(2)}$ in this segment
must be a $u$. Continuing this argument around the cylinder, implies that all the $g$'s
must be $u$'s. If, instead, $g_{k_1}^{(1)}=d$, this 3-tangle could be followed,
on the other side of the $(ud)$ to its right, by a $g$ being a $u$. 
However, this $u$ would force every $g$ to its right to be
a $u$, leading to an inconsistency with $g_{k_1}^{(1)}=d$ once the cycle has been traversed.
We thus conclude that all $g$'s must be $u$'s or all $g$'s must be $d$'s, implying that
$\Jb_{\!A_\ell',A_\ell}=(-1)^{\frac{\ell(A_\ell)-\ell(A_\ell')}{2}}((-1)^\ell+1)$.

For $h(A_\ell')=2$, there is at least one arc of height 2. In the contributing terms in the
expansion of $\Jb$, the 3-tangle containing the leftmost node of this arc must correspond
to a $u$ followed by a $u$ to its immediate right, 
while the 3-tangle containing the rightmost node corresponds to a $d$ preceded by a $d$
to its immediate left, that is,
\be
 \ldots f_{\kappa_\nu}^{(n)}(uu\ldots dd)f_1^{(1)}\ldots f_{\kappa_1}^{(1)}(uu\ldots dd)f_1^{(2)}
   \ldots f_{\kappa_2}^{(2)}(uu\ldots dd)f_1^{(3)}\ldots
\label{gudg2}
\ee
where every $f$ is a $u$ or a $d$, while $\kappa_j,\nu\geq0$. 
For $\nu=0$, which presupposes $\ell$ even, there is only 
one contributing term, and we see that $\Jb_{\!A_\ell',A_\ell}=(-1)^{\frac{\ell}{2}}$.
For $\nu>0$, every segment $dd f_1^{(j)}\ldots f_{\kappa_1}^{(j)}uu$
expands as
\be
 ddg_1^{(j,1)}\ldots g_{k_1}^{(j,1)}(ud)g_1^{(j,2)}\ldots g_{k_2}^{(j,2)}\ldots
  (ud)g_1^{(j,n)}\ldots g_{k_n}^{(j,n)}uu
\label{gg}
\ee
where $k_i,n\geq0$. As in the argument for $h(A_\ell')=1$, a $g=u$ in (\ref{gg})
cannot be followed by a $g=d$ in (\ref{gg}).
Ignoring the 4-tangles $(ud)$, the sequence of $g$'s in (\ref{gg}) can therefore be any of the 
$\kappa_j+1$ sequences
of the form $d\ldots du\ldots u$. Adding up these contributions, while taking into account
the `boundary effects' coming from the sandwiching by $dd\ldots uu$ in (\ref{gg}),
we see that all but one of the terms cancel out pairwise leaving
\be
 \frac{1+(-1)^{\kappa_j}}{2}dd\big(d\ldots d(ud)d\ldots d\ldots (ud)d\ldots d\big)uu
\ee
For $\nu>0$, the `value' for $f_{\kappa_j}^{(j)}$ (that is, $u$ or $d$) does not affect the value
for $f_1^{(j+1)}$ ($f_1^{(1)}$ for $j=\nu=1$).
This independence of the segments implies that $\Jb_{A_\ell',A_\ell}$ is given by the product
of the $\nu$ contributions just obtained (being 0 or 1)
\be
 \Jb_{A_\ell',A_\ell}\;=\;\begin{cases}
  0,\ &\mbox{if any $\kappa_j$ is odd}
 \\[4pt]
  1,\ &\mbox{if all $\kappa_j$ are even}
 \end{cases}
\ee

In summary, $\Jb=0$ for $N$ odd, while, for $N$ even, $\Jb$ is a sparse upper-triangular matrix
whose non-vanishing entries are given by
\be
 \Jb_{A',A}\;=\;(-1)^{\frac{N-\ell'}{2}}\begin{cases}
  2+(\al^2-4)\delta_{\ell',0},\quad & \mathrm{if}\quad A'=A
 \\[2pt]
  2+(\al-2)\delta_{\ell',0},\quad & \mathrm{if}\quad h_{A',A}=1
 \\[2pt]
  1,\quad & \mathrm{if}\quad h_{A',A}=2,\ 
      s_{A',A}\subset 2\mathbb{N}_0
  \end{cases}
\ee
where $\ell'=\ell(A')$ and $h_{A',A}=h(arc(A')\setminus arc(A))$, 
while $s_{A',A}$ is the set of numbers of nodes
between the rightmost node of a height-2 arc of $arc(A')\!\setminus\!arc(A)$ and the leftmost 
node of the next height-2 arc to its right (which, due to the cylinder topology, can be the same
height-2 arc). It is noted, that $h_{A',A}=1,2$ requires that $arc(A)\subsetneq arc(A')$.

For $N$ even, we observe that the sum of the entries of a column of $\Jb$ is given by
\be
 C\;=\;(-1)^{\frac{N}{2}}\begin{cases}
  \al^2-2, &\ \ell=0 \\[2pt]
  2\al-2, &\ \ell=2\ \mathrm{mod}\ N
\end{cases}
\ee
For given $N$, these sums are all equal if and only if $\al=0,2$.

\end{document}